\documentclass[10pt,aps,axodraw,showpacs,nofootinbib,prd,aps,epsf,floats,
               amsmath,amssymb,amsfonts,graphics]{revtex4}
\usepackage{amsmath, amssymb, graphics,axodraw}
\newcommand{\mathsym}[1]{{}}

\usepackage{graphicx}
\usepackage{amsmath}
\usepackage{amssymb}
\usepackage{bm}
\newcommand{\be}{\begin{equation}}
\newcommand{\ee}{\end{equation}}
\newcommand{\bea}{\begin{eqnarray}}
\newcommand{\eea}{\end{eqnarray}}

\newcommand{\rem}[1]{}
\newsavebox{\PSLASH}
 \sbox{\PSLASH}{$p$\hspace{-1.8mm}/}
 
\renewcommand{\theequation}{\thesection.\arabic{equation}}
\newcounter{saveeqn}
\newcommand{\add}{\addtocounter{equation}{1}}
\newcommand{\alpheqn}{\setcounter{saveeqn}{\value{equation}}%
\setcounter{equation}{0}%
\renewcommand{\theequation}{\mbox{\thesection.\arabic{saveeqn}{\alph{equation}}}}}
\newcommand{\reseteqn}{\setcounter{equation}{\value{saveeqn}}%
\renewcommand{\theequation}{\thesection.\arabic{equation}}}

 \newsavebox{\notrightarrow}
 \sbox{\notrightarrow}{$\to$\hspace{-4mm}/}
 
 \newsavebox{\PARTIALSLASH}
 \sbox{\PARTIALSLASH}{$\partial$\hspace{-1.6mm}/}
 
 \newsavebox{\ASLASH}
 \sbox{\ASLASH}{$A$\hspace{-2.1mm}/}
 
 \newsavebox{\KSLASH}
 \sbox{\KSLASH}{$k$\hspace{-1.8mm}/}
 
 \newsavebox{\LSLASH}
 \sbox{\LSLASH}{$\ell$\hspace{-1.8mm}/}
 
 \newsavebox{\QSLASH}
 \sbox{\QSLASH}{$q$\hspace{-1.8mm}/}
 
 \newsavebox{\DSLASH}
 \sbox{\DSLASH}{$D$\hspace{-2.2mm}/}
 
 \newsavebox{\DbfSLASH}
 \sbox{\DbfSLASH}{${\mathbf D}$\hspace{-2.8mm}/}
 
 \newsavebox{\DELVECRIGHT}
 \sbox{\DELVECRIGHT}{$\stackrel{\rightarrow}{\partial}$}
 
 \newcommand{\blue}{\IfColor{\textCadetBlue}{}}
\newcommand{\black}{\IfColor{\textBlack}{}}
\newcommand{\red}{\IfColor{\textRed}{}}
\newcommand{\green}{\IfColor{\textOliveGreen}{}}
\newcommand{\lila}{\IfColor{\textRedViolet}{}}

\begin{document}
\begin{flushright}
0905.2097 [hep-ph]
\end{flushright}
\title{Chiral MHD description of a perfect magnetized QGP using\\ the effective NJL model in a strong magnetic field}

\author{N\'eda Sadooghi}\email{sadooghi@physics.sharif.ir}
\affiliation{Department of Physics, Sharif University of Technology,
P.O. Box 11155-9161, Tehran-Iran}
\begin{abstract}
To study the effect of a strong magnetic field $B$ on the sound
velocity $v_{s}$ of plane waves propagating in a strongly magnetized
quark-gluon plasma (QGP), a chiral magnetohydrodynamical (MHD)
description of a perfect (non-dissipative) QGP exhibiting dynamical
chiral symmetry breaking (D$\chi$SB) is developed using the
effective action of the Nambu-Jona-Lasinio (NJL) model of QCD at
finite temperature, finite baryon chemical potential and in the
presence of a strong magnetic field. Here, the D$\chi$SB arises due
to the phenomenon of magnetic catalysis. Apart from an interesting
frequency dependence, for plane waves propagating in the transverse
or longitudinal direction with respect to the $B$ field, the sound
velocity is anisotropic and depends on the angle between the
corresponding wave vectors and the direction of the $B$ field.
Moreover, for plane waves propagating in the transverse
(longitudinal) direction to the external $B$ field, the sound
velocity has a maximum (minimum) at $T<T_{c}$, reaches a local
minimum (maximum) at $T\sim T_{c}$ and remains constant at $T\gtrsim
T_{c}$. Here, $T_{c}$ is the critical temperature of the chiral
phase transition. Thus, the constant value $v_{s}\sim 1.5 c_{s}$ at
$T \gtrsim T_{c}$ turns out to be a lower (upper) bound for waves
propagating in the transverse (longitudinal) direction with respect
to the external $B$ field. Here, $c_{s}=1/\sqrt{3}$ is the sound
velocity in an ideal gas.
\end{abstract}  \maketitle
\section{Introduction}\label{introduction}
\par\noindent
Hydrodynamics is a framework to describe the time evolution of a
system under local thermal equilibrium (LTE). Its relativistic
version is used extensively in the past years to describe the
phenomena observed in the heavy ion experiments at BNL-RHIC, where
excited nuclear matter has been already created in the Au-Au
collisions with the collision energy of $\sim 100$ GeV per nucleon
\cite{ollitrault, vanderkolk, schaefer}. As it is known from lattice
QCD simulations, for temperatures larger than $T_{c}\sim 170$ MeV,
confined nuclear matter shall build a phase of deconfined plasma of
quarks and gluons - the QGP phase \cite{shuryak80}. Although the
collision energies at RHIC are sufficient to build this phase, the
main goals of the physics of heavy ion collision (HIC) are far more
than only to create the QGP. One would like to discover the
deconfined QGP under thermal and chemical equilibrium in order to
study its static properties such as equation of state, temperature,
transport coefficients, etc. But, since it is known that the system
at RHIC evolves within time duration of the order of $10-100$ fm/c,
it is necessary to study at the same time its dynamical properties.
Filling the large gap between the static and the dynamical aspects
of HIC, relativistic hydrodynamics is one of the most effective
methods able to describe the time evolution of the expanding QGP
\cite{hydro}, and to explore the behavior of nuclear matter in the
vicinity of phase transition point \cite{kodama}. In order for
hydrodynamics to be applicable, the duration of a heavy ion event,
$\tau$, has to be large compared to the equilibration time. It is
widely believed that the QGP produced at RHIC behaves as a nearly
perfect fluid: The transverse radius of an Au nucleus is
approximately $6$ fm and on the order of 7000 particles are produced
overall \cite{schaefer}. The motion of particles is relativistic and
the nuclei are Lorentz contracted by a factor of $\gamma\sim 100$.
The duration of a heavy ion event $\tau\sim 6$ fm. As it is shown in
\cite{hydro}, there is indeed a fascinating agreement between
predictions from ideal relativistic hydrodynamic models with the
experimental data. Whereas plasma instabilities, such as
electromagnetic Weibel instabilities in relativistic shocks can be
made responsible for the possible short time thermalization of the
QGP at RHIC \cite{iwazaki09}, methods based on gauge-string duality
starting from collisions of gravitational shock waves \cite{pufu3-4}
can describe the enormous entropy production in the heavy ion
experiments \cite{gravity-shock}.
\par
Apart from the confinement-deconfinement phase transition, the
chiral symmetry restoration plays an important role in the QCD phase
transition. Recent lattice QCD results \cite{karsch} predict a
critical temperature $\sim 180-200$ MeV for the transition from a
chirally broken to a chirally symmetric phase, the chiral phase
transition. At this critical temperature, following a first order
phase transition, the energy density as a function of temperature is
shown to increase suddenly by $\sim 1$ GeV/fm$^3$. The same
phenomenon is believed to occur during the QCD phase transition in
the Early Universe \cite{kolb}. Here, the phase transition, being of
first order, is usually followed by a mechanism of supercooling
\cite{kolb}. Depending on the degree of supercooling the kinetics of
domain formation and growth can be described by the mechanism of
nucleation or spinodal decomposition. In contrast to the Early
Universe, where the QCD phase transition was driven by nucleation of
bubbles of true vacuum inside the metastable phase, the QGP created
in RHIC expands several orders of magnitude faster than the
primordial Universe, and, bypassing the nucleation process enters
the domain of spinodal decomposition \cite{fraga}. The phenomenon of
\textit{sudden hadronization} is also suggested by several results
from CERN-SPS and BNL-RHIC \cite{sudden-hadron}.
\par
The mechanism responsible for the onset of instabilities in an
expanding QGP at nonzero temperature and baryon chemical potential
is studied in \cite{fraga}. Using the effective potential of the
linear $\sigma$-model as the thermodynamic (effective) potential, a
\textit{chiral hydrodynamic} description of the expanding QGP near
the chiral phase transition is derived (see also \cite{paech}). The
effective potential is determined by integrating out the quark
degrees of freedom that play the role of a heat bath for the chiral
fields $\sigma$ and $\pi$ with temperature $T$ and baryon chemical
potential $\mu$. Instead of using the thermodynamic potential to
calculate the total pressure and energy density and to obtain the
conserved energy-momentum tensor for the expanding perfect fluid,
the authors in \cite{fraga} adopt the variational formulation to
obtain the hydrodynamic equations for their system \cite{koide}.
This approach provides a natural way to merge chiral and fluid
dynamics. Performing a stability analysis up to first order, the
dispersion relation for a plane wave propagating in the above QGP
coupled to chiral fields is derived. Solving the dispersion
relation, the pressure (sound) modes and chiral modes are further
determined.
\par
In the present paper, we will use the same variational method to
study the effect of large magnetic fields on the sound propagation
of a plane wave propagating in an expanding \textit{magnetized} QGP
coupled to chiral fields at finite $(T,\mu)$ and in a strong
magnetic field. For a magnetic field aligned in the third direction,
we will, in particular, determine the \textit{anisotropy in the
velocity} of a plane wave propagating at angles $(\varphi, \theta)$
of spherical coordinates from the external magnetic field. Similar
frequency dependent anisotropy was previously observed in magnetic
fluids in condensed matter physics \cite{parsons}.\footnote{Magnetic
fluid is a colloid of tiny ($100 {\AA}$) magnetic particles or
grains suspended in a carrier fluid such as water. The magnetization
of the fluid varies with the applied magnetic field, typically
reaching a saturation of $10^{2}-10^{3}$ Gau\ss$\  $
\cite{parsons}.} To mimic the hot and dense QGP in a strong magnetic
field, we will use the one-loop effective action of the Nambu-Jona
Lasinio (NJL) model of QCD \cite{njl, klevansky} at finite $(T,\mu)$
and in a strong magnetic field $\mathbf{B}$. As it is shown in
\cite{miransky-NPB-95}, the one-loop effective action of the NJL
model exhibits a dynamical chiral symmetry breaking due to the
well-known phenomenon of magnetic catalysis. According to this
phenomenon, in the limit of strong magnetic fields, NJL dynamics is
dominated by the lowest Landau level (LLL), where even at the
weakest attractive interaction between fermions the chiral symmetry
of the theory is broken by a dynamically generated fermion mass.
This mass is shown to depend on $(T,\mu)$ and the strength of the
applied magnetic field.
\par
The magnetic catalysis has applications in both cosmology
\cite{cosmology} and condensed matter physics \cite{condensed}, and
seems to be also relevant in the physics of heavy ion collision: As
it is reported in \cite{warringa, kharzeev-chiral,
kharzeev-summary}, in off-central collisions, heavy ions possess a
very large relative angular momentum and create very strong magnetic
fields. In this situation, the presence of topological charge was
predicted to induce the charge separation with respect to the
reaction plane. Large magnetic fields play therefore an important
role in the physics of non-central heavy ion collisions and provide
a possible signature of the presence of CP-odd domains in the
presumably formed QGP phase \cite{warringa, kharzeev-chiral,
kharzeev-summary}. The detailed theory of this \textit{chiral
magnetic effect} describing the interplay between the chiral charge
and the background magnetic field has been developed in
\cite{kharzeev-chiral}.\footnote{In \cite{kharzeev50-voloshin} an
experimental observable sensitive to the chiral magnetic effect has
been proposed and the first preliminary results have been reported
by STAR Collaborations \cite{kharzeev51-STAR}. As it is suggested in
\cite{kharzeev-summary}, the generation of chirality in the QGP,
responsible for the charge separation, is the QCD counterpart of the
generation of baryon asymmetry in the electroweak phase transition
in the Early Universe \cite{sakharov, shaposhnikov}.} Relativistic
shock waves, that are believed to be build in the heavy ion
collisions, can be viewed as a possible origin of the magnetic field
generation in the heavy ion collision. The mechanism is well-known
from astrophysics, where the generation of large magnetic fields in
relativistic shocks plays an important role in the fire-ball model
for Gamma-ray Bursts \cite{wiersma}. This model proposes that the
non-thermal radiation observed in the prompt and afterglow emission
from the Gamma-ray Bursts is synchrotron radiation from
collisionless relativistic shocks. In collisionless shocks, plasma
instabilities can generate magnetic fields. Within the shock
transition layer the relative motion of the mixing pre- and
post-shock plasma produces \textit{very anisotropic velocity
distributions} for all particle species concerned. Fluctuating
electromagnetic fields deflect the incoming charged particles and
act as the effective collisional process needed to complete the
shock transition. These fluctuating fields occur naturally because
anisotropic velocity distributions are unstable against several
plasma instabilities, such as electromagnetic Weibel instability
\cite{weibel}. The latter is an instability of the currents that
result from charge bunching in the beams, and leads to spontaneously
growing transverse waves with $|\textbf{B}|\geq |\textbf{E}|$. In a
relativistic shock, where the velocity of the pre- and post-shock
plasma approaches the velocity of light, the Weibel instability
dominates, because it has the largest growth rate. In heavy ion
collisions, Weibel instabilities are made responsible for the short
equilibration time of the system from the initial colliding stage
\cite{iwazaki2}.\footnote{Recently, an alternative scenario based on
Nielsen-Olesen instability \cite{nielsen-olesen} which is
characteristic for the configuration of a uniform magnetic field is
introduced in \cite{iwazaki09}.}
\par
The effect of a strong magnetic field to modify the nature of the
chiral phase transition in QCD is recently studied in
\cite{fraga-magnetic, ayala09, campanelli}. In
\cite{fraga-magnetic}, using the one-loop effective potential of the
linear $\sigma$-model, the hot and dense QGP is simulated and it is
shown that for high enough magnetic fields, comparable to the ones
expected to be created in non-central high energy heavy ion
collisions at RHIC, the original crossover is turned into a first
order transition. In \cite{ayala09}, the finite-temperature
effective potential of linear $\sigma$-model in the presence of
constant and weak magnetic field, including the contribution of the
pion ring diagrams in the background of constant classical
$\sigma$-fields is calculated. It is shown that there is a region of
the parameter space where the effect of ring diagrams is to preclude
the phase transition from happening. Inclusion of magnetic field has
small effects and becomes more important as the system evolves to
the lowest temperatures.\footnote{The effect of ring diagrams on the
dynamical chiral phase transition of QED in the presence of strong
magnetic field is studied in \cite{sadooghi-sohrabi}.} In
\cite{campanelli}, the response of the QCD vacuum to an external
Abelian chromomagnetic field in the framework of a nonlocal NJL
model with Polyakov loop is studied and a linear relationship
between the deconfinement temperature and the squared root of the
applied magnetic field is found.
\par
The next question is how the magnetic field would affect the
hydrodynamical quantities of an expanding \textit{magnetized} QGP.
One of these quantities is the velocity of sound modes of a
propagating plane wave in this medium. We investigate this question
in the present paper. Further, we are interested on the
hydrodynamical signatures of a chiral phase transition once the
temperature reaches the critical temperature $T_{c}$. Previous
studies on the temperature dependence of sound velocity $v_{s}$
shows that at $T=T_{c}$ the sound waves has minimum velocity
\cite{florkowski}. As it will be shown in Sect. \ref{numerics}, at
finite $(T,\mu)$, apart from an anisotropy in the sound velocity
with respect to the direction of the magnetic field, the same
behavior occurs for a plane wave propagating in the transverse
direction to the external magnetic field. In contrast, for the waves
propagating in the longitudinal direction to the magnetic field, the
sound velocity has a minimum for $T<T_{c}$, reaches its maximum at
$T\sim T_{c}$ and remains constant after the temperature passes the
chiral critical point, i.e. for $T\gtrsim T_{c}$. On the other hand,
at nonzero baryon density, the sound waves seems to die out, as
$v_{s}$ has a real and an imaginary part. Whereas
$\mathfrak{Re}(v_{s})$ has the same behavior as $v_{s}$ for $\mu=0$,
the $\mathfrak{Im}(v_{s})$ is several orders of magnitude smaller
than $\mathfrak{Re}(v_{s})$ and oscillates. This behavior which can
be interpreted as the onset of the aforementioned spinodal effects
is also observed recently in \cite{minami}, where the dynamical
density fluctuations are studied around the QCD critical point using
dissipative relativistic fluid dynamics with no chiral fields
included. This is interpreted experimentally as a possible fate of
Mach cone at the chiral critical point. The disappearance or
suppression of the Mach cone would be a signal that the created
matter has passed through the critical region, showing the existence
of the QCD critical point \cite{minami}.
\par
The paper is organized as follows: In the first part of the paper
(Sects. \ref{njl} and \ref{effective}), we will derive the effective
Lagrangian density of the NJL model, consisting of the effective
kinetic and the one-loop effective potential of the theory at finite
$(T,\mu)$ and in the presence of strong magnetic field $\mathbf{B}$.
This part can be viewed as a generalization of the results presented
in \cite{miransky-NPB-95, ebert} and the methods introduced in
\cite{meisinger} and \cite{mellin} to the case of nonzero chemical
potential. The effective potential and effective kinetic terms are
determined in Sects. \ref{effpot} and  \ref{kinetic}, respectively.
The effective kinetic term is in particular determined using a
derivative expansion following the standard effective field theory
methods presented in \cite{miransky-old}. Solving the corresponding
gap equation that arises from one-loop effective potential of Sect.
\ref{effpot}, the dynamical mass of the NJL model generated in the
presence of strong magnetic fields at finite $(T,\mu)$ is calculated
in Sect. \ref{dynamical}. Note that the dynamical mass, being the
configuration that minimizes the effective potential, can be viewed
as the equilibrium configuration, once the instabilities in the
magnetized QGP are set on. Comparing to the effective action of the
linear $\sigma$-model used in \cite{fraga}, both the effective
kinetic term and the minimum of the effective potential of the NJL
model in a strong magnetic field depend on $(T,\mu)$ and the
constant magnetic field. To have a link to hydrodynamics, we will
determine in Sect. \ref{effective} the energy-momentum tensor
$T^{\mu\nu}$ of the effective NJL model in the presence of a strong
magnetic field using the effective Lagrangian density from Sect.
\ref{njl}. Here, we will introduce a polarization tensor
$M^{\mu\nu}$ containing the magnetization $M$ of the medium. The
same tensor appears also in \cite{de-groot, sachdev}, where a MHD
description of an expanding dissipative fluid is introduced to study
the Nernst effect in the vicinity of superfluid-insulator transition
of condensed matter physics.\footnote{In condensed matter physics,
the Nernst coefficient measures the transverse voltage arising in
response to an applied thermal gradient in the presence of a
magnetic field \cite{sachdev}.}
\par
In the second part of the paper (Sects. \ref{hydro}-\ref{numerics}),
we will use the method introduced in \cite{fraga} and the results
from the first part of the paper to present a \textit{chiral
magnetohydrodynamic} description of a magnetized perfect
(non-dissipative) QGP coupled to chiral fields of the effective NJL
model. This is in contrast to the work done in \cite{sachdev} and
\cite{buchel} where in the presence of a weak magnetic field, the
magnetized fluid does not involve any chiral fields. In Sect.
\ref{thermo}, we will first generalize the thermodynamic relations
to the case of nonzero magnetic field. In Sect. \ref{mhd}, we
describe the variational method used in \cite{fraga} by comparing
first the polarization tensor $M^{\mu\nu}$ of the NJL model with the
expected magnetization of the magnetized medium (Sect. \ref{pol}),
and then the energy-momentum tensor of the effective NJL model with
the corresponding $T^{\mu\nu}$ of an expanding QGP coupled to chiral
fields (Sect. \ref{energy}). In Sect. \ref{stability}, using the
hydrodynamical equations, the energy-momentum conservation relation
and the conservation relations for the number and entropy densities,
we will perform a first order stability analysis and derive the
corresponding dispersion relation of the magnetized QGP coupled to
chiral fields $\sigma$ and $\pi$. The sound velocity is then derived
in Sect. \ref{sound}. We will use the method in \cite{parsons} to
determine the anisotropy in the sound velocity. In our specific
model studied in this paper, a wave propagating in a plane
transverse to the external magnetic field, the anisotropy is
independent of the angle $\varphi$ of the spherical coordinate
system. On the other hand, a $\theta$-dependence arises in the
anisotropy function, when the propagating wave is in the
longitudinal plane with respect to the direction of the magnetic
field. In Sect. \ref{numerics}, performing a numerical analysis, we
will visualize our results in several figures. Our results are
summarized in Sect. \ref{discussion}. In Appendix \ref{appA}, we
will generalize the Bessel-function identities appearing in the high
temperature expansion of Matsubara sums of finite temperature field
theory presented in \cite{meisinger} to the case of finite chemical
potential. In Appendix \ref{mellin} a generalization of the Mellin
transformation \cite{mellin} for the case of nonzero chemical
potential is presented. A detailed derivation of thermodynamic
relations used in Sect. \ref{stability} is presented in Appendix
\ref{derivation}.
\section{NJL model at finite $T,\mu$ and in a strong magnetic
field}\label{njl}
\par\noindent
\setcounter{equation}{0} We start with the action of the NJL model
in $3+1$ dimensions at zero temperature and zero baryonic density
\begin{eqnarray}\label{N1}
{\cal{L}}=\frac{1}{2}\big[\bar{\psi},(i\gamma^{\mu}D_{\mu})
\psi\big]+\frac{G}{2}\big[\left(\bar{\psi}\psi\right)^{2}+\left(\bar{\psi}i\gamma^{5}\psi\right)^{2}\big]
-{\cal{F}},
\end{eqnarray}
where ${\cal{F}}\equiv \frac{1}{4}(F_{\mu\nu})^{2}$ is the gauge
kinetic term. Defining the covariant derivative by $D_{\mu}\equiv
\partial_{\mu}-ieA_{\mu}^{ext}$, with
the gauge field in the symmetric gauge
$A_{\mu}^{ext}=\frac{1}{2}\left(0,-Bx_{2},Bx_{1},0\right)$, a
constant magnetic field aligned in the $x_{3}$-direction is
generated. In the above symmetric gauge, the only non-vanishing
elements of the field-strength tensor $F_{\mu\nu}$ are
$F_{12}=-F_{21}=B$ and $F_{\mu\nu}=F^{\mu\nu}$. We have therefore
${\cal{F}}=\frac{B^{2}}{2}$. Note, that (\ref{N1}) is chirally
invariant under $U_{L}(1)\times U_{R}(1)$ group transformation. As
it is described in \cite{miransky-NPB-95}, the above theory is
equivalent to the bosonized Lagrangian density
\begin{eqnarray}\label{N3}
{\cal{L}}=\frac{1}{2}\big[\bar{\psi},(i\gamma^{\mu}D_{\mu})
\psi\big]-\bar{\psi}\left(\sigma+i\gamma^{5}\pi\right)\psi-\frac{1}{2G}\left(\sigma^{2}+\pi^{2}\right)-{\cal{F}},
\end{eqnarray}
where the Euler-Lagrange equations for the auxiliary fields $\sigma$
and $\pi$ take the form of constraint equations
\begin{eqnarray}\label{N4}
\sigma=-G\left(\bar{\psi}\psi\right),\qquad\mbox{and}\qquad
\pi=-G\left(\bar{\psi}i\gamma^{5}\psi\right).
\end{eqnarray}
Integrating out the fermion degrees of freedom, we obtain the
effective action for the composite fields $\vec{\rho}=(\sigma,\pi)$
at finite temperature $T$ and density $\mu$ and in the presence of a
constant magnetic field $eB$.\footnote{It can be shown that the
presence of a constant magnetic field, defines a natural scale
$\ell_{B}\equiv \frac{1}{\sqrt{eB}}$, where $e$ is the
electromagnetic coupling constant.} It is given by
\begin{eqnarray}\label{N5}
\Gamma[\vec{\rho}; eB,\mu,T]=\int d^{4}x\
{\cal{L}}_{\mbox{\tiny{eff}}}(\vec{\rho};eB,T,\mu)=\int d^{4}x
\left({\cal{L}}_{k}-\Omega-{\cal{F}}\right).
\end{eqnarray}
In the following, we will first introduce the effective potential
$\Omega$, and then determine the kinetic term ${\cal{L}}_{k}$ using
the method introduced in \cite{miransky-NPB-95}.
\subsection{The effective potential of the NJL model at finite $T,\mu$  in a strong magnetic
field}\label{effpot}
\par\noindent
The effective potential of the NJL model consists of a tree level
part and a one-loop part. The latter is determined in
\cite{miransky-NPB-95} at zero temperature and density, and in the
presence of a constant magnetic field. It is expressed through the
path integral over fermions
\begin{eqnarray}\label{N6}
\exp\left({i\tilde{\Gamma}}(\vec{\rho};eB)\right)&=&\int{\cal{D}}\bar{\psi}{\cal{D}}\psi
\exp\left(\frac{i}{2}\big[\bar{\psi},(i\gamma^{\mu}D_{\mu}-\sigma-i\gamma_{5}\pi)
\psi\big]\right)\nonumber\\
&=&\exp\left(\mbox{Tr}\ln\left(i\gamma^{\mu}D_{\mu}-\sigma-i\gamma_{5}\pi\right)\right).
\end{eqnarray}
Using the definition of Tr$\left(\cdots\right)$, $\tilde{\Gamma}$
can be given by
\begin{eqnarray}\label{N7}
\tilde{\Gamma}=\int d^{4}x\ V^{(1)}(\vec{\rho}; eB, T=\mu=0)
,\qquad\mbox{where}\qquad
V^{(1)}=-i\mbox{tr}\ln\left(i\gamma^{\mu}D_{\mu}-\sigma-i\gamma_{5}\pi\right).
\end{eqnarray}
Here, tr$\left(\cdots\right)$ is to be built only over the internal
degrees of freedom. It can be further evaluated using the
Schwinger's proper-time formalism \cite{schwinger, miransky-NPB-95}.
The effective potential of the NJL model at zero $T$ and $\mu$,
$\Omega$ including the tree level contribution $V^{(0)}$, and
one-loop effective potential $V^{(1)}$, is therefore given by
\begin{eqnarray}\label{N8}
\Omega(\vec{\rho};eB, T=\mu=0)=V^{(0)}(\vec{\rho};eB,
T=\mu=0)+V^{(1)}(\vec{\rho};eB, T=\mu=0),
\end{eqnarray}
where
\begin{eqnarray}\label{N9}
V^{(0)}=\frac{\rho^{2}}{2G},\qquad\mbox{and}\qquad
V^{(1)}=\frac{eB}{8\pi^{2}}\int_{\frac{1}{\Lambda^{2}}}^{\infty}\frac{ds}{s^{2}}
\coth(eBs)e^{-s\rho^{2}}.
\end{eqnarray}
Here, $\rho^{2}=\sigma^{2}+\pi^{2}$. In the limit of very large
magnetic field $eB\to\infty$,\footnote{The strong magnetic field
limit is characterized by a comparison between $eB$ and the momenta
of the particles included in the theory. In this limit
$|{\mathbf{k}}_{\|}|, |{\mathbf{k}}_{\perp}|\ll \sqrt{eB}$, where,
$|{\mathbf{k}}_{\|}| and |{\mathbf{k}}_{\perp}|$ are the
longitudinal and transverse momenta with respect to the direction of
the constant magnetic field, respectively.} the UV cutoff $\Lambda$
can be replaced by $\Lambda\to \Lambda_{B}\equiv \sqrt{eB}$, and
$\coth(eBs)\approx 1$. The integral over $s$ in (\ref{N9}) can
therefore be performed using
\begin{eqnarray}\label{N10}
\Gamma(n,z)=\int_{z}^{\infty} dt\ t^{n-1} e^{-t}.
\end{eqnarray}
It leads to
\begin{eqnarray}\label{N11}
\Omega(\vec{\rho};eB,
T=\mu=0)&=&\frac{\rho^{2}}{2G}+\frac{\rho^{2}eB}{8\pi^{2}}\Gamma\left(-1,\frac{\rho^{2}}{\Lambda_{B}^{2}}\right)
\nonumber\\
&\stackrel{eB\to\infty}{\approx}&\frac{\rho^{2}}{2G}+
\frac{(eB)^{2}}{8\pi^{2}}-\frac{\rho^{2}eB}{8\pi^{2}}\left(1-\gamma_{E}-\ln\left(\frac{\rho^{2}}{eB}\right)\right)+{\cal{O}}
\left(\frac{\rho^{2}}{eB}\right),
\end{eqnarray}
where the asymptotic expansion of $\Gamma(-1,z)\approx
\frac{1}{z}-1+\gamma_{E}+\ln z+{\cal{O}}(z)$ for $z\to\infty$ is
used. Here, $\gamma_{E}\simeq 0.577$ is the Euler-Mascheroni number.
\par
Similar methods can be used to determine the effective potential of
the NJL model at finite temperature and density, and in the presence
of a constant magnetic field \cite{miransky-1,sato}
\begin{eqnarray}\label{N12}
\Omega\left(\vec{\rho};eB,T,\mu\right)=\frac{\rho^{2}}{2G}
+\frac{\mbox{2 eB}}{\beta}\int\limits_{0}^{\infty}ds
\frac{\Theta_{2}\left(\frac{2\pi\mu s}{\beta}|\frac{4\pi
is}{\beta^{2}}\right)}{\left(4\pi s\right)^{(D-1)/2}}\
\coth\left(seB\right)e^{-s\left(\rho^{2}-\mu^{2}\right)}.
\end{eqnarray}
Here, $\beta$ is the inverse temperature, $\beta\equiv \frac{1}{T}$,
and
\begin{eqnarray}\label{N13}
\Theta_{2}(u|\tau)\equiv
2\sum\limits_{n=0}^{\infty}e^{i\pi\tau\left(n+\frac{1}{2}\right)^{2}}\cos\left(\left(2n+1\right)u\right),
\end{eqnarray}
is the elliptic $\Theta$-function of second kind. Using the identity
\cite{miransky-1}
\begin{eqnarray}\label{N14}
\Theta_{2}(u|\tau)=\left(\frac{i}{\tau}\right)^{1/2}e^{-\frac{iu^{2}}{\pi\tau}}\Theta_{4}\left(\frac{u}{\tau}|
-\frac{1}{\tau}\right),
\end{eqnarray}
where
\begin{eqnarray}\label{N15}
\Theta_{4}(u|\tau)=1+2\sum\limits_{n=1}^{\infty}(-1)^{n}\ e^{i\pi
n^{2}\tau}\cos\left(2nu\right),
\end{eqnarray}
is the fourth Jacobian $\Theta$-function, and setting
$\coth(eBs)\approx 1$ in the strong magnetic field limit $eB\to
\infty$,\footnote{At finite temperature $T$, the strong magnetic
field limit is characterized by $T^{2}\ll eB$.} the one-loop
effective potential in (\ref{N12}) can be separated into two parts,
\begin{eqnarray}\label{N16}
V^{(1)}(\vec{\rho};eB,T,\mu)=V^{(1)}(\vec{\rho};eB,T=\mu=0)+V^{(1)}(\vec{\rho};eB,T\neq
0, \mu\neq 0).
\end{eqnarray}
The effective potential $\Omega(\vec{\rho};eB,T,\mu)$ including the
tree level $V^{(0)}$, and the one-loop effective potential $V^{(1)}$
is therefore given by
\begin{eqnarray}\label{N17}
\Omega(\vec{\rho};eB,T,\mu)=\frac{\rho^{2}}{2G}+\frac{eB}{8\pi^{2}}\int_{0}^{\infty}
\frac{ds}{s^{2}}\
e^{-s\rho^{2}}\bigg[1+2\sum\limits_{n=1}^{\infty}(-1)^{n}e^{-\frac{\beta^{2}n^{2}}{4s}}\cosh(n\mu\beta)\bigg].
\end{eqnarray}
To evaluate the integration over $s$, the integral (\ref{N10}) and
\begin{eqnarray}\label{N18}
\int\limits_{0}^{\infty}dx\
x^{\nu-1}\exp\left(-\frac{\beta}{x}-\gamma
x\right)=2\left(\frac{\beta}{\gamma}\right)^{\frac{\nu}{2}}K_{\nu}\left(2\sqrt{\beta\gamma}\right),
\end{eqnarray}
can be used. The effective potential (\ref{N17}) is therefore given
by
\begin{eqnarray}\label{N19}
\Omega(\vec{\rho};eB,T,\mu)&=&\frac{\rho^{2}}{2G}+\frac{\rho^{2}eB}{8\pi^{2}}
\Gamma\left(-1,\frac{\rho^{2}}{\Lambda_{B}^{2}}\right)+\frac{\rho
eB}{\beta\pi^{2}}
\sum\limits_{\ell=1}^{\infty}\frac{(-1)^{\ell}}{\ell}K_{1}\left(\beta\ell\rho\right)
\coth\left(\ell\mu\beta\right),
\end{eqnarray}
where $\rho\equiv |\vec{\rho}|$. The $(T,\mu)$ independent part of
(\ref{N19}) is exactly the same as (\ref{N11}) and can be similarly
evaluated in the strong magnetic field limit. As for the $(T,\mu)$
dependent part, it can be expanded in the orders of $(\rho\beta)$,
using the Bessel function identities from Appendix \ref{appA}. In
particular, we will use the identity
\begin{eqnarray}\label{N20}
\lefteqn{\sum\limits_{\ell=1}^{\infty}\frac{(-1)^{\ell}}{\ell}K_{1}(\ell
z)\cosh(\ell z')=\frac{1}{8}z\bigg[1-2\gamma_{E}-2\ln\left(\frac{z}{\pi}\right)}\nonumber\\
&&-\sum\limits_{n=1}^{\infty}
\frac{(-1)^{n}(1+n)}{(\Gamma(2+n))^{2}}
\left(\frac{z}{4\pi}\right)^{2n}\left\{
\bigg|\psi^{(2n)}\left(\frac{1}{2}-\frac{iz'}{2\pi}\right)\bigg|+
\bigg|\psi^{(2n)}\left(\frac{1}{2}+\frac{iz'}{2\pi}\right)\bigg|
\right\}\bigg]\nonumber\\
&&-\frac{z^{2}}{3}\sum\limits_{k=1}^{\infty}\frac{(-1)^{k}(2^{2k+1}-1)}{2^{2k+1}}\left(\frac{z'}{\pi}\right)^{2k}\zeta(2k+1)
+\frac{1}{2z}\left\{\mbox{Li}_{2}\left(e^{i(\pi+iz')}\right)+\mbox{Li}_{2}\left(e^{-i(\pi+iz')}\right)\right\},
\end{eqnarray}
with $z=\rho\beta$ and $z'=\mu\beta$ from (\ref{AA23}) to evaluate
the sum over $\ell$ in (\ref{N19}). In (\ref{N20}), the polygamma
function $\psi(z)\equiv \frac{d}{dz}\ln\Gamma(z)$, and
$\psi^{(m)}(z)\equiv \frac{d^{m}\psi(z)}{dz^{m}}$ is the
$(m+1)^{th}$ logarithmic derivative of the $\Gamma(z)$-function.
Furthermore, the dilogarithm function $\mbox{Li}_{2}(z)$ is defined
by
\begin{eqnarray}\label{N21}
\mbox{Li}_{2}(z)\equiv\sum_{k=1}^{\infty}\frac{z^{k}}{k^{2}},
\end{eqnarray}
and satisfies
\begin{eqnarray}\label{N22}
\mbox{Li}_{2}(z)=-\int\limits_{0}^{z}\frac{\ln(1-t)}{t}\ dt.
\end{eqnarray}
The effective potential of the NJL model (\ref{N19}) at finite
$(T,\mu)$ and in the limit of $eB\to \infty$ is therefore given by
\begin{eqnarray}\label{N23}
\lefteqn{\Omega(\vec{\rho};eB,T,\mu)\stackrel{eB\to\infty}{\approx}\frac{\rho^{2}}{2G}+
\frac{(eB)^{2}}{8\pi^{2}}-\frac{\rho^{2}eB}{8\pi^{2}}\left(\gamma_{E}+\ln\left(\frac{eB\beta^{2}}{\pi^{2}}\right)
\right)
}\nonumber\\
&&-\frac{eB}{8\pi^{2}\beta^{2}}\sum\limits_{n=1}^{\infty}
\frac{(-1)^{n}}{(16\pi^{2})^{n}}\frac{(n+1)(\rho\beta)^{2(n+1)}}{(\Gamma(2+n))^{2}}
\left\{
\bigg|\psi^{(2n)}\left(\frac{1}{2}-\frac{i\mu\beta}{2\pi}\right)\bigg|+
\bigg|\psi^{(2n)}\left(\frac{1}{2}+\frac{i\mu\beta}{2\pi}\right)\bigg|
\right\}\nonumber\\
&&-\frac{eB(\rho\beta)^{3}}{3\pi^{2}\beta^{2}}\sum\limits_{k=1}^{\infty}
\frac{(-1)^{k}(2^{2k+1}-1)}{2^{2k+1}}\left(\frac{\mu\beta}{\pi}\right)^{2k}\zeta(2k+1)
+\frac{eB}{2\pi^{2}\beta^{2}}\left\{\mbox{Li}_{2}\left(e^{i(\pi+i\mu\beta)}\right)
+\mbox{Li}_{2}\left(e^{-i(\pi+i\mu\beta)}\right)\right\}.\nonumber\\
\end{eqnarray}
It exhibits a spontaneous symmetry breaking, which is mainly due to
a dynamical mass generation in the regime of lowest Landau level
(LLL) dominance. The type of the phase transition as well as the
critical temperature depends on the strength of the magnetic field
$eB$. Fig. 1 shows two examples of $\Omega$ for a) $eB=10$ GeV$^2$
which is equivalent to $B\sim 10^{21}$ Gau\ss,\footnote{As it is
shown in \cite{sadooghi-sohrabi}, eB in GeV$^{2}$ is equivalent to
$B=1.691\times 10^{20}$ Gau\ss.} and for b) $eB=10^{-4}$ GeV$^2$
which is equivalent to $B\sim 10^{16}$ Gau\ss. In both cases the
potential $\Omega$ from (\ref{N23}) is calculated up to
${\cal{O}}((\rho\beta)^{7})$ and ${\cal{O}}((\mu\beta)^{7})$, and
$\mu\beta$ is chosen to be $\mu=10^{-3}$ GeV. Let us note that the
qualitative picture that arises here does not depend too much on
$\mu$.
\par
In the next section, the effective potential (\ref{N23}) will be
used to determine the dynamical mass, that is generated due to the
phenomenon of magnetic catalysis in the strong magnetic field limit
\cite{miransky-NPB-95}.
\begin{center}
\begin{figure}[hbt]
\includegraphics[width=7.5cm, height=5.5cm]{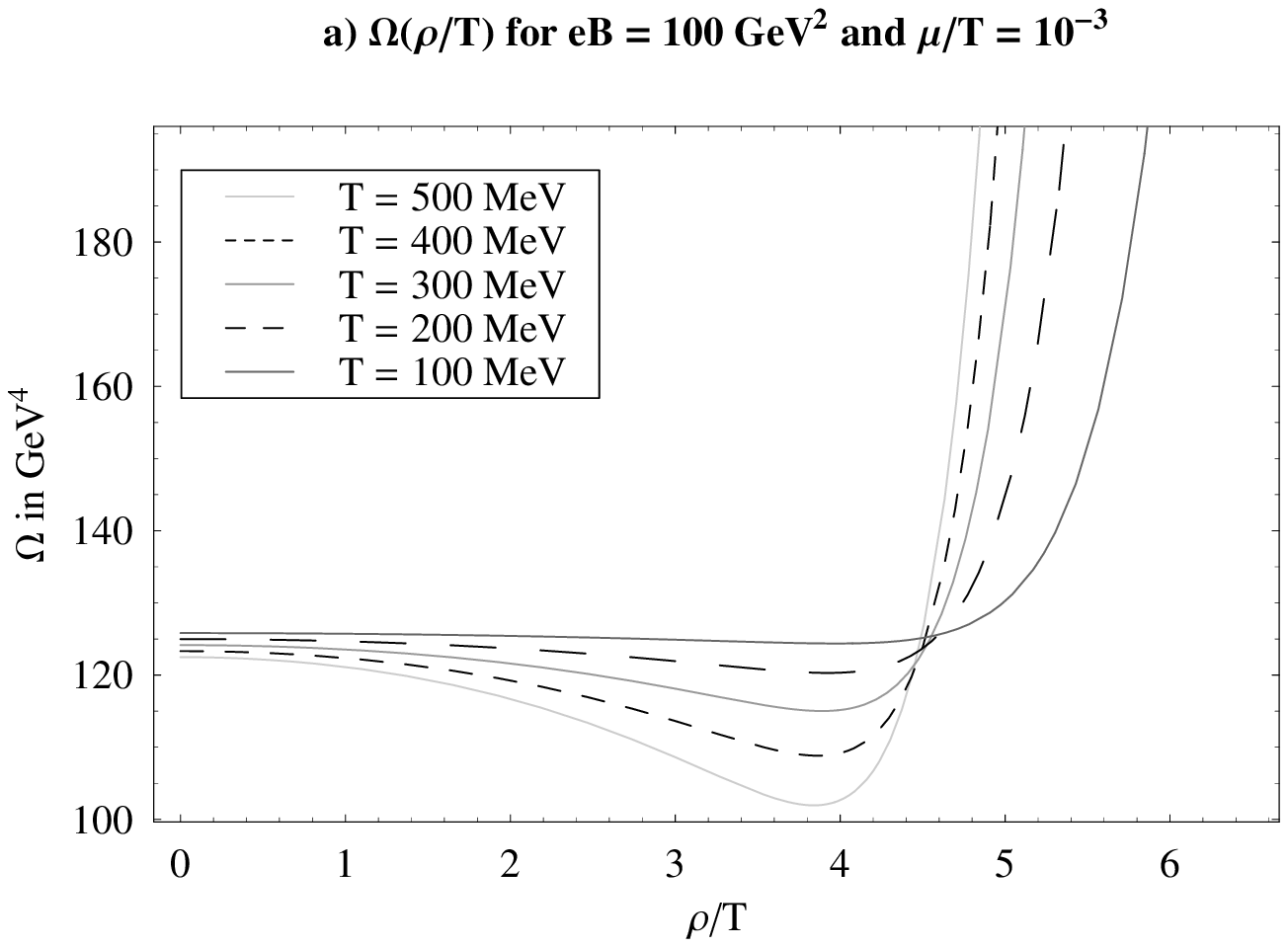}
\includegraphics[width=7.5cm, height=5.5cm]{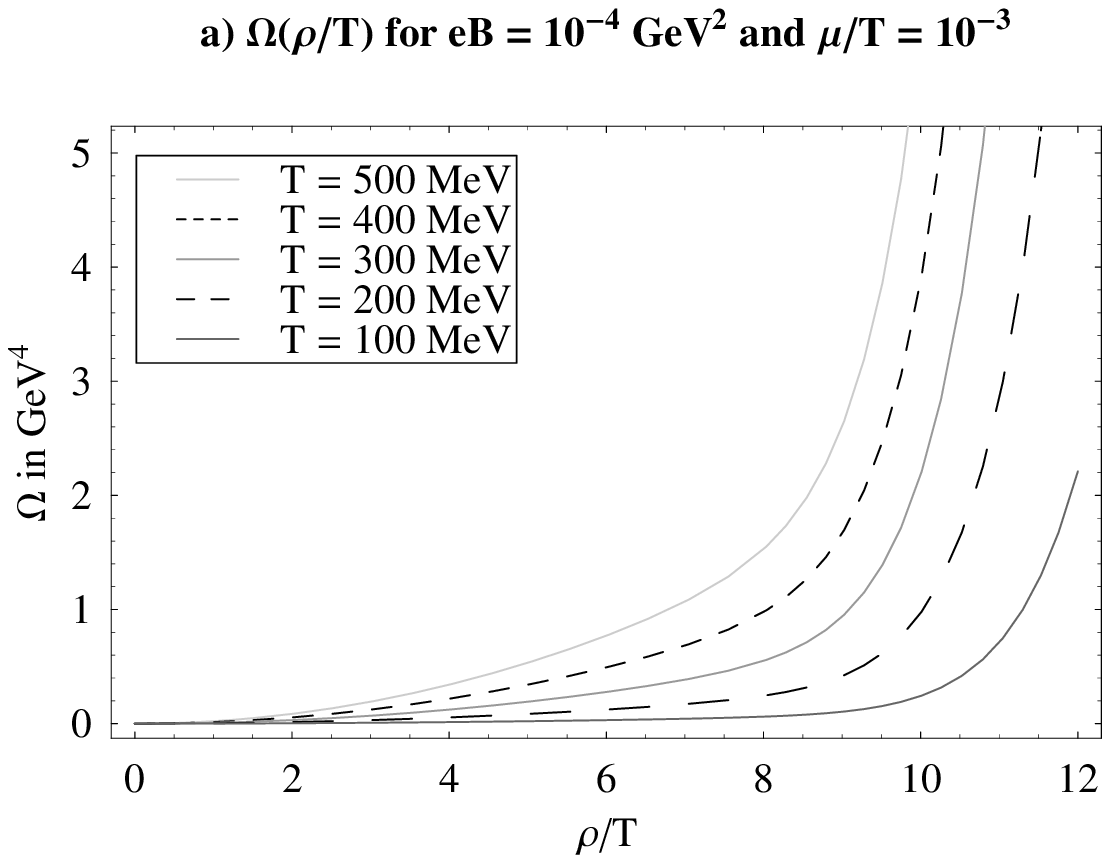}
\caption{a) Effective potential $\Omega$ of the NJL model as a
function of $\rho/T$ for $eB=10$ GeV$^{2}$ and $\mu\beta=10^{-3}$.
Here, a second order chiral phase transition occurs at $T_{c}\simeq
100$ MeV (far right). b) Effective potential $\Omega$ of the NJL
model as a function of $\rho/T$ for $eB=10^{-4}$ GeV$^{2}$ and
$\mu\beta=10^{-3}$. Here, a smooth crossover occurs at $T_{c}\simeq
100$ MeV (far right). Here, the NJL coupling $G=5.86$ GeV$^{-2}$ and
the sum over $n$ in (\ref{N23}) is up to $n,k=7$. }
\end{figure}
\end{center}
\subsection{Dynamical mass of the NJL model at finite $T,\mu$  in a strong magnetic
field}\label{dynamical}
\par\noindent
As it is shown in \cite{miransky-NPB-95}, a constant magnetic field
in $3+1$ dimensions is a strong catalyst of dynamical chiral
symmetry breaking. This phenomenon leads to the generation of a
fermion dynamical mass even at the weakest attractive interaction
between the fermions. The essence of this effect is a dimensional
reduction from $D\to D-2$ in the dynamics of fermion pairing in a
magnetic field. At zero temperature and vanishing chemical
potential, the dynamical mass can be determined using the effective
potential
$$\Omega(\vec{\rho};eB,T=\mu=0)=\frac{\rho^{2}}{2G}+\frac{\rho^{2}eB}{8\pi^{2}}\left(-1,\frac{\rho^{2}}{eB}\right),$$
from (\ref{N11}) through the
gap equation
\begin{eqnarray}\label{N24}
\frac{\partial \Omega(\vec{\rho};eB,T=\mu=0)}{\partial
\sigma}\bigg|_{\sigma_{0}\neq 0, \pi_{0}=0}=0.
\end{eqnarray}
Here, the minimum of the potential is assumed to be at unknown but
constant $\sigma_{0}$ and vanishing $\pi_{0}$. The gap equation
(\ref{N24}) leads to
\begin{eqnarray}\label{N25}
\frac{\sigma_{0}}{G}=\frac{\sigma_{0}eB}{4\pi^{2}}\Gamma\left(0,\frac{\sigma_{0}^{2}}{eB}\right),
\end{eqnarray}
that, after excluding the trivial solution $\sigma_{0}=0$, and using
the approximation $\lim\limits_{z\to 0}\Gamma(0,z)\simeq
-\gamma_{E}-\ln z$, leads to the non-vanishing dynamical mass at
zero temperature and chemical potential
\begin{eqnarray}\label{N26}
\sigma_{0}(eB,T=\mu=0)={\cal{C}}_{m}\sqrt{eB}\exp\left(-\frac{4\pi^{2}}{GeB}\right),\qquad\mbox{with}\qquad
{\cal{C}}_{m}=e^{-\gamma_{E}/2}\simeq 0.74.
\end{eqnarray}
The same result arises also in \cite{ebert}. Next, we will determine
the mass gap at finite $(T,\mu)$ and in the strong magnetic field
limit. Using $\Omega\left(\vec{\rho};eB,T,\mu\right)$ from
(\ref{N17}), we arrive first at the gap equation
\begin{eqnarray}\label{N27}
0=\frac{\partial \Omega(\vec{\rho};eB,T,\mu)}{\partial
\sigma}\bigg|_{\sigma_{0}\neq 0,
\pi_{0}=0}=\frac{\sigma_{0}}{G}-\frac{\sigma_{0}eB}{4\pi^{2}}\left\{
\Gamma\left(0,\frac{\sigma_{0}^{2}}{eB}\right)+4\sum\limits_{\ell=1}^{\infty}(-1)^{\ell}\cosh(\mu\beta\ell)
K_{0}(\beta\ell\sigma_{0}) \right\}.\nonumber\\
\end{eqnarray}
The temperature independent part of (\ref{N27}) can be evaluated as
in (\ref{N25}), leading to the iterative solution
\begin{eqnarray}\label{N28}
\sigma_{0}(eB,T,\mu)={\cal{C}}_{m}\sqrt{eB}\exp\left(-\frac{4\pi^{2}}{GeB}+4\sum\limits_{\ell=1}^{\infty}(-1)^{\ell}\cosh(\mu\beta\ell)
K_{0}(\beta\ell\sigma_{0})\right),
\end{eqnarray}
where the numerical factor ${\cal{C}}_{m}\simeq 0.74$ is in
(\ref{N26}). Setting $\mu=0$, the above result (\ref{N28}) coincides
with the dynamical mass determined in \cite{ebert}. At this stage,
we would however use the Bessel-function identity (\ref{AA16}) from
Appendix \ref{appA} that includes a complete expansion in the orders
of $(\sigma_{0}\beta)$ and $(\mu\beta)$,\footnote{This step turns
out to be useful in Sect. \ref{stability}.}
\begin{eqnarray}\label{N29}
\lefteqn{\hspace{-2cm}\sum\limits_{\ell=1}^{\infty}(-1)^{\ell}\cosh(\mu\beta\ell)K_{0}(\beta\ell\sigma_{0})=\frac{1}{2}
\left(\gamma_{E}+\ln\frac{\sigma_{0}\beta}{\pi}\right)
+\sum\limits_{k=1}^{\infty}\frac{(-1)^{k}\left(2^{2k+1}-1\right)}{2^{2k+1}}\left(\frac{\mu\beta}{\pi}\right)^{2k}\zeta(2k+1)}\nonumber\\
&&+\sum\limits_{n=1}^{\infty}\frac{(-1)^{n}}{4^{2n+1}(n!)^{2}}\left(\frac{\sigma_{0}\beta}{\pi}\right)^{2n}
\left\{
\bigg|\psi^{(2n)}\left(\frac{1}{2}-\frac{i\mu\beta}{2}\right)\bigg|+\bigg|\psi^{(2n)}\left(\frac{1}{2}
+\frac{i\mu\beta}{2}\right)\bigg| \right\}.
\end{eqnarray}
We arrive at the dynamical mass of the NJL model at finite
temperature and density and in the strong magnetic field limit
\begin{eqnarray}\label{N30}
\sigma_{0}(eB,T,\mu)&=&{\cal{C}}_{m}\sqrt{eB}\exp\bigg[-\frac{4\pi^{2}}{GeB}+2
\left(\gamma_{E}+\ln\frac{\sigma_{0}\beta}{\pi}\right)
+\sum\limits_{k=1}^{\infty}\frac{(-1)^{k}\left(2^{2k+1}-1\right)}{2^{2k-1}}\left(\frac{\mu\beta}{\pi}\right)^{2k}\zeta(2k+1)\nonumber\\
&&+\sum\limits_{n=1}^{\infty}\frac{(-1)^{n}}{4^{2n}(n!)^{2}}\left(\frac{\sigma_{0}\beta}{\pi}\right)^{2n}
\left\{
\bigg|\psi^{(2n)}\left(\frac{1}{2}-\frac{i\mu\beta}{2}\right)\bigg|+\bigg|\psi^{(2n)}\left(\frac{1}{2}
+\frac{i\mu\beta}{2}\right)\bigg| \right\} \bigg].
\end{eqnarray}
In (\ref{N29}), the polygamma function $\psi(z)$ is defined as in
(\ref{N20}). The identity (\ref{N29}) is a generalization of
Bessel-function identities from \cite{meisinger} for $\mu\neq 0$ and
will be proved in Appendix \ref{appA}.
\par
\begin{center}
\begin{figure}[hbt]
\includegraphics[width=5.5cm, height=5cm]{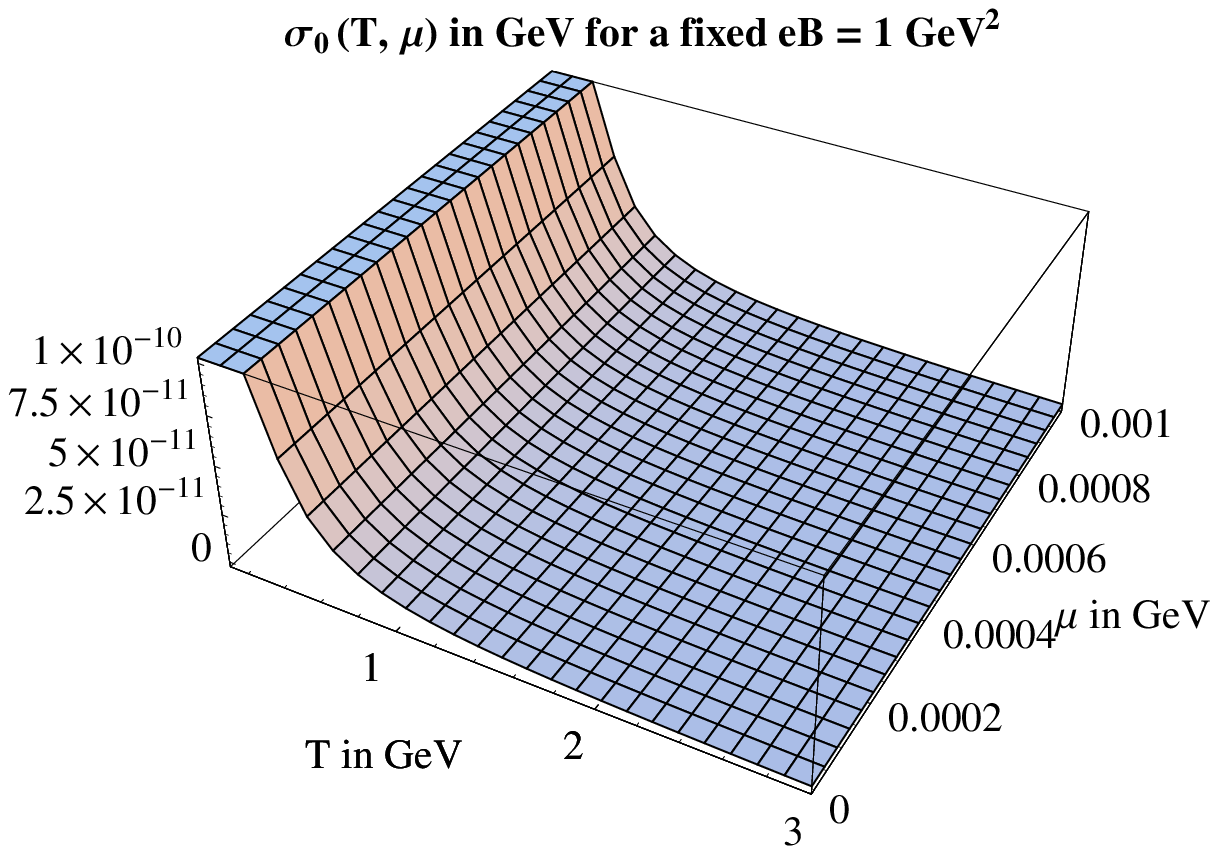}
\includegraphics[width=5.5cm, height=5cm]{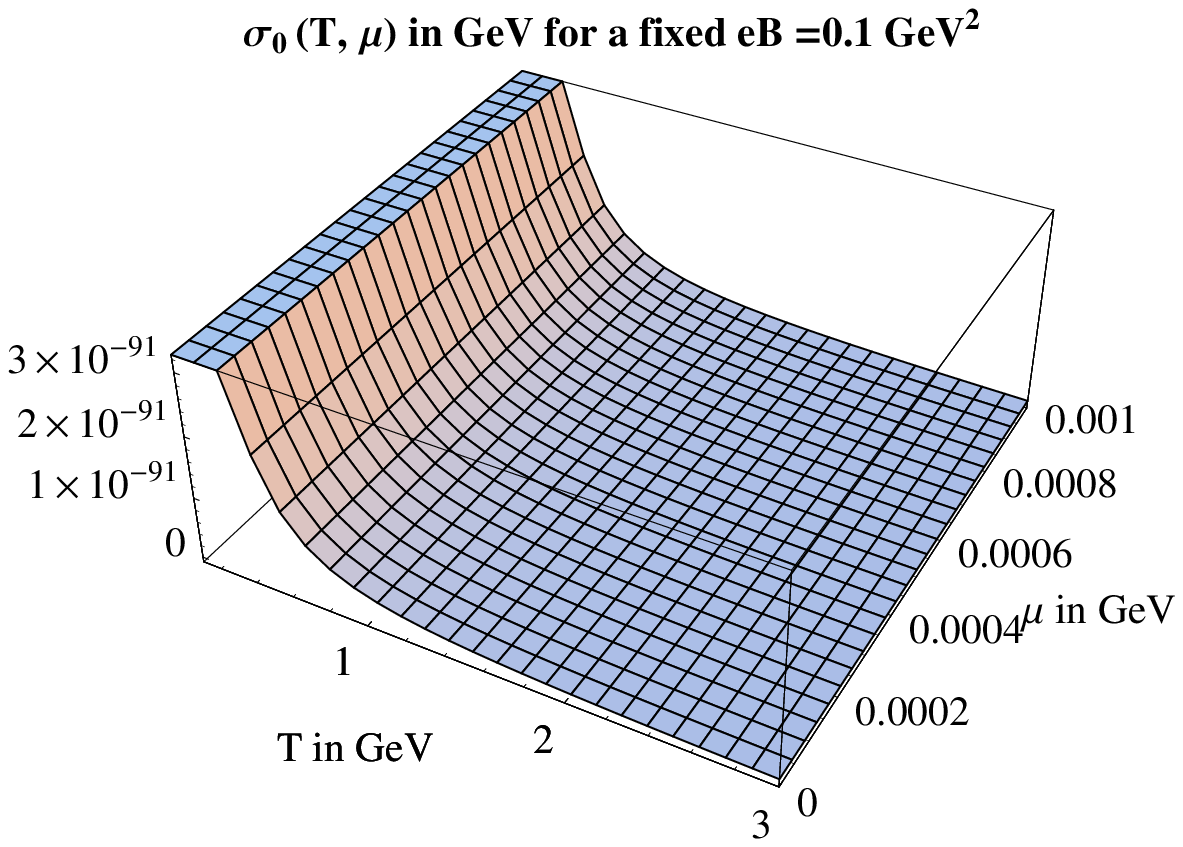}
\caption{Qualitative dependence of NJL dynamical mass $\sigma_{0}$
from (\ref{N30}) on $T$ and $\mu$ in a strong magnetic field
$eB=0.1$ and $eB=1$ GeV$^{2}$. The dynamical mass abruptly decreases
by a change of the magnetic field of one order of magnitude.}
\end{figure}
\end{center}
\vspace{-1cm}\par Fig. 2 shows qualitatively the $(T,\mu)$
dependence of $\sigma_{0}(eB,T,\mu)$ from (\ref{N30}) for $eB=0.1,
1$ GeV$^{2}$, corresponding to $B\simeq 10^{19}, 10^{20}$ Gau\ss,
respectively. Since the solution (\ref{N30}) is an iterative one, it
is necessary to choose an initial value for $\sigma_{0}$ on the
r.h.s. of (\ref{N30}). In Fig. 2, this initial value is chosen to be
$\sigma_{0}(eB,T=\mu=0)$ from (\ref{N26}). Further, the sum on the
r.h.s. of (\ref{N30}) is up to $n=k=7$. The NJL coupling $G$ is
chosen to be $G=5.86$ GeV$^{-2}$. Fig. 2 shows that the dynamical
mass vanishes for a fixed value of $eB$ by increasing the
temperature. Its qualitative behavior does not depend too much on
the chemical potential $\mu$. In the next section, the kinetic term
of the effective action (\ref{N5}) will be determined in a high
temperature expansion.
\subsection{The kinetic term of the NJL model at finite $T,\mu$ and in a strong magnetic
field}\label{kinetic}
\par\noindent
The kinetic term ${\cal{L}}_{k}$ of the effective action (\ref{N5})
at \textit{zero} temperature and density, and in the presence of a
constant magnetic field is previously determined in
\cite{miransky-NPB-95}. In the first part of this section, we will
briefly outline the arguments in \cite{miransky-NPB-95}. We then
generalize them for the case of non-vanishing temperature and
chemical potential.
\par
Let us start with the general structure of the kinetic term of the
effective action in a derivative expansion
\begin{eqnarray}\label{N31}
{\cal{L}}_{k}=\frac{1}{2}F_{1}^{\mu\nu}\partial_{\mu}\rho_{j}\partial_{\nu}\rho_{j}+
\frac{1}{\rho^{2}}F_{2}^{\mu\nu}\left(\rho_{j}\partial_{\mu}\rho_{j}\right)\left(\rho_{i}
\partial_{\nu}\rho_{i}\right),
\end{eqnarray}
that is implied by the $U_{L}(1)\times U_{R}(1)$ symmetry of the
original Lagrangian density. Here, $\vec{\rho}=(\sigma,\pi)$, and
$F_{i}^{\mu\nu}, i=1,2$ are structure constants depending
nontrivially on the external magnetic field $B$. Following the
general field theory arguments,\footnote{See e.g. in
\cite{miransky-old} for more details.} and the definitions
\begin{eqnarray}\label{N32}
\frac{\delta^{2}\tilde{\Gamma}}{\delta\sigma(x)\delta\sigma(0)}\bigg|_{\sigma_{0}\neq
0,
\pi_{0}=0}&=&-\left(F_{1}^{\mu\nu}+2F_{2}^{\mu\nu}\right)\bigg|_{\sigma_{0}\neq
0,
\pi_{0}=0}\partial_{\mu}\partial_{\nu}\delta^{4}(x),\nonumber\\
\frac{\delta^{2}\tilde{\Gamma}}{\delta\pi(x)\delta\pi(0)}\bigg|_{\sigma_{0}\neq
0, \pi_{0}=0}&=&-F_{1}^{\mu\nu}\bigg|_{\sigma_{0}\neq 0,
\pi_{0}=0}\partial_{\mu}\partial_{\nu}\delta^{4}(x),
\end{eqnarray}
the structure functions can be given as
\begin{eqnarray}\label{N33}
F_{1}^{\mu\nu}&=&-\frac{1}{2}\int d^{4}x
x^{\mu}x^{\nu}\frac{\delta^{2}\tilde{\Gamma}}{\delta\pi(x)\delta\pi(0)}\nonumber\\
F_{2}^{\mu\nu}&=&-\frac{1}{4}\int d^{4}x
x^{\mu}x^{\nu}\frac{\delta^{2}\tilde{\Gamma}}{\delta\sigma(x)\delta\sigma(0)}-\frac{1}{2}F_{1}^{\mu\nu},
\end{eqnarray}
where
$\tilde{\Gamma}=-i\mbox{Tr}\ln(i\gamma^{\mu}D_{\mu}-\sigma-i\gamma_{5}\pi)$
is the effective action from (\ref{N6}). In (\ref{N32}) the minimum
of the potential is chosen to be at $\sigma_{0}=\mbox{const.}$ and
$\pi_{0}=0$. To calculate $F_{i}^{\mu\nu}, i=1,2$ from (\ref{N33}),
we use the definition of the fermion propagator
$iS^{-1}=i\gamma^{\mu}D_{\mu}-\sigma_{0}$,\footnote{Note that here,
$\sigma_{0}=\mbox{const.}$ plays the role of the dynamical mass. It
will be determined analytically by solving the corresponding gap
equation in Sect. \ref{dynamical}.} and arrive therefore at
\begin{eqnarray}\label{N34}
\frac{\delta^{2}\tilde{\Gamma}}{\delta\sigma(x)\delta\sigma(0)}\bigg|_{\sigma_{0}\neq
0,
\pi_{0}=0}&=&-i\mbox{tr}\left(S(x,0)S(0,x)\right)\nonumber\\
\frac{\delta^{2}\tilde{\Gamma}}{\delta\pi(x)\delta\pi(0)}\bigg|_{\sigma_{0}\neq
0,
\pi_{0}=0}&=&+i\mbox{tr}\left(S(x,0)\gamma_{5}S(0,x)\gamma_{5}\right).
\end{eqnarray}
Plugging (\ref{N34}) back in (\ref{N33}), the structure functions in
the momentum space are given by
\begin{eqnarray}\label{N35}
F_{1}^{\mu\nu}&=&+\frac{i}{2}\int\frac{d^{4}k}{(2\pi)^{4}}\mbox{tr}\left(\tilde{S}(k)\gamma_{5}\frac{\partial^{2}\tilde{S}}{\partial
k_{\mu}\partial k_{\nu}}\gamma_{5}\right)\nonumber\\
F_{2}^{\mu\nu}&=&-\frac{i}{4}\int\frac{d^{4}k}{(2\pi)^{4}}\mbox{tr}\left(\tilde{S}(k)\frac{\partial^{2}\tilde{S}}{\partial
k_{\mu}\partial k_{\nu}}\right)-\frac{1}{2}F_{1}^{\mu\nu}.
\end{eqnarray}
As it is shown in \cite{miransky-NPB-95}, the fermion propagator in
a constant magnetic field and at zero $(T,\mu)$ is given by
\begin{eqnarray}\label{N36}
S(x,y)=\exp\left(\frac{ie}{2}\left(x-y\right)^{\mu}A_{\mu}^{ext}(x+y)\right)
\tilde{S}(x-y),
\end{eqnarray}
with the Fourier transform of $\tilde{S}$
\begin{eqnarray}\label{N37}
\tilde{S}(k)=2i\exp\left(-\frac{{\mathbf{k}}^{2}_{\perp}}{eB}\right)\sum\limits_{n=0}^{\infty}(-1)^{n}
\frac{D_{n}(eB,k)}{{\mathbf{k}}_{\|}^{2}-\sigma_{0}^{2}-2eB n},
\end{eqnarray}
where $n$ labels the Landau levels that arise in the presence of a
constant magnetic field, and ${\mathbf{k}}_{\|}$ as well as
${\mathbf{k}}_{\perp}$ are the longitudinal and transverse momenta
with respect to the external magnetic field, i.e. for
${\mathbf{B}}=B{\mathbf{e}}_{3}$ we get
${\mathbf{k}}_{\|}=(k_{0},k_{3})$,
${\mathbf{k}}_{\perp}=(k_{1},k_{2})$. Further, $D_{n}(eB,k)$ is
given in terms of Laguerre polynomials $L_{n}$ by
\begin{eqnarray}\label{N38}
D_{n}(eB,k)=\left({\mathbf{k}}_{\|}\cdot
\gamma_{\|}+\sigma_{0}\right)\left\{{\cal{O}}_{-}L_{n}\left(2\frac{{\mathbf{k}}^{2}_{\perp}}{eB}
\right)-{\cal{O}}_{+}L_{n-1}\left(2\frac{{\mathbf{k}}^{2}_{\perp}}{eB}
\right)\right\}-2{\mathbf{k}}_{\perp}\cdot\gamma_{\|}
L_{n-1}^{(1)}\left(2\frac{{\mathbf{k}}^{2}_{\perp}}{eB}\right),
\end{eqnarray}
where ${\cal{O}}_{\pm}\equiv \frac{1}{2}\left(1\pm
i\gamma_{1}\gamma_{2} \mbox{sign}(eB)\right)$. In
\cite{miransky-NPB-95}, the fermion propagator in a constant
magnetic field (\ref{N36})-(\ref{N37}) is used to determine
$F_{i}^{\mu\nu}, i=1,2$ from (\ref{N35}). But, since we are
interested only on the strong magnetic field limit $eB\to\infty$,
where the fermion dynamics is dominated by the lowest Landau level
(LLL) pole of the fermion propagator, it is enough to use the
fermion propagator in the LLL approximation
\begin{eqnarray}\label{N39}
S(x)=p_{\perp}({\mathbf{x}}_{\perp})s_{\|}({\mathbf{x}}_{\|}){\cal{O}}_{-},
\end{eqnarray}
with
\begin{eqnarray}\label{N40}
p_{\perp}({\mathbf{x}}_{\perp})\equiv
\frac{ieB}{2\pi}e^{-\frac{eB}{4}{\mathbf{x}}^{2}_{\perp}},\qquad\mbox{and}\qquad
s_{\|}({\mathbf{x}}_{\|})\equiv\int\frac{d^{2}{\mathbf{k}}_{\|}}{(2\pi)^{2}}\frac{e^{-i{\mathbf{k}}_{\|}\cdot
{\mathbf{x}}_{\|}}}{\left({\mathbf{k}}_{\|}\cdot
\gamma_{\|}-\sigma_{0}\right)}.
\end{eqnarray}
It arises from (\ref{N36})-(\ref{N38}) by setting $n=0$. In Sect.
\ref{hydro}, a hydrodynamical description of the NJL model in the
presence of a strong magnetic field will be presented, where the
combination $G^{\mu\nu}\equiv F_{1}^{\mu\nu}+2F_{2}^{\mu\nu}$ is
shown to be relevant. Plugging therefore (\ref{N39})-(\ref{N40}) in
(\ref{N35}), the structure function $G^{\mu\nu}$ in the LLL
approximation are given by
\begin{eqnarray}\label{N41}
G^{11}&=&G^{22}=-\frac{i}{4\pi}\int\frac{d^{2}k_{\|}}{(2\pi)^{2}}\mbox{tr}
\left(\tilde{s}_{\|}({\mathbf{k}}_{\|}){\cal{O}}_{-}\tilde{s}_{\|}({\mathbf{k}}_{\|}){\cal{O}}_{-}
\right),\nonumber\\
G^{ab}&=&\frac{ieB}{4\pi}\int\frac{d^{2}{k_{\|}}}{(2\pi)^{2}}
\mbox{tr}\left(\tilde{s}_{\|}({\mathbf{k}}_{\|}){\cal{O}}_{-}\frac{\partial^{2}}{\partial
k_{a}\partial
k_{b}}\tilde{s}_{\|}({\mathbf{k}}_{\|}){\cal{O}}_{-}\right), \qquad
a,b\in\{0,3\},
\end{eqnarray}
where
$\tilde{s}_{\|}({\mathbf{k}}_{\|})=\left({\mathbf{k}}_{\|}\cdot
\gamma_{\|}-\sigma_{0}\right)^{-1}$ is the Fourier transform of
$s_{\|}({\mathbf{x}}_{\|})$. Plugging
$\tilde{s}_{\|}({\mathbf{k}}_{\|})$ in $G^{\mu\nu}$ and using the
identities tr$\left(1\right)=4$,
tr$\left(\gamma^{a}\gamma^{b}\right)=4g^{ab},
{\cal{O}}_{-}^{2}={\cal{O}}_{-},{\cal{O}}_{-}\gamma_{a}=\gamma_{a}{\cal{O}}_{-}$,
$\mbox{tr}({\cal{O}}_{-}\gamma_{a}{\cal{O}}_{-}\gamma_{b})=2g_{ab},$
and  $\mbox{tr}({\cal{O}}_{-}\gamma_{a})=0$ for $a,b\in\{0,3\}$, as
well as $\mbox{tr}({\cal{O}}_{-})=2$, we arrive first at
\begin{eqnarray}\label{N42}
G^{11}=G^{22}=-\frac{i}{2\pi}\int \frac{d^{2}k_{\|}}{(2\pi)^{2}}\
\frac{({\mathbf{k}}_{\|}^{2}+\sigma_{0}^{2})}{({\mathbf{k}}_{\|}^{2}-\sigma_{0}^{2})^{2}},
\end{eqnarray}
and
\begin{eqnarray}\label{N43}
G^{aa}=-\frac{ieB}{\pi}\int \frac{d^{2}k_{\|}}{(2\pi){2}}\
\frac{1}{({\mathbf{k}}_{\|}^{2}-\sigma_{0}^{2})^{3}}\left(2k_{a}^{2}+({\mathbf{k}}_{\|}^{2}+\sigma_{0}^{2})-4\frac{k_{a}^{2}({\mathbf{k}}_{\|}^{2}+
\sigma_{0}^{2})}{({\mathbf{k}}_{\|}^{2}-\sigma_{0}^{2})}\right),
\end{eqnarray}
whereas the non-diagonal terms identically vanish, i.e.
\begin{eqnarray}\label{N44}
G^{03}=G^{30}=-\frac{2ieB}{\pi}\int
\frac{d^{2}k_{\|}}{(2\pi)^{2}}\frac{k_{0}k_{3}}{({\mathbf{k}}_{\|}^{2}-\sigma_{0}^{2})^{3}}\left(1-2\frac{({\mathbf{k}}_{\|}\cdot
\gamma_{\|}+\sigma_{0}^{2})}{({\mathbf{k}}_{\|}^{2}-\sigma_{0}^{2})}\right)=0.
\end{eqnarray}
To calculate $G^{\mu\nu}$ from (\ref{N41}) at finite temperature and
density, we will first perform the following necessary replacements
\begin{eqnarray}\label{N45}
k_{0}\to i\tilde{k}_{0}\qquad \mbox{with}\qquad
\tilde{k}_{0}=\left(\omega_{\ell}-i\mu\right),\qquad\mbox{and}\qquad
\omega_{\ell}=\frac{(2\ell+1)\pi}{\beta},
\end{eqnarray}
as well as
\begin{eqnarray}\label{N46}
\int d^{2}k_{\|}\to iT\sum\limits_{\ell=-\infty}^{\infty}\int
dk_{3},
\end{eqnarray}
and arrive at
\begin{eqnarray}\label{N47}
G^{11}=G^{22}&=&-\frac{1}{4\pi^{2}\beta}\sum\limits_{\ell=-\infty}^{\infty}\int
\frac{dk_{3}}{2\pi}\frac{\tilde{k}_{0}^{2}+k_{3}^{2}-\sigma_{0}^{2}}{(\tilde{k}_{0}^{2}+k_{3}^{2}
+\sigma_{0}^{2})^{2}}\nonumber\\
G^{00}&=&\frac{eB}{2\pi^2\beta}\sum\limits_{\ell=-\infty}^{\infty}\int\frac{dk_{3}}{2\pi}
\frac{(k_{3}^{4}-\tilde{k}_{0}^{4}-\sigma_{0}^{4}+6\sigma_{0}^{2}\tilde{k}_{0}^{2})}{(\tilde{k}_{0}^{2}+
k_{3}^{2}+\sigma_{0}^{2})^{4}},\nonumber\\
G^{33}&=&\frac{eB}{2\pi^{2}\beta}\sum\limits_{\ell=-\infty}^{\infty}\int\frac{dk_{3}}{2\pi}\frac{(3k_{3}^{4}-6k_{3}^{2}\sigma_{0}^{2}-4k_{3}^{2}\tilde{k}_{0}^{2}+
\tilde{k}_{0}^{4}-\sigma_{0}^{4})}{(\tilde{k}_{0}^{2}+k_{3}^{2}+\sigma_{0}^{2})^{4}}.
\end{eqnarray}
To evaluate the $k_{3}$-integration as well as the sum over the
Matsubara frequencies $\ell$, we use the Mellin transformation
technique. This technique, which is originally introduced in
\cite{mellin} for finite temperature and zero density, is
generalized in Appendix \ref{mellin} for finite temperature and
nonzero chemical potential. Using in particular the main identity
including an expansion in the orders of $(m\beta)$,
\begin{eqnarray}\label{N48}
\lefteqn{\frac{1}{\beta}\sum\limits_{\ell=-\infty}^{\infty}\int\frac{d^{d}k}{(2\pi)^{d}}\
\frac{\left({\mathbf{k}}^{2}\right)^{a}\tilde{k}_{0}^{2t}}{(\tilde{k}_{0}^{2}+{\mathbf{k}}^{2}+m^{2})^{\alpha}}
=\frac{1}{2
(4\pi)^{d/2}\Gamma(\alpha)\beta}\frac{\Gamma\left(\frac{d}{2}+a\right)}{\Gamma\left(\frac{d}{2}\right)}
\left(\frac{2\pi}{\beta}\right)^{-2\alpha+d+2(t+a)}
}\nonumber\\
&&\times\sum\limits_{k=0}^{\infty}\frac{(-1)^{k}}{k!}\Gamma\left(\alpha-a+k-\frac{d}{2}\right)\bigg[
\zeta\left(2\left(\alpha+k-t-a\right)-d;\frac{1}{2}-
\frac{i\mu\beta}{2\pi}\right)+(\mu\to
-\mu)\bigg]\left(\frac{m\beta}{2\pi}\right)^{2k},\nonumber\\
\end{eqnarray}
and neglecting the temperature independent singularities, we arrive
at
\begin{eqnarray}\label{N49}
G^{11}=G^{22}&=&\frac{1}{32\pi^{3}}\bigg[\psi\left(\frac{1}{2}-\frac{i\mu\beta}{2\pi}\right)+(\mu\to
-\mu)\bigg]+\frac{3\left(\sigma_{0}\beta\right)^{2}}{256\pi^{5}}\bigg[\zeta\left(3;\frac{1}{2}-\frac{i\mu\beta}{2\pi}\right)+
(\mu\to
-\mu)\bigg]\nonumber\\
&&-\frac{15\left(\sigma_{0}\beta\right)^{4}}{4096\pi^{7}}\bigg[\zeta\left(5;\frac{1}{2}-\frac{i\mu\beta}{2\pi}\right)
+(\mu\to -\mu)\bigg]
 +{\cal{O}}\left((\sigma_{0}\beta)^{6}\right),
\end{eqnarray}
and
\begin{eqnarray}\label{N50}
G^{00}&=&-\frac{eB\beta^{2}}{256\pi^{5}}\left\{
\bigg[\zeta\left(3;\frac{1}{2}-\frac{i\mu\beta}{2\pi}\right)+(\mu\to
-\mu)\bigg]-\frac{23(\beta\sigma_{0})^{2}}{8\pi^{2}}\bigg[\zeta\left(5;\frac{1}{2}-\frac{i\mu\beta}{2\pi}\right)+(\mu\to
-\mu)\bigg]\right.\nonumber\\
&&\left.+\frac{295(\sigma_{0}\beta)^{4}}{128\pi^{4}}\bigg[\zeta\left(7;\frac{1}{2}-\frac{i\mu\beta}{2\pi}\right)+(\mu\to
-\mu)\bigg]\right\}+{\cal{O}}\left((\sigma_{0}\beta)^{6}\right),
\end{eqnarray}
and
\begin{eqnarray}\label{N51}
G^{33}&=&\frac{5eB\beta^{2}}{512\pi^{5}}\left\{
\bigg[\zeta\left(3;\frac{1}{2}-\frac{i\mu\beta}{2\pi}\right)+(\mu\to
-\mu)\bigg]-\frac{7(\beta\sigma_{0})^{2}}{8\pi^{2}}\bigg[\zeta\left(5;\frac{1}{2}-\frac{i\mu\beta}{2\pi}\right)+(\mu\to
-\mu)\bigg]\right.\nonumber\\
&&\left.+\frac{55(\sigma_{0}\beta)^{4}}{128\pi^{4}}\bigg[\zeta\left(7;\frac{1}{2}-\frac{i\mu\beta}{2\pi}\right)+(\mu\to
-\mu)\bigg]\right\}+{\cal{O}}\left((\sigma_{0}\beta)^{6}\right).
\end{eqnarray}
In (\ref{N48})-(\ref{N51}), $\zeta(s;a)$ is the generalized Riemann
zeta function given by $\zeta(s;a)=\sum_{k=0}^{\infty}(k+a)^{-s}$,
where any term with $(k+a)=0$ is excluded. In (\ref{N49}), the
digamma function $\psi(z)\equiv \frac{d}{dz}\ln\Gamma(z)$ arises by
regularizing $\zeta(1;a)$,
\begin{eqnarray}\label{N52}
\lim\limits_{\epsilon\to
0}\zeta(1+\epsilon;a)=\lim\limits_{\epsilon\to
0}\frac{1}{\epsilon}-\psi(a).
\end{eqnarray}
\begin{center}
\begin{figure}[hbt]
\includegraphics[width=5.5cm, height=5cm]{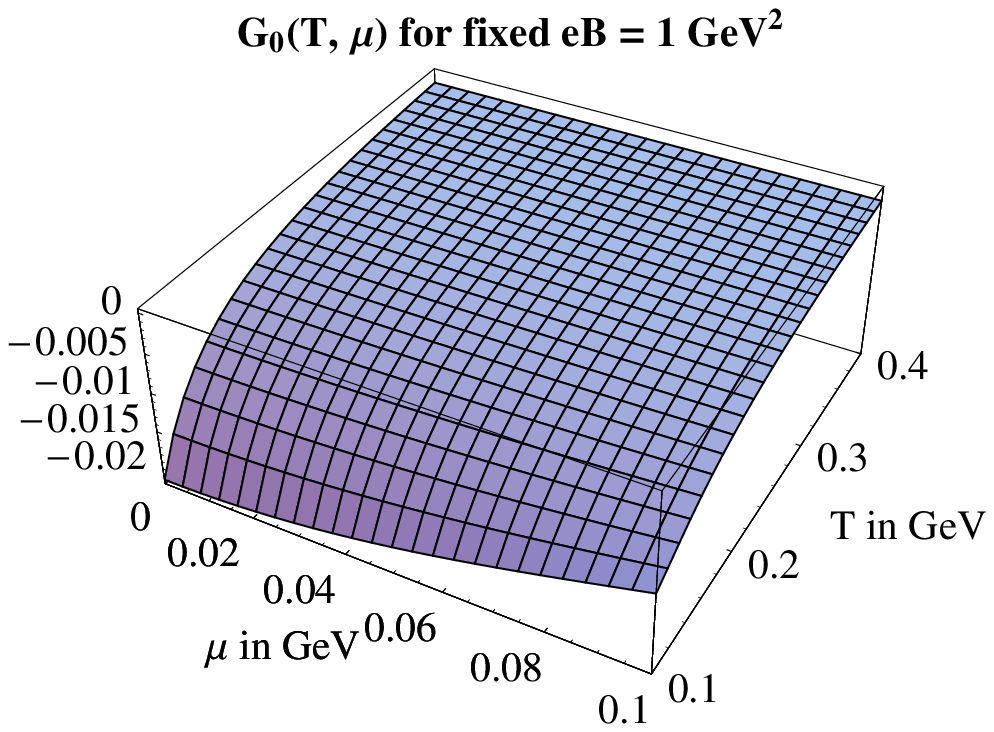}
\includegraphics[width=5.5cm, height=5cm]{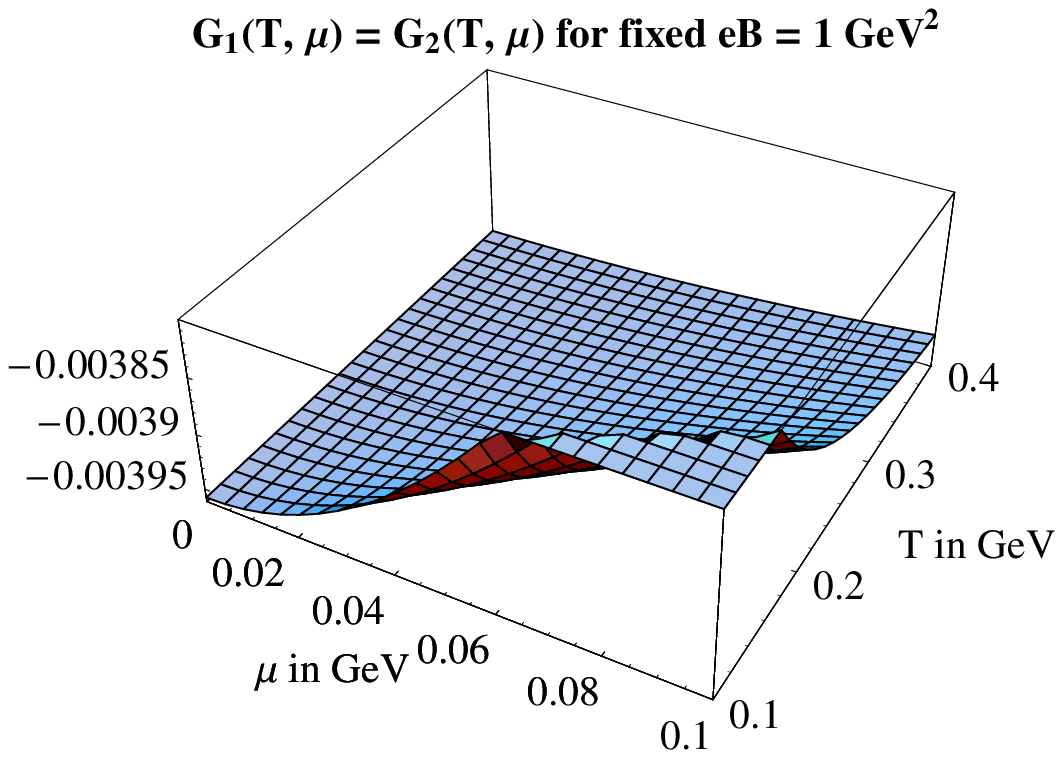}
\includegraphics[width=5.5cm, height=5cm]{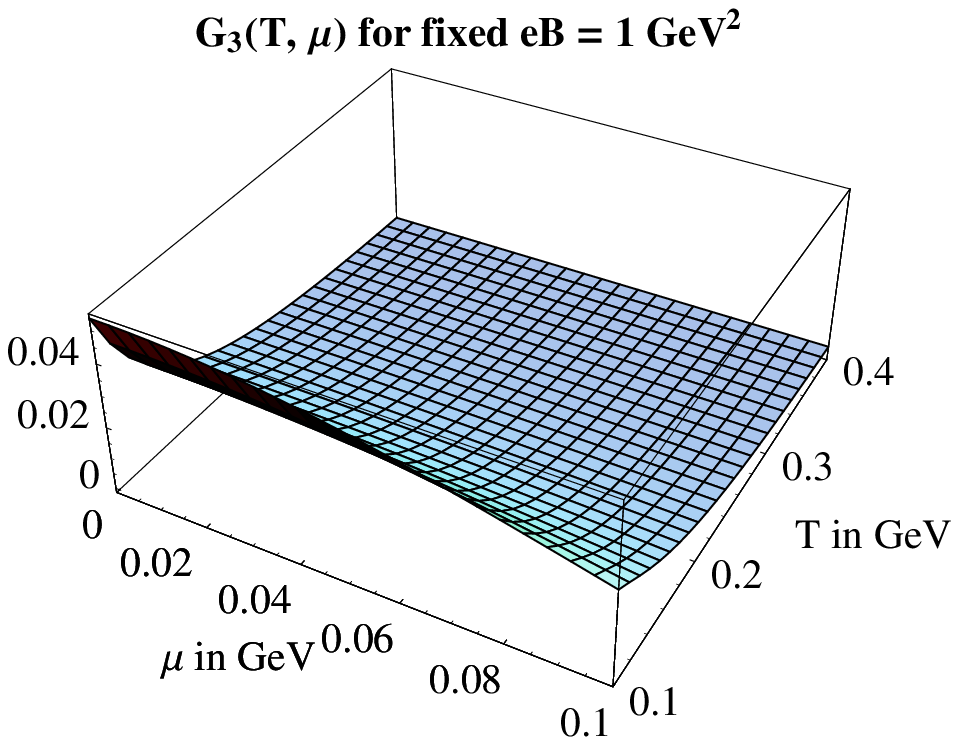}
\caption{Qualitative dependence of $G^{i}\equiv G^{ii}, i=0,\cdots,
3$ on $(T,\mu)$ for fixed $eB=1$ GeV$^{2}$ corresponding to
$B=10^{20}$ Gau\ss. They exhibit a qualitative change in the
transition region $T_{c}\sim 100$ MeV and small $\mu$.}
\end{figure}
\end{center}
Fig. 3 shows $G^{i}\equiv G^{ii}, i=0,\cdots,3$ in the $(T,\mu)$
phase space for $eB=1$ GeV$^{2}$ or equivalently $B\sim 10^{20}$
Gau\ss. To determine $G^{i}, i=,0,\cdots,3$, we have used the
expression (\ref{N30}) for $\sigma_{0}(eB,T,\mu)$. The iteration is
started as in the previous section with the value of the dynamical
mass at zero temperature and density $\sigma_{0}(eB,T=\mu=0)$ from
(\ref{N26}). The sum in (\ref{N30}) is taken up to $n=k=7$ and
$G=5.86$ GeV$^{-2}$. To determine $G^{i}, i=0,\cdots 3$ from
(\ref{N49})-(\ref{N51}), we have used (\ref{N48}). For the structure
functions in Fig. 3, the high temperature approximation in
(\ref{N48}) is up to order $(\sigma_{0}\beta)^{6}$. All three
structure functions $G^{i}, i=0,\cdots,3$ exhibit a qualitative
change in the vicinity of the critical temperature $T_{c}$.
\par
In the next section, we will use the effective kinetic term
(\ref{N31}) and the effective potential (\ref{N23}) to determine the
energy-momentum tensor of an effective NJL model described by the
effective action (\ref{N5}).
\section{The effective NJL model at finite $T,\mu$ and in a strong magnetic field}\label{effective}
\setcounter{equation}{0} 
\par\noindent
The main goal of this paper is to present a hydrodynamical
description of an expanding chiral \textit{and} magnetized QGP at
finite temperature $T$ and baryon chemical potential $\mu$. It will
be described using the NJL model of QCD in the presence of a
constant magnetic field, that exhibits, as we have shown in the
previous section, a chiral symmetry breaking due to a dynamically
generated fermion mass. The thermodynamics of the NJL model has been
extensively studied in the past few decades \cite{njl, klevansky}. A
systematic study concerning the existence, location and properties
of the critical/tricritical point of various NJL type models is
recently presented in \cite{costa}. In all previous treatments, the
effective potential of the theory, $\Omega$, was of fundamental
interest. Within a mean field approximation, $\Omega$ can indeed be
identified with the thermodynamical partition function of the
theory, $Z$, through the relation
$\Omega(\sigma_{0},T,\mu)=-\frac{1}{\beta V}\ln Z$, where
$\beta=T^{-1}$ is the inverse of the temperature and $V$ is the
volume. Using this partition function, it is then easy to derive all
thermodynamic (static) quantities, like pressure, entropy, etc (see
next section for more details on thermodynamics).
\par
We, however, are interested in the dynamical properties of an
expanding magnetized QGP near the chiral critical point. To study
the time evolution of such a system, we shall go beyond the mean
field approximation and to consider the full effective action of the
NJL model (\ref{N5}) including the effective potential (\ref{N23})
as well as the kinetic term (\ref{N31}). The latter depends, in
contrast to the linear $\sigma$ model used in \cite{fraga, paech},
on $(T,\mu)$ and the external magnetic field. The linear $\sigma$
model is in particular used in \cite{fraga, paech} to describe an
expanding QGP coupled to chiral fields.
\par
In this section, using the full effective action of the NJL model,
we will derive first the equation of motion (EoM) and eventually
determine the energy-momentum tensor of the effective NJL model at
finite $(T,\mu)$ and in the presence of constant magnetic field $B$.
The energy-momentum tensor will be then used in Sect. \ref{hydro} to
present a magnetohydrodynamical (MHD) description of an expanding
perfect QGP near the chiral critical point.
\par
Using the effective action (\ref{N5}), and in particular the general
structure of the kinetic term (\ref{N31}), the EoM of the chiral
fields $\vec{\rho}=(\sigma,\pi)$ of the strongly magnetized NJL
effective model can be derived from the ordinary Euler-Lagrange
equation
\begin{eqnarray}\label{M1}
\frac{\partial {\cal{L}}_{\mbox{\tiny{eff}}}}{\partial
\rho_{k}}-\partial_{\lambda}\frac{\partial
{\cal{L}}_{\mbox{\tiny{eff}}}}{\partial\left(\partial_{\lambda}\rho_{k}\right)}=0,
\qquad\mbox{with}\qquad \vec{\rho}=(\sigma,\pi).
\end{eqnarray}
It is given by
\begin{eqnarray}\label{M2}
F_{1}^{\mu}\partial_{\mu}\partial^{\mu}\rho_{k}+\frac{2\rho_{k}}{\rho^{2}}F_{2}^{\mu}\left(
\rho_{i}\partial_{\mu}\partial^{\mu}\rho_{i}+\partial_{\mu}\rho_{i}\partial^{\mu}\rho_{i}-\frac{1}{\rho^{2}}
\left(\rho_{i}\partial_{\mu}\rho_{i}\right)\left(\rho_{j}\partial^{\mu}\rho_{j}\right)\right)
= -R_{k},
\end{eqnarray}
where
\begin{eqnarray}\label{M3}
R_{k}\equiv \frac{\partial\Omega}{\partial\rho_{k}}.
\end{eqnarray}
Here, the effective potential $\Omega=V^{(0)}+V^{(1)}$ includes the
tree level potential $V^{(0)}$ and the one loop effective potential
$V^{(1)}$ presented in (\ref{N23}). In (\ref{M2}), we have used the
property $F_{i}^{\mu\nu}=F_{i}^{\mu\nu}g^{\mu\nu}$ (no summation
over $\mu,\nu$ is considered) and denoted as in the previous section
$F_{i}^{\mu\mu}\equiv F_{i}^{\mu}$, with $i=1,2$ and $\mu=0,\cdots
3$. For $F_{1}^{\mu}=1$ and $F_{2}^{\mu}=0$, this result is indeed
comparable with the results presented in \cite{fraga, paech}, where
the equation of motion of a linear $\sigma$ model at finite
$(T,\mu)$ and with no magnetic field is given by
$\Box\rho_{k}=-R_{k}$.
\par
Let us now look at the symmetric energy-momentum tensor
corresponding to the effective action (\ref{N5}), that arises from
the standard relation
\begin{eqnarray}\label{M4}
T^{\mu\nu}_{(tot)}=g^{\mu\nu}{\cal{L}}_{\mbox{\tiny{eff}}}-\frac{\partial{\cal{L}}_{\mbox{\tiny{eff}}}}{\partial
(\partial_{\mu}\xi_{\ell})}\partial^{\nu}\xi_{\ell},
\end{eqnarray}
where the dynamical fields $\xi$ are given by $\xi=\{\rho_{k},
A_{\mu}^{ext}\}$. The energy-momentum tensor will then consist of
two parts:
$T^{\mu\nu}_{(tot)}=T^{\mu\nu}_{(\rho)}+T^{\mu\nu}_{(\cal{F})}$.
Here, the matter part $T^{\mu\nu}_{(\rho)}$, and the gauge part
$T^{\mu\nu}_{({\cal{F}})}$ are given by
\begin{eqnarray}\label{M5}
T^{\mu\nu}_{(\rho)} =\frac{1}{2}\Theta^{\mu\nu\lambda\sigma}_{1}
\left(\partial_{\lambda}\rho_{k}\partial_{\sigma}\rho_{k}\right)+\frac{1}{\rho^{2}}
\Theta_{2}^{\mu\nu\lambda\sigma}
\left(\rho_{i}\partial_{\lambda}\rho_{i}\right)\left(\rho_{j}\partial_{\sigma}\rho_{j}\right)
-g^{\mu\nu}\Omega,
\end{eqnarray}
with
\begin{eqnarray}\label{M6}
\Theta_{i}^{\mu\nu\lambda\sigma}\equiv
-2F_{i}^{\mu}g^{\mu\lambda}g^{\nu\sigma}+F_{i}^{\lambda}g^{\lambda\sigma}g^{\mu\nu},
\qquad i=1,2,
\end{eqnarray}
and
\begin{eqnarray}\label{M7}
T^{\mu\nu}_{(\cal{F})}=-g^{\mu\nu}{\cal{F}}-2F^{\nu}_{\
\lambda}\frac{\partial \cal{L}_{\mbox{\tiny{eff}}}}{\partial
F_{\mu\lambda}}=-g^{\mu\nu}{\cal{F}}-F^{\nu}_{\
\lambda}F^{\mu\lambda}\frac{\partial{\cal{L}}_{\mbox{\tiny{eff}}}}{\partial{\cal{F}}},
\end{eqnarray}
with ${\cal{F}}\equiv \frac{1}{4}(F_{\mu\nu})^{2}$, respectively.
The relation (\ref{M7}) can be brought in a more familiar form
\begin{eqnarray}\label{M8}
T^{\mu\nu}_{(\cal{F})}=\left(F^{\mu}_{\
\lambda}F^{\lambda\nu}+g^{\mu\nu}\frac{1}{4}F_{\rho\sigma}F^{\sigma\rho}\right)\frac{{\partial\cal{L}}_{\mbox{\tiny{eff}}}}{\partial
{\cal{F}}}+
g^{\mu\nu}\left(-{\cal{F}}-{\cal{F}}\frac{{\partial\cal{L}}_{\mbox{\tiny{eff}}}}{\partial
{\cal{F}}}\right),
\end{eqnarray}
that appears also in \cite{schwinger} and reduces to the well-known
Maxwell tensor
\begin{eqnarray}\label{M9}
T^{\mu\nu}_{M}=-F^{\mu}_{\
\lambda}F^{\lambda\nu}-g^{\mu\nu}\frac{1}{4}F_{\rho\sigma}F^{\sigma\rho},
\end{eqnarray}
once ${\cal{L}}_{\mbox{\tiny{eff}}}=-{\cal{F}}$. For $F_{1}^{\mu}=1$
and $F_{2}^{\mu}=0$, (\ref{M5}) is comparable with the matter part
of the energy-momentum tensor $T_{(\rho)}^{\mu\nu}$ of the linear
$\sigma$ model
\begin{eqnarray}\label{M10}
T^{\mu\nu}_{(\rho)}=\frac{1}{2}\left(-2g^{\mu\lambda}g^{\nu\rho}+g^{\mu\nu}g^{\lambda\sigma}\right)
\left(\partial_{\lambda}\rho_{k}\partial_{\sigma}\rho_{k}\right)-\Omega(\rho)g^{\mu\nu}.
\end{eqnarray}
The energy-momentum tensor (\ref{M10}) is used in \cite{fraga,
paech} to present a hydrodynamical description of the linear
$\sigma$ model. To derive the energy-momentum conservation relation,
we first identify the term $F^{\mu\lambda}\frac{\partial
{\cal{L}}_{\mbox{\tiny{eff}}}}{\partial{\cal{F}}}$ in (\ref{M7})
with the {\textit{polarization tensor}} $M^{\mu\lambda}$, that
appears also in \cite{sachdev},\footnote{See next section for more
detail on the polarization tensor $M^{\mu\nu}$.}
\begin{eqnarray}\label{M11}
M^{\mu\lambda}\equiv F^{\mu\lambda}\frac{\partial
{\cal{L}}_{\mbox{\tiny{eff}}}}{\partial{\cal{F}}}.
\end{eqnarray}
The total energy-momentum tensor can then be given as
\begin{eqnarray}\label{M12}
T^{\mu\nu}_{(tot)}=T^{\mu\nu}_{k}-F^{\nu}_{\ \lambda}M^{\mu\lambda},
\end{eqnarray}
with
\begin{eqnarray}\label{M13}
T^{\mu\nu}_{k}=T^{\mu\nu}_{(\rho)}-g^{\mu\nu}{\cal{F}},
\end{eqnarray}
where $T^{\mu\nu}_{(\rho)}$ is from (\ref{M5}) and
$-g^{\mu\nu}{\cal{F}}$ is the first term on the r.h.s. of
(\ref{M7}). Using the EoM of the effective NJL model (\ref{M2}) and
the fact that for constant magnetic field
$\partial_{\mu}{\cal{F}}=0$, it is easy to show that
\begin{eqnarray}\label{M14}
\partial_{\mu}T^{\mu\nu}_{k}=0,
\end{eqnarray}
and that the energy-momentum conservation equation reduces to
\begin{eqnarray}\label{M15}
\partial_{\mu}T^{\mu\nu}_{(tot)}=-F^{\nu}_{\ \lambda}\partial_{\mu}M^{\mu\lambda}.
\end{eqnarray}
The same relation appears also in \cite{sachdev}, where a
magnetohydrodynamical description of the Nernst effect in condensed
matter physics is presented. In the next section, we will identify
the above energy-momentum tensor (\ref{M12}) arising from the
effective NJL model in a strong magnetic field with the
energy-momentum tensor of a perfect {\it chiral} and {\it
magnetized} fluid. In contrast to the description in \cite{sachdev},
the magnetohydrodynamical description of this theory will include a
chiral symmetry breaking arising from the dynamical mass generation
in the presence of a strong magnetic field.
\section{Magnetohydrodynamic description of the effective NJL model at finite $T,\mu$}\label{hydro}
\setcounter{equation}{0} \par\noindent In the first part of this
section, we recall standard identities of thermodynamics
\textit{without matter fields} which are often used in
hydrodynamical models with no background magnetic field. In the
second part, generalizing the method introduced in \cite{fraga}, we
will implement the chiral fields in a hydrodynamical description of
an expanding magnetized QGP including a spontaneous chiral symmetry
breaking in the presence of a strong magnetic field.
\subsection{Thermodynamics in a constant magnetic field; General
identities}\label{thermo}
\par\noindent
Let us start with the identity
\begin{eqnarray}\label{E1}
\textsf{U}(\textsf{V},\textsf{N},\textsf{S},\textsf{M})=-P\textsf{V}+T\textsf{S}+\mu
\textsf{N}+{\mathbf{B}}\cdot \textsf{M},
\end{eqnarray}
that defines the internal energy, which is an extensive function of
extensive variables; the volume $V$, the entropy $S$, the baryon
number $N$, and the magnetization vector $\textsf{M}$. Here, $P$ is
the pressure, $\mu$ is the chemical potential and ${\mathbf{B}}$ is
the constant induced magnetic field. Note that in nonrelativistic
systems, $N$ is generally the number of particles, which is
conserved. In a relativistic system, the number of particles are not
conserved: it is always possible to create a particle-antiparticle
pair, provided the energy is available. In this case $N$ no longer
denotes a number of particles, but a conserved quantity, such as the
baryon number (see \cite{ollitrault} for a recent review on
relativistic hydrodynamics). In what follows, we will assume that
the magnetization vector is parallel to the external ${\mathbf{B}}$
field, i.e. in our setup ${\mathbf{B}}=B{\mathbf{e}}_{3}$, we get
${\mathbf{B}}\cdot \textsf{M}=B\textsf{M}$. The differential of
internal energy is given by the thermodynamic identity
\begin{eqnarray}\label{E2}
d\textsf{U}=-Pd\textsf{V}+Td\textsf{S}+\mu d\textsf{N}+B
d\textsf{M}.
\end{eqnarray}
The identity (\ref{E1}) can be derived by differentiating the
relation
\begin{eqnarray}\label{E3}
\textsf{U}(\lambda \textsf{V},\lambda \textsf{N}, \lambda
\textsf{S}, \lambda\textsf{M})=\lambda
\textsf{U}(\textsf{V},\textsf{N},\textsf{S},\textsf{M}),
\end{eqnarray}
with respect to $\lambda$, taking $\lambda=1$, and using (\ref{E2}).
Differentiating (\ref{E1}) and using again (\ref{E2}), we obtain the
Gibbs-Duhem relation
\begin{eqnarray}\label{E4}
\textsf{V}dP=\textsf{S}dT+\mu d\textsf{N}+\textsf{M} dB,
\end{eqnarray}
that reduces to
\begin{eqnarray}\label{E5}
dP=s dT+n d\mu+M dB,
\end{eqnarray}
where $s\equiv \frac{\textsf{S}}{\textsf{V}}$, $n\equiv
\frac{\textsf{N}}{\textsf{V}}$, and $M\equiv
\frac{\textsf{M}}{\textsf{V}}$, the entropy density, the baryon
number density and the magnetization density are introduced.
Defining further the energy density $\epsilon\equiv
\frac{\textsf{U}}{\textsf{V}}$, (\ref{E1}) and (\ref{E2}) are given
by
\begin{eqnarray}\label{E6}
\epsilon+P=Ts+\mu n+BM,
\end{eqnarray}
and
\begin{eqnarray}\label{E7}
d\epsilon=Tds+\mu dn+B dM.
\end{eqnarray}
Using the Gibbs-Duhem relation (\ref{E5}), the thermodynamical
variables are given by
\begin{eqnarray}\label{E8}
s=\left(\frac{\partial P}{\partial T}\right)_{\mu,B}, \qquad
n=\left(\frac{\partial P}{\partial \mu}\right)_{T,B}, \qquad
M=\left(\frac{\partial P}{\partial{B}}\right)_{\mu,T}.
\end{eqnarray}
Similarly the temperature $T$, the chemical potential $\mu$ and the
magnetic field $B$ can be defined by (\ref{E7}) using
\begin{eqnarray}\label{E9}
T=\left(\frac{\partial \epsilon}{\partial s}\right)_{n,M}, \qquad
\mu=\left(\frac{\partial \epsilon}{\partial n}\right)_{s,M}, \qquad
B=\left(\frac{\partial \epsilon}{\partial{M}}\right)_{s,n}.
\end{eqnarray}
In the next section, we will generalize the above thermodynamic
relations to the case where the energy density $\epsilon$ depends
explicitly on the chiral fields $\vec{\rho}=(\sigma,\pi)$.
\subsection{Chiral magnetohydrodynamics in and out of thermal
equilibrium}\label{mhd}
\subsubsection*{Extended thermodynamic identities}
\par\noindent
Following the variational method presented in \cite{fraga}, we treat
the gas of quarks for the chiral fields $\sigma$ and $\pi$ as a heat
bath with temperature $T$ and baryon chemical potential $\mu$. In
thermal equilibrium, this identification is justified by the
relation
\begin{eqnarray}\label{E10}
P_{0}=-\Omega(\vec{\rho};eB,T,\mu)|_{\sigma_{0}\neq 0, \pi_{0}=0},
\end{eqnarray}
which is often used in the standard thermal field theory. Here,
$P_{0}$ is the thermodynamic pressure given by
\begin{eqnarray}\label{E11}
P_{0}=\frac{1}{\beta V}\ln Z,
\end{eqnarray}
where $\beta=T^{-1}$ is the inverse temperature, $V$ the volume and
$Z$ the thermodynamic partition function. The effective potential
$\Omega$ in (\ref{E10}) arises by integrating out the fermions from
the original theory. The configuration $(\sigma_{0}\neq 0,
\pi_{0}=0)$ builds the stationary point of the effective potential
or equivalently describes the thermodynamic equilibrium. In Sect.
\ref{stability}, where the sound velocity of a plane wave
propagating in the magnetized QGP will be calculated, we will assume
that the same relation between the pressure and the effective action
as in (\ref{E10}), i.e. $P=-\Omega(\vec{\rho};eB,T,\mu)$, still
holds in a system which is out of equilibrium. Here,
$\vec{\rho}=(\sigma,\pi)$ is the out of equilibrium configuration of
the chiral fields and depends in particular on the space-time point
$x$.
\par
As we have seen in the previous section, the thermodynamic pressure
is related to the energy density $\epsilon$ through the relation
(\ref{E6}),\footnote{In the stability analysis, that will be
performed in Sect. \ref{stability}, we will assume that $B,T,\mu$
and $M$ are always constant. We will vary only
$\rho_{k}=(\sigma,\pi),n$ and $s$ and consequently $P$ and
$\epsilon$ around their equilibrium configurations $\sigma_{0},
n_{0}, s_{0}$ as well as $P_{0}$ and $\epsilon_{0}$. }
$$
P=-\epsilon+Ts+\mu n+BM.
$$
In the above merging process of thermodynamics and quantum effective
field theory, expressed by (\ref{E10}) or more generally by
$P=-\Omega(\vec{\rho};eB,T,\mu)$, it is therefore natural that the
energy density $\epsilon(\vec{\rho};n,s,M)$ depends not only on
$(n,s,M)$, as in the ordinary thermodynamics, but also on the chiral
field $\vec{\rho}=(\sigma,\pi)$. For $\epsilon\equiv
\epsilon(\vec{\rho};n,s,M)$, (\ref{E7}) is therefore generalized as
\cite{fraga}
\begin{eqnarray}\label{E12}
d\epsilon=Tds+\mu dn+BdM+R_{k}d\rho_{k},
\end{eqnarray}
with
\begin{eqnarray}\label{E13}
T=\left(\frac{\partial \epsilon}{\partial s}\right)_{n,M,\rho_{k}},
\qquad \mu=\left(\frac{\partial \epsilon}{\partial
n}\right)_{s,M,\rho_{k}}, \qquad B=\left(\frac{\partial
\epsilon}{\partial{M}}\right)_{n,s,\rho_{k}},\qquad R_{k}\equiv
\left(\frac{\partial\epsilon}{\partial \rho_{k}}\right)_{n,s,M}.
\end{eqnarray}
In thermal equilibrium, where the only relevant configuration is
$(\sigma_{0}\neq 0, \pi_{0}=0)$, we have in particular,
$\epsilon_{0}\equiv \epsilon(\sigma_{0};n_{0},s_{0},M)$. In this
case, the last relation in (\ref{E13}) is modified as
\begin{eqnarray}\label{E14}
R_{\sigma_{0}}\equiv \left(\frac{\partial\epsilon_{0}}{\partial
\sigma_{0}}\right)_{n_{0},s_{0},M}, \qquad\mbox{and}\qquad
R_{\pi_{0}}\equiv \left(\frac{\partial\epsilon_{0}}{\partial
\pi_{0}}\right)_{n_{0},s_{0},M}.
\end{eqnarray}
Using further (\ref{E6}) and (\ref{E12}), the corresponding
Gibbs-Duhem relation (\ref{E5}) becomes
\begin{eqnarray}\label{E15}
dP=sdT+nd\mu+MdB-R_{k}d\rho_{k}.
\end{eqnarray}
This implies the definitions
\begin{eqnarray}\label{E16}
s=\left(\frac{\partial P}{\partial T}\right)_{\mu,B,\rho_{k}},
\qquad n=\left(\frac{\partial P}{\partial
\mu}\right)_{T,B,\rho_{k}}, \qquad M=\left(\frac{\partial
P}{\partial{B}}\right)_{T,\mu,\rho_{k}},\qquad R_{k}\equiv
-\left(\frac{\partial P}{\partial \rho_{k}}\right)_{T,\mu,B}.
\end{eqnarray}
In thermal equilibrium, $P_{0}=P(\sigma_{0};eB,T,\mu)$, the last
relation in (\ref{E16}) is modified, upon using (\ref{E10}), as
\begin{eqnarray}\label{E17}
R_{\sigma_{0}}\equiv \left(\frac{\partial
\Omega(\sigma_{0};eB,T,\mu)}{\partial \sigma_{0}}\right)_{T,\mu,B},
\qquad R_{\pi_{0}}\equiv
\left(\frac{\partial\Omega(\sigma_{0};eB,T,\mu)}{\partial
\pi_{0}}\right)_{T,\mu,B}.
\end{eqnarray}
Note that the definitions (\ref{E17}) are indeed equivalent with
(\ref{M3}) for $\sigma=\sigma_{0}$ and $\pi=\pi_{0}=0$. Using the
EoM (\ref{M2}) for this constant field configuration,
$R_{\sigma_{0}}$ and $R_{\pi_{0}}$ from (\ref{E17}) as well as
(\ref{E14}) vanish, i.e. $R_{\sigma_{0}}=R_{\pi_{0}}=0$.
\subsubsection*{MHD description of the magnetized QGP}
\par\noindent
Using the above generalized thermodynamic relations including the
chiral matter fields in thermal equilibrium, it is indeed possible
to derive all the thermodynamic (static) quantities of the effective
NJL model in the presence of a strong magnetic field. It is,
however, the goal of this paper to go beyond the thermal equilibrium
to study dynamical properties of a magnetized QGP near the chiral
critical point. To this purpose, we will use relativistic
hydrodynamics which can be used as an effective description of the
theory in the same footing as all the other effective descriptions
including the time evolution of the fluid. There is only one
assumption, which is associated with hydrodynamics: the local
thermal equilibrium (LTE). No other assumption is made concerning
the nature of the particles and fields included in the fluid. Their
interactions and the classical/quantum nature of the phenomena
involved are fully encoded in the thermodynamic properties, i.e. in
the equation of state (see \cite{ollitrault} for a recent review on
relativistic hydrodynamics).
\par
The fundamental ingredients of a hydrodynamic analysis at LTE are
the conserved quantities and their equations of motion. Here,
generalizing a similar treatment from \cite{sachdev}, where a
similar MHD theory in the vicinity of superfluid-insulator
transition in two spatial dimensions is evaluated, we introduce the
four velocity $u^{\mu}\equiv \frac{dx^{\mu}}{d\tau}$, that
represents the velocity of the system in LTE with respect to the
laboratory frame. In the rest frame $u^{\mu}=(1,0,0,0)$. It
satisfies the normalization $u_{\mu}u^{\mu}=1$. Further, we will use
the standard definition of the metric tensor
$g^{\mu\nu}=\mbox{diag}\left(1,-1,-1,-1\right)$.
\par
Let us start with the conservation laws of total energy-momentum
tensor \cite{sachdev}
\begin{eqnarray}\label{E18}
u_{\nu}\partial_{\mu}T^{\mu\nu}_{(tot)}=u_{\mu}F^{\mu\nu}J_{\nu}^{(tot)},
\end{eqnarray}
where the hydrodynamic energy-momentum tensor in the presence of a
background magnetic field is defined by
\begin{eqnarray}\label{E19}
T^{\mu\nu}_{(tot)}\equiv(\epsilon+P)u^{\mu}u^{\nu}-Pg^{\mu\nu}+T^{\mu\nu}_{E}-F^{\nu}_{\
\lambda}M^{\mu\lambda}.
\end{eqnarray}
Here, $T^{\mu\nu}_{E}$ is the {\textit{energy magnetization
current}}. It appears also in \cite{sachdev}, where the authors do
not determine it explicitly, and use only its property
$\partial_{\mu}T^{\mu\nu}_{E}=0$. The latter will be shown to be
only valid in the field free case and for $B=0$. In our case,
however, where, the perfect magnetized fluid is coupled to chiral
fields $\sigma$ and $\pi$, the combination ${\cal{T}}^{\mu\nu}\equiv
(\epsilon+P)u^{\mu}u^{\nu}-Pg^{\mu\nu}+T^{\mu\nu}_{E}$ satisfies
$\partial_{\mu}{\cal{T}}^{\mu\nu}=0$.
\par
On the r.h.s. of (\ref{E18}), the total baryon current density
$J^{\mu}_{(tot)}$ is defined by
\begin{eqnarray}\label{E20}
J^{\mu}_{(tot)}\equiv nu^{\mu}+\partial_{\nu}M^{\mu\nu},
\end{eqnarray}
and satisfies the total current conservation law
\begin{eqnarray}\label{E21}
\partial_{\mu}J^{\mu}_{(tot)}=0.
\end{eqnarray}
In (\ref{E20}), $n$ is the baryon number density, as is defined in
the previous section,\footnote{This is in contrast to
\cite{sachdev}, where $J^{\mu}_{(tot)}$ is the electrical current
and $n$ is the charge density.} and $M^{\mu\nu}$ is the polarization
tensor \cite{de-groot, sachdev}
\begin{eqnarray}\label{E22}
M^{\mu\nu}\equiv \left(\begin{array}{cccc}
0&0&0&0\\
0&0&M_{s}&0\\
0&-M_{s}&0&0\\
0&0&0&0
\end{array}\right).
\end{eqnarray}
Note that due to the antisymmetry of $M^{\mu\nu}$, (\ref{E21})
implies the third hydrodynamic relation
\begin{eqnarray}\label{E23}
\partial_{\mu}(nu^{\mu})=0,
\end{eqnarray}
which is then completed with the fourth hydrodynamic relation
\begin{eqnarray}\label{E24}
\partial_{\mu}(su^{\mu})=0.
\end{eqnarray}
The above entropy current density conservation (\ref{E24}) is valid
only in a perfect (non-dissipative) fluid. Next, we have to find a
link between the hydrodynamic quantities defined in the above
relations and the field theory identities derived in Sect.
\ref{njl}, as well as the thermodynamics relations from Sect.
\ref{mhd}. In what follows, we will first define $M_{s}$ in terms of
magnetization density $M$ that appears in the thermodynamic
relations, (\ref{E5}), (\ref{E12}) and (\ref{E15}). This will be
done by identifying the polarization tensor $M^{\mu\nu}$ with the
tensor defined in (\ref{M11}) using the standard field theoretical
methods. We will then compare the hydrodynamic energy-momentum
tensor (\ref{E19}), with the total energy-momentum tensor of an
effective NJL model in a strong magnetic field presented in
(\ref{M12}), and determine $T_{E}^{\mu\nu}$ explicitly.
\subsubsection{The polarization tensor $M^{\mu\nu}$}\label{pol}
\par\noindent
Let us start by understanding the relation between $M_{s}$ appearing
in (\ref{E22}) with the magnetization density $M$ appearing in the
thermodynamic relations (\ref{E5}), (\ref{E12}) and (\ref{E15}). To
this purpose, we define a Gibbs free energy density ${\cal{G}}$ in
the presence of a constant magnetic field $B$ \cite{shovkovy}
\begin{eqnarray}\label{E25}
{\cal{G}}(\vec{\rho};eB,T,\mu)=\frac{B^{2}}{2}+\Omega(\vec{\rho};eB,T,\mu)-H
B,
\end{eqnarray}
where $\Omega$ is the free energy density, and $H$ is the external
magnetic field. Whereas in vacuum $H=B$, in a medium with finite
magnetization density, the external magnetic field $H$ is different
from the induced magnetic field $B$ \cite{shovkovy}. Using
thermodynamical argument, we will prove the relation $B=M+H$, that,
apart from a normalization factor, appears also in \cite{shovkovy}.
To do this, let us evaluate ${\cal{G}}$ at its stationary point with
respect to $\vec{\rho}=(\sigma, \pi)$ and $B$. At this point,
${\cal{G}}$ describes all the thermodynamic properties of the system
at thermal equilibrium. The stationary point, which was in the
previous sections characterized by the configuration
$(\sigma,\pi)\to(\sigma_{0}, 0)$, corresponds to the solution of the
gap equations
\begin{eqnarray}\label{E26}
\frac{\partial{\cal{G}}}{\partial \sigma}\bigg|_{\sigma_{0}\neq 0,
\pi_{0}=0}=0,
&\qquad\mbox{or}\qquad&\frac{\partial\Omega}{\partial\sigma}\bigg|_{\sigma_{0}\neq
0, \pi_{0}=0}=0,\nonumber\\
\frac{\partial{\cal{G}}}{\partial B}\bigg|_{\sigma_{0}\neq 0,
\pi_{0}=0}=0,&\qquad\mbox{or}\qquad&\frac{\partial \Omega}{\partial
B}\bigg|_{\sigma_{0}\neq 0, \pi_{0}=0}-H+B=0.
\end{eqnarray}
The first equation in (\ref{E26}) is the same as (\ref{N27}). In the
second equation, we set, $P_{0}=-\Omega|_{\sigma_{0}\neq 0,
\pi_{0}=0}$ from (\ref{E10}). Using now $-\frac{\partial
\Omega}{\partial B}|_{\sigma_{0}\neq 0,\pi_{0}=0}=\frac{\partial
P_{0}}{\partial B}=M$ from (\ref{E16}), we arrive at
\begin{eqnarray}\label{E27}
B=H+M,
\end{eqnarray}
as was claimed. Comparing now the expression on the r.h.s. of
(\ref{E25}) with the definition of the effective action from
(\ref{N5}), we get
\begin{eqnarray}\label{E28}
{\cal{G}}(\vec{\rho};eB,T,\mu)=-{\cal{L}}_{\mbox{\tiny{eff}}}+{\cal{L}}_{k}-H
B.
\end{eqnarray}
At the stationary point, the effective kinetic term vanishes, i.e.
${\cal{L}}_{k}|_{\sigma_{0}\neq 0,\pi_{0}=0}=0$. The gap equation
$\frac{\partial{\cal{G}}}{\partial B}|_{\sigma_{0}\neq
0,\pi_{0}=0}=0$ from (\ref{E26}) leads therefore to
\begin{eqnarray}\label{E29}
\frac{\partial {\cal{L}}_{\mbox{\tiny{eff}}}}{\partial
B}\bigg|_{\sigma_{0}\neq 0, \pi_{0}=0}=-H=M-B.
\end{eqnarray}
Using the above relations, we will find in what follows the relation
between $M_{s}$ from (\ref{E22}) and the magnetization density $M$
from (\ref{E29}). To do this, we will identify the polarization
tensor $M^{\mu\nu}$ from (\ref{E22}) with the polarization tensor
defined in the field theory treatment of the effective NJL model
(\ref{M11}), i.e. $M^{\mu\lambda}\equiv F^{\mu\lambda}\frac{\partial
{\cal{L}}_{\mbox{\tiny{eff}}}}{\partial {\cal{F}}}$. For
${\mathbf{B}}=B{\mathbf{e}}_{3}$, we arrive at
\begin{eqnarray}\label{E30}
M^{\mu\lambda}\equiv \left(\begin{array}{cccc}
0&0&0&0\\
0&0&M_{s}&0\\
0&-M_{s}&0&0\\
0&0&0&0
\end{array}\right)=\left(
\begin{array}{cccc}
0&0&0&0\\
0&0&B&0\\
0&-B&0&0\\
0&0&0&0
\end{array}
\right)\frac{\partial
{\cal{L}}_{\mbox{\tiny{eff}}}}{\partial{\cal{F}}},
\end{eqnarray}
that upon using ${\cal{F}}=\frac{1}{2}B^{2}$, leads to $
M_{s}=\frac{\partial {\cal{L}}_{\mbox{\tiny{eff}}}}{\partial B}$. At
the stationary point, this relation can be compared with (\ref{E29})
and leads to the definition of $M_{s}$ in terms of the magnetization
$M$ and the induced magnetic field $B$
\begin{eqnarray}\label{E31}
M_{s}|_{\sigma_{0}\neq 0, \pi_{0}=0}=M-B.
\end{eqnarray}
Note that without considering the gauge kinetic term $-{\cal{F}}$ in
the effective action (\ref{N5}), $M_{s}|_{\sigma_{0}\neq 0,
\pi_{0}=0}$ would be the magnetization $M$ at thermal equilibrium,
as is also introduced in \cite{sachdev}.
\subsubsection{The total energy-momentum tensor
$T^{\mu\nu}_{(tot)}$}\label{energy}
\par\noindent
Our next task will be to compare the energy-momentum tensor of the
effective NJL model in the presence of a strong magnetic field
(\ref{M12}) with the energy-momentum tensor of a magnetized fluid
consisting of chiral fields (\ref{E19}). First, let us show that the
energy-momentum conservation law, (\ref{M15}), on the field theory
side reduces to (\ref{E18}) in the hydrodynamics side. To do this,
we consider (\ref{M12}), that satisfies
\begin{eqnarray}\label{E32}
u_{\nu}\partial_{\mu}T^{\mu\nu}_{(tot)}&=&-u_{\nu}F^{\nu}_{\
\lambda}M^{\mu\lambda},\nonumber\\
&=&-u_{\nu}F^{\nu}_{\
\lambda}\left(nu^{\lambda}-J^{\lambda}_{(tot)}\right),\nonumber\\
&=&u_{\nu}F^{\nu}_{\ \lambda}J_{(tot)}^{\lambda},
\end{eqnarray}
as expected. To derive (\ref{E32}), we have used the facts from
previous section, and replaced $M^{\mu\nu}$ from (\ref{M11})
appearing in the energy-momentum tensor in the field theory side,
with $M^{\mu\nu}$ that appears in the definition of
$J^{\mu}_{(tot)}$ from (\ref{E20}) on the hydrodynamics side.
Further, $\partial_{\mu}T^{\mu\nu}_{k}=0$ with $T^{\mu\nu}_{k}$ from
(\ref{M13}), and the antisymmetry of the field strength tensor
$F^{\mu\nu}$ are used.
\par
Next, we will determine $T^{\mu\nu}_{E}$, that appears in
(\ref{E19}). To do this, let us use the thermodynamic relations
(\ref{E6}), (\ref{E12}) and (\ref{E15}), and the hydrodynamic
equations (\ref{E23}) and (\ref{E24}) to determine first
$u_{\nu}\partial_{\mu}\left(wu^{\mu}u^{\nu}\right)$, where $w\equiv
\epsilon+P$,
\begin{eqnarray}\label{E33}
u_{\nu}\partial_{\mu}\left(w
u_{\mu}u_{\nu}\right)&=&\left(\partial_{\mu}w\right)u^{\mu}+w(\partial_{\mu}u^{\mu})\nonumber\\
&=&\partial_{\mu}\left(\mu n+Ts+BM\right)u^{\mu}+\left(Ts+\mu
n+BM\right)\partial_{\mu}u^{\mu}\nonumber\\
&=&u^{\mu}\partial_{\mu}P+u^{\mu}R_{k}\partial_{\mu}\rho_{k}+u_{\nu}\partial_{\mu}\left(BMu^{\mu}u^{\nu}\right).
\end{eqnarray}
On the first line, we have used $u_{\nu}u^{\nu}=1$, and set
$u_{\nu}\partial_{\mu}u^{\nu}=0$. The expression on the last line
arises by making use of  (\ref{E23}), (\ref{E24}), and (\ref{E15}),
as well as the fact that for constant magnetic field
$\partial_{\mu}B=0$. The relation
\begin{eqnarray}\label{E34}
u_{\nu}\partial_{\mu}\left((\epsilon+P-BM)u^{\mu}u^{\nu}\right)=u^{\mu}\partial_{\mu}P+u^{\mu}R_{k}\partial_{\mu}\rho_{k},
\end{eqnarray}
with zero magnetic field appears also in \cite{fraga}. It
corresponds to the Euler's equation of energy-momentum conservation
in the non-relativistic hydrodynamics, and will be used in Sect.
\ref{stability}, to derive the dispersion relation and the sound
velocity in the vicinity of chiral critical point. Plugging this
relation back in (\ref{E19}), and using the energy-momentum
conservation (\ref{E18}), we arrive first at
\begin{eqnarray}\label{E35}
u_{\nu}\partial_{\mu}T^{\mu\nu}_{E}=-u^{\mu}R_{k}\partial_{\mu}\rho_{k}-u_{\nu}\partial_{\mu}\left(BM
u^{\mu}u^{\nu}\right).
\end{eqnarray}
Using now the definition of $R_{k}=\frac{\partial\Omega}{\partial
\rho_{k}}$ from (\ref{M3}) and assuming constant $T,\mu$ and $B$, we
then get
\begin{eqnarray}\label{E36}
T^{\mu\nu}_{E}=Pg^{\mu\nu}-BM u^{\mu}u^{\nu}+C^{\mu\nu},
\end{eqnarray}
where $P=-\Omega$  is used. Here, the constant tensor $C^{\mu\nu}$
satisfies $u_{\nu}\partial_{\mu}C^{\mu\nu}=0$. Plugging (\ref{E36})
back in (\ref{E19}), the total energy-momentum tensor is given by
\begin{eqnarray}\label{E37}
T^{\mu\nu}_{(tot)}=(\epsilon+P-BM)u^{\mu}u^{\nu}-F^{\nu}_{\lambda}M^{\mu\lambda}+C^{\mu\nu}.
\end{eqnarray}
where $C^{\mu\nu}$ can be determined by comparing
$T^{\mu\nu}_{(tot)}$ from the effective field theory (\ref{E12})
with the total energy-momentum tensor from the hydrodynamic side
(\ref{E19}). It turns out to be arbitrary. To show this, we will
define, as in \cite{fraga}, an energy-momentum tensor of the fluid
$T^{\mu\nu}_{\mbox{\tiny{fluid}}}$. This will done for the effective
field theory, as well as for hydrodynamics. Only under the
assumption that these two tensors are equal, it can be shown that
$C^{\mu\nu}$ in (\ref{E37}) vanishes.
\par
Let us start by defining, as in \cite{fraga}, the energy-momentum
tensor of the fluid $T^{(1)\ \mu\nu}_{\mbox{\tiny{fluid}}}$ using
the notation of the hydrodynamics
\begin{eqnarray}\label{E38}
T^{(1)\
\mu\nu}_{\mbox{\tiny{fluid}}}=wu^{\mu}u^{\nu}-P_{\mbox{\tiny{fluid}}}g^{\mu\nu}-BM
u^{\mu}u^{\nu},
\end{eqnarray}
where $P_{\mbox{\tiny{fluid}}}\equiv -V^{(1)}=P+V^{(0)}$, with
$P=-\Omega=-(V^{(0)}+V^{(1)})$. Here, $V^{(0)}$ and $V^{(1)}$ are
the tree level and one-loop effective potentials, respectively. They
are defined in (\ref{N8}) generally and for our specific model in
(\ref{N23}). In (\ref{E38}), $T^{\mu\nu}_{\mbox{\tiny{fluid}}}$
satisfies
\begin{eqnarray}\label{E39}
u_{\nu}\partial_{\mu}T^{(1)\
\mu\nu}_{\mbox{\tiny{fluid}}}=u_{\mu}R_{k}\partial^{\mu}\rho_{k}-u_{\mu}\partial^{\mu}V_{0}=u_{\mu}\partial^{\mu}V_{1},
\end{eqnarray}
where (\ref{E34}) and the definition $R_{k}$ from (\ref{M3}) are
used. On the other hand, let us consider the energy-momentum tensor
(\ref{E12}) of the effective field theory, and define\footnote{See
\cite{paech} for a similar definition.}
\begin{eqnarray}\label{E40}
T_{\mbox{\tiny{fluid}}}^{(2)\
\mu\nu}\equiv T_{k}+V_{1}g^{\mu\nu},
\end{eqnarray}
where $T_{k}$ is defined in (\ref{E13}). Using the EoM, we have
already shown that $\partial_{\mu}T_{k}^{\mu\nu}=0$. We arrive
therefore at
\begin{eqnarray}\label{E41}
u_{\nu}\partial_{\mu}T^{(2)\
\mu\nu}_{\mbox{\tiny{fluid}}}=u_{\mu}\partial^{\mu}V_{1}.
\end{eqnarray}
Comparing (\ref{E41}) with (\ref{E39}) and assuming that $T^{(1)\
\mu\nu}_{\mbox{\tiny{fluid}}}=T^{(2)\
\mu\nu}_{\mbox{\tiny{fluid}}}$, we arrive at the relation
\begin{eqnarray}\label{E42}
T_{k}=(\epsilon+P-BM)u^{\mu}u^{\nu},
\end{eqnarray}
where $P_{\mbox{\tiny{fluid}}}\equiv -V^{(1)}=P+V^{(0)}$ and
$P=-\Omega=-(V^{(0)}+V^{(1)})$ are used. Plugging finally
(\ref{E42}) back in (\ref{E12}) and comparing the resulting
expression with (\ref{E37}), we get $C^{\mu\nu}=0$, and
\begin{eqnarray*}
T^{\mu\nu}_{(tot)}=(\epsilon+P-BM)u^{\mu}u^{\nu}-F^{\nu}_{\lambda}M^{\mu\lambda},\qquad\mbox{with}\qquad
u_{\nu}\partial_{\mu}T^{\mu\nu}_{(tot)}=u_{\nu}F^{\nu}_{\lambda}J^{\lambda}_{(tot)},
\end{eqnarray*}
as expected. Note that the above assumption  $T^{(1)\
\mu\nu}_{\mbox{\tiny{fluid}}}=T^{(2)\ \mu\nu}_{\mbox{\tiny{fluid}}}$
leading in particular to the relation (\ref{E42}) is the link
between effective field theory and hydrodynamics. Whereas $T_{k}$,
on the effective field theory side of (\ref{E42}), includes the
detailed information concerning the matter content of the model and
the interactions involved, the expression
$(\epsilon+P-BM)u^{\mu}u^{\nu}$ on the hydrodynamic side describes
effectively the state of a perfect magnetized fluid through its
energy density $\epsilon$, its pressure $P$ and its magnetization
$M$. In contrast to \cite{sachdev}, the present MHD description of a
magnetized fluid consists of chiral field or equivalently
quasiparticles $\sigma$ and $\pi$ in the language of condensed
matter physics, and exhibits through the special construction of our
model a spontaneous chiral symmetry breaking.
\par
In the next section, using the above MHD description and performing
a first order stability analysis, we will study the effect of the
external magnetic field on the sound modes propagating in an
expanding magnetized QGP including the chiral fields.
\section{Stability analysis and sound velocity in a strong magnetic
field}\label{stability} 
\setcounter{equation}{0}\par\noindent In this section, we will first
determine the dispersion relation of a perfect magnetized QGP at
finite $(T,\mu)$ coupled to chiral fields. Using the dispersion
relation, we will then determine the sound velocity of plane waves
propagating in this hot and dense medium. Due to the broken
rotational symmetry in the presence of a constant magnetic field, we
expect the sound velocity to have an unisotropic distribution. In
particular, for plane waves propagating in the transverse as well as
longitudinal directions with respect to the external magnetic field,
different dependence on the angle $(\theta,\varphi)$ of the
spherical coordinate system will arise.
\subsection{The dispersion relation in a perfect magnetized QGP coupled to chiral fields}\label{dispersion}
\par\noindent
To study the effect of the magnetic field on the magnetized plasma
modeled in this paper via the effective NJL model in the presence of
a strong magnetic field, we will study the onset of instabilities by
performing a first order stability analysis. In Table \ref{table1},
a list of all the necessary relations from field theory,
thermodynamics and hydrodynamics is presented. They are assumed to
be valid in a state out of equilibrium.
\begin{center}
\begin{table}[h]
{\small{
\begin{tabular}{l|ccll}\hline\hline
Field
theory&$F_{1}^{\mu}\partial_{\mu}\partial^{\mu}\rho_{k}+\frac{2\rho_{k}}{\rho^{2}}
                        F_{2}^{\mu}\left(\rho_{i}\partial_{\mu}\partial^{\mu}\rho_{i}+\partial_{\mu}\rho_{i}\partial^{\mu}\rho_{i}-\frac{1}{\rho^{2}}
                        \left(\rho_{i}\partial_{\mu}\rho_{i}\right)\left(\rho_{j}\partial^{\mu}\rho_{j}\right)\right)
                        = -R_{k},
                        $&&from& (\ref{M2})\\
\hline
Thermodynamics&$\epsilon+P=Ts+\mu n+BM,$&&from& (\ref{E6})\\
                       &$d\epsilon=Tds+\mu dn+BdM+R_{k}d\rho_{k},$&&from& (\ref{E12})\\
                       &$dP=sdT+nd\mu+MdB-R_{k}d\rho_{k},$&&from& (\ref{E15})\\
\hline
Hydrodynamics&$\partial_{\mu}(nu^{\mu})=0$,&&from& (\ref{E23})\\

                        &$\partial_{\mu}(su^{\mu})=0$,&&from& (\ref{E24})\\
                        &$u_{\nu}\partial_{\mu}\left((\epsilon+P-BM)u^{\mu}u^{\nu}\right)=u^{\mu}\partial_{\mu}P+u^{\mu}R_{k}\partial_{\mu}\rho_{k},$&&from& (\ref{E34})\\
\hline\hline
\end{tabular}
}} \caption{}\label{table1}
\end{table}
\end{center}
As was indicated in the previous section, we will assume that
$B,T,\mu$, and $M$ remain constant (in the space-time $x$) and will
expand $n,s,\vec{\rho}=(\sigma,\pi)$, and $u_{\mu}$ around their
equilibrium configurations $n_{0}, s_{0}, \vec{\rho}_{0}$ and
$v^{\mu}_{0}=(1,{\mathbf{0}})$ to first order
\begin{eqnarray}\label{D1}
n(x)&=&n_{0}+n_{1}(x),\nonumber\\
s(x)&=&s_{0}+s_{1}(x),\nonumber\\
\sigma(x)&=&\sigma_{0}+\sigma_{1}(x),\nonumber\\
\pi(x)&=&\pi_{0}+\pi_{1}(x),\nonumber\\
u^{\mu}(x)&=&v^{\mu}_{0}+v^{\mu}_{1}(x).
\end{eqnarray}
Here, $\pi_{0}=0$, $v^{\mu}_{0}=(1,{\mathbf{0}})$ and
$v^{\mu}_{1}=(0,{\mathbf{v}}_{1})$. Further, according to the
arguments in the paragraph following (\ref{E17}), in the thermal
equilibrium $R_{\sigma_{0}}=R_{\pi_{0}}=0$. Perturbing first the EoM
(\ref{M2}), we arrive at
\begin{eqnarray}\label{D2}
F_{1}^{\mu}\partial_{\mu}\partial^{\mu}\rho_{1,k}+\frac{2\rho_{0,k}\rho_{0,j}}{\rho_{0}^{2}}
F_{2}^{\mu}\partial_{\mu}\partial^{\mu}\rho_{1,j}=-n_{1}\left(\frac{\partial
R_{k}}{\partial
n}\right)_{s_{0},M,\vec{\rho}_{0}}-s_{1}\left(\frac{\partial
R_{k}}{\partial s}\right)_{n_{0},M,\vec{\rho}_{0}}-m_{k\ell}^{2}\
\rho_{1,\ell},
\end{eqnarray}
with the mass matrix
\begin{eqnarray}\label{D3}
m_{k\ell}^{2}\equiv
\left(\frac{\partial^{2}\epsilon}{\partial\rho_{k}\partial\rho_{\ell}}\right)_{n_{0},s_{0},M}.
\end{eqnarray}
Here, we have used the definition of $R_{k}=\frac{\partial
\epsilon}{\partial\rho_{k}}$, which we assume to be also valid in a
state out of equilibrium. Expanding $\epsilon(\vec{\rho};n,s,M)$ up
to second order in the thermodynamic quantities, we get
\begin{eqnarray}\label{D4}
\epsilon=\epsilon_{0}+n_{1}\left(\frac{\partial\epsilon_{0}}{\partial
n_{0}}\right)_{s_{0},M,\rho_{0}}+s_{1}\left(\frac{\partial\epsilon_{0}}{\partial
s_{0}}\right)_{n_{0},M,\rho_{0}}+\rho_{1,\ell}\left(\frac{\partial\epsilon_{0}}{\partial\rho_{0,\ell}}\right)_{n_{0},s_{0},M},
\end{eqnarray}
with
$\epsilon_{0}=\epsilon_{0}\left(\vec{\rho_{0}},n_{0},s_{0},M\right)$.
Differentiating (\ref{D4}) with respect to $\rho_{0,k}$ and using
$R_{\sigma_{0}}=R_{\pi_{0}}=0$, we arrive at (\ref{D2}). Note, that
$F_{i}^{\mu}(\sigma_{0};eB,T,\mu), i=1,2$ are already defined at the
vacuum configuration $(\sigma_{0}\neq 0, \pi_{0}=0)$ [see Sect.
\ref{kinetic}]. For $k=\sigma$, (\ref{D2}) reduces to
\begin{eqnarray}\label{D5}
\big[G^{\mu}\partial_{\mu}\partial^{\mu}+m_{\sigma}^{2}\big]\sigma_{1}=
-n_{1}\left(\frac{\partial R_{\sigma}}{\partial
n}\right)_{s_{0},M,\sigma_{0}}-s_{1}\left(\frac{\partial
R_{\sigma}}{\partial s}\right)_{n_{0},M,\sigma_{0}},
\end{eqnarray}
with $ m^{2}_{\sigma\sigma}\equiv m_{\sigma}^{2}$ and
$G^{\mu}(\sigma_{0};eB,T,\mu)\equiv
 F_{1}^{\mu}+2F_{2}^{\mu}$, whereas for $k=\pi$, we arrive at
\begin{eqnarray}\label{D6}
F_{1}^{\mu}\partial_{\mu}\partial^{\mu}\pi_{1}=0.
\end{eqnarray}
The conservation laws (\ref{E23}) and (\ref{E24}) lead to
\begin{eqnarray}\label{D7}
\partial_{0}n_{1}(x)+n_{0}\mbox{\textbf{$\nabla$}}\cdot
{\mathbf{v}}_{1}&=&0,\nonumber\\
\partial_{0}s_{1}(x)+s_{0}\mbox{\textbf{$\nabla$}}\cdot
{\mathbf{v}}_{1}&=&0.
\end{eqnarray}
Finally, perturbing the energy-momentum conservation law
(\ref{E34}), we get
\begin{eqnarray}
W_{0}\nabla\cdot{\mathbf{v}}_{1}&=&-\partial_{0}\epsilon_{1}, \label{D8}\\
 W_{0}\
\partial_{0}{\mathbf{v}}_{1}&=&-\nabla P_{1},\label{D9}
\end{eqnarray}
where $W_{0}\equiv \epsilon_{0}+P_{0}-BM=Ts+\mu n$. Using
\begin{eqnarray}\label{D10}
\epsilon_{1}=\mu n_{1}+Ts_{1},
\end{eqnarray}
that arises from (\ref{D4}), and the conservation laws (\ref{D7}),
the time component of the Euler's equation, (\ref{D8}), is
automatically satisfied. As for (\ref{D9}), $P_{1}$ is given by
\begin{eqnarray}\label{D11}
P_{1}=n_{1}\frac{\partial P_{0}}{\partial n_{0}}+s_{1}\frac{\partial
P_{0}}{\partial s_{0}}+\sigma_{1}\frac{\partial P_{0}}{\partial
\sigma_{0}},
\end{eqnarray}
where $P_{0}\equiv P(\epsilon_{0},n_{0}, s_{0}, \sigma_{0},M)$.
Using now $P_0=-\epsilon_0+Ts_0+\mu n_0+BM$, and the thermodynamic
relations (\ref{D14}) in thermal equilibrium, we have
\begin{eqnarray}\label{D12}
\frac{\partial P_{0}}{\partial
n_{0}}&=&n_{0}\frac{\partial^{2}\epsilon_{0}}{\partial
n_{0}^{2}}+s_{0}\frac{\partial^{2}\epsilon_{0}}{\partial
s_{0}\partial n_{0}}+M\frac{\partial^{2}\epsilon_{0}}{\partial M\partial n_{0}}\nonumber\\
\frac{\partial P_{0}}{\partial
s_{0}}&=&n_{0}\frac{\partial^{2}\epsilon_{0}}{\partial n_{0}\partial
s_{0}}+s_{0}\frac{\partial^{2}\epsilon_{0}}{\partial
s_{0}^{2}}+M\frac{\partial^{2}\epsilon_{0}}{\partial M\partial s_{0}}\nonumber\\
\frac{\partial P_{0}}{\partial
\sigma_{0}}&=&R''_{\sigma}\left(W_{0}+BM\right)=R''_{\sigma}\left(\epsilon_{0}+P_{0}\right),
\end{eqnarray}
with
\begin{eqnarray}\label{D13}
R''_{\sigma}\equiv
\frac{1}{(W_{0}+BM)}\bigg[n_{0}\left(\frac{\partial
R_{\sigma}}{\partial
n}\right)_{s_{0},M,\sigma_{0}}+s_{0}\left(\frac{\partial
R_{\sigma}}{\partial
s}\right)_{n_{0},M,\sigma_{0}}+M\left(\frac{\partial
R_{\sigma}}{\partial M}\right)_{n_{0},s_{0},\sigma_{0}}\bigg].
\end{eqnarray}
To determine the dispersion law in the above magnetized QGP, we
consider a plane wave of the form
$\xi_{1}(x)=\tilde{\xi}_{1}e^{-ikx}$, with $\xi_{1}=\{n_{1},
s_{1},\sigma_{1}, \pi_{1}, {\mathbf{v}}_{1}\}$ and
$k^{\mu}=(\omega,{\mathbf{k}})$ in the medium and rewrite the
relations (\ref{D5})-(\ref{D7}), and (\ref{D9}) in the momentum
space as
\begin{eqnarray}\label{D14}
\left(F_{1}^{0}\omega^{2}-F_{1}^{i}k_{i}^{2}\right)\tilde{\pi}_{1}&=&0,\nonumber\\
\left(G^{0}\omega^{2}-G^{i}k_{i}^{2}-m_{\sigma}^{2}\right)\tilde{\sigma}_{1}&=&\frac{W_{0}}{\omega}R'_{\sigma}
{\mathbf{k}}\cdot\tilde{\mathbf{v}}_{1},\nonumber\\
\left(\omega^{2}-{\mathbf{k}}^{2}P'\right){\mathbf{k}}\cdot
\tilde{\mathbf{v}}_{1}&=&\frac{\omega}{W_{0}}\left(W_{0}+BM\right){\mathbf{k}}^{2}R''_{\sigma}\tilde{\sigma}_1,
\end{eqnarray}
where,
\begin{eqnarray}\label{D15}
R'_{\sigma}&\equiv& \frac{1}{W_{0}}\bigg[n_{0}\left(\frac{\partial
R_{\sigma}}{\partial
n}\right)_{s_{0},M,\sigma_{0}}+s_{0}\left(\frac{\partial
R_{\sigma}}{\partial s}\right)_{n_{0},M,\sigma_{0}}\bigg],\nonumber\\
P'&\equiv& \frac{1}{W_{0}}\bigg[n_{0}\left(\frac{\partial
P}{\partial n}\right)_{s_{0},M,\sigma_{0}}+s_{0}\left(\frac{\partial
P}{\partial s}\right)_{n_{0},M,\sigma_{0}}\bigg].
\end{eqnarray}
In Appendix \ref{derivation}, we present a derivation of the last
two equations in (\ref{D14}), using the relations
\begin{eqnarray}\label{D16}
\frac{\tilde{n}_{1}}{n_{0}}=\frac{\tilde{s}_{1}}{s_{0}}=\frac{{\mathbf{k}}\cdot
\tilde{\mathbf{v}}_{1}}{\omega},
\end{eqnarray}
that arise by plugging $n_{1}(x)=\tilde{n}_{1}e^{-ikx}$ as well as
$s_{1}(x)=\tilde{s}_{1}e^{-ikx}$ in (\ref{D7}). Multiplying the last
two equations in (\ref{D14}), we arrive at the dispersion relation
\begin{eqnarray}\label{D18}
\left(G^{0}\omega^{2}-G^{i}k_{i}^{2}-m_{\sigma}^{2}\right)\left(\omega^{2}-{\mathbf{k}}^{2}P'\right)=
{\mathbf{k}}^{2}R_{\sigma}^{'}R_{\sigma}^{''}\left(W_{0}+BM\right).
\end{eqnarray}
In what follows, we will linearize the above dispersion relation in
$k$ and determine the sound velocity of a plane wave propagating in
the magnetized QGP coupled to the chiral field $\sigma$. Note that,
in the above first order perturbation, the pion field $\pi$ is
decoupled from the dynamics of the expanding magnetized QGP. Similar
phenomenon is also observed in \cite{fraga}, where the effective
action of the linear $\sigma$ model is used to present a chiral
hydrodynamic description of an expanding QGP.
\subsection{Sound velocity in a perfect magnetized QGP coupled to chiral fields}\label{sound}
\par\noindent
Sound is defined as a small disturbance propagating in a uniform,
field free fluid at rest. Perturbing the energy density
$\epsilon(x)$ and the pressure $P(x)$ around their equilibrium
values $\epsilon_{0}$ and $P_{0}$, up to small $\delta\epsilon$ and
$\delta P$, and linearizing the energy-momentum conservation
equation, we arrive at
\begin{eqnarray}\label{D19}
\frac{\partial^{2}(\delta\epsilon)}{\partial
t^{2}}-c_{s}^{2}\Delta(\delta\epsilon)=0,
\end{eqnarray}
where $c_{s}^{2}\equiv \frac{\partial P}{\partial \epsilon}$. For an
ideal classical gas, consisting of independent and massless
particles without interaction, $c_{s}^{2}=1/3$. This can be shown
using the identity $P=\frac{\epsilon}{3}$, which is derived using
the kinetic pressure and the energy density in terms of
Maxwell-Boltzmann statistics, that replaces the Bose-Einstein and
Fermi-Dirac statistics in a classical limit. Note that the relation
$P=\frac{\epsilon}{3}$, leading to $c_{s}^{2}=1/3$, arises only by
assuming the rotational symmetry. It holds approximately for a
uniform QGP at high temperature, where interaction are small due to
asymptotic freedom \cite{ollitrault}.
\par
In \cite{fraga}, the sound velocity of a perfect chiral fluid is
determined using the effective potential of a linear $\sigma$ model
in terms of chiral fields. The dispersion relation of this model is
similar to (\ref{D18}), where $G^{i}=1, i=0,\cdots 3$ and $B=0$.
Assuming the rotational symmetry, the dispersion relation is then
solved. For long wavelength fluctuations, the roots are
approximately given by \cite{fraga}
\begin{eqnarray}\label{D20}
\omega_{\sigma}^{2}&=&m_{\sigma}^{2}+{\cal{O}}\left(k^{2}\right),\nonumber\\
\omega_{p}^{2}/k^{2}&=&\left(P'-\frac{w_{0}R^{'2}_{\sigma}}{m_{\sigma}^{2}}\right)+{\cal{O}}\left(k^{2}\right),
\end{eqnarray}
where $w_{0}\equiv \epsilon_{0}+P_{0}$. Note, however, that in the
presence of a constant magnetic field which is aligned e.g. in the
third direction, the rotational symmetry is \textit{a priori}
broken, and the sound velocity will be therefore different from
$c_{s}^{2}=1/3$ of an ideal non-interacting QGP. It will be also
different from (\ref{D20}) for a perfect QGP coupled to chiral
fields.
\par
To determine the sound velocity in a perfect fluid in the presence
of a constant magnetic field and consisting of chiral fields, we
will use a method introduced in \cite{parsons}, where a linearized
hydrodynamical theory of magnetic fluids\footnote{In \cite{parsons},
a magnetic fluid is defined as a colloid of tiny (100 {\AA})
magnetic particles or grains suspended in a carrier fluid as water.}
in a strong magnetic field is presented. Solving the equations for a
sound wave propagating at angle $(\theta,\varphi)$ with respect to
the external magnetic field direction, the sound velocity is shown
to be anisotropic. Here, $(\theta,\varphi)$ are the angle in
spherical coordinate system. In the next paragraphs, we will
consider two different cases separately: a) propagation in the
${\mathbf{e}}_1-{\mathbf{e}}_2$ plane, and b) propagation in the
${\mathbf{e}}_1-{\mathbf{e}}_3$ plane. We will show that whereas the
the sound velocity for a propagation transverse to the external
magnetic field ${\mathbf{B}}=B{\mathbf{e}}_{3}$ (case a) is
independent of the angle $\varphi$, for the propagation longitudinal
to ${\mathbf{B}}$ (case b), it depends on the angle $\theta$ between
the wave vector ${\mathbf{k}}$ and the direction of the external
magnetic field.
\subsubsection{Propagation in ${\mathbf{e}}_1-{\mathbf{e}}_2$
plane}\label{prop12}
\par\noindent
Let us begin our derivation, by choosing a plane wave with the wave
vector ${\mathbf{k}}=(k\sin\theta\cos\varphi,
k\sin\theta\sin\varphi, k\cos\theta)$ in the
${\mathbf{e}}_1-{\mathbf{e}}_2$ plane, transverse to the direction
of the magnetic field ${\mathbf{B}}=B{\mathbf{e}}_{3}$. This
corresponds to $\theta=\pi/2$ and $\varphi\neq 0$, where $0\leq
\theta\leq \pi$ and $0\leq \varphi\leq 2\pi$ are angles in spherical
coordinate system. The wave vector is therefore given by
${\mathbf{k}}_{12}=\left(k_{12}\sin\varphi,
k_{12}\cos\varphi,0\right)$, where the superscripts correspond to a
propagation in the ${\mathbf{e}}_1-{\mathbf{e}}_2$ plane. For a
given frequency $\omega$, equation (\ref{D18}) determines $k$. In a
linearized hydrodynamics, we will evaluate $k$ around
$k_{0}\equiv\frac{\omega}{c_{s}}$, where $c_{s}=1/\sqrt{3}$ is the
sound velocity in an ideal classical gas. Thus plugging, the
relation
\begin{eqnarray}\label{D21}
k_{12}=k_{0}+k'_{12},
\end{eqnarray}
in the dispersion relation (\ref{D18}) and expanding the resulting
expression in the order of $k'_{12}$ up to ${\cal{O}}(k'_{12})$, we
get
\begin{eqnarray}\label{D22}
k'_{12}=\frac{{\cal{N}}_{12}}{{\cal{D}}_{12}},
\end{eqnarray}
with the numerator
\begin{eqnarray}\label{D23}
{\cal{N}}_{12}&=&-2c_{s}^{2}
[m_{\sigma}^{2}(c_{s}^{2}-P')+R^{'}_{\sigma}R''_{\sigma}w_{0}]+\omega^{2}(2c_{s}^{2}G^{0}-G^{1}-G^{2})(c_{s}^{2}-P')\nonumber\\
&&-\omega^{2}(G^{1}-G^{2})(c_{s}^{2}-P')\cos(2\varphi),
\end{eqnarray}
with $w_{0}\equiv W_{0}+BM=\epsilon_{0}+P_{0}$ and the denominator
\begin{eqnarray}\label{D24}
{\cal{D}}_{12}&=&2c_{s}^{2}[m_{\sigma}^{2}(c_{s}^{2}+P')-R'_{\sigma}R''_{\sigma}w_{0}]-
\omega^{2}[2c_{s}^{4}G^{0}-3(G^{1}+G^2)P'\nonumber\\
&&+c_{s}^{2}\left(G^{1}+G^{2}+2G^{0}P'\right)]-(G^{1}-G^{2})\omega^{2}\cos(2\varphi).
\end{eqnarray}
Setting $G^{i}=1, i=0,1,2$, $P'=c_{s}^{2}$ and $B=0$ as well as
$R'_{\sigma}=R''_{\sigma}=0$, we get $k'_{12}=0$, as expected. To
determine the frequency dependent anisotropy $\Delta$, we use the
identity $\omega\approx k_{12} v_{12}$ with
$k_{12}=\frac{\omega}{c_{s}}+k_{12}'$ and $v_{12}\equiv
c_{s}\left(1+\Delta_{12}\right)$. Using $k'_{12}$ from
(\ref{D22})-(\ref{D24}), we arrive at
\begin{eqnarray}\label{D25}
\Delta_{12}(G^{0}, G^{1}, G^{2};
\varphi,\omega)=-\frac{k'_{12}c_{s}}{k'_{12}c_{s}+\omega},
\end{eqnarray}
where $G^{i}, i=0,1,2$ are functions of $eB, T$ and $\mu$. In our
specific model with $G^{1}=G^{2}$ [see Sect. \ref{kinetic}], the
$\varphi$-dependence in the anisotropy $\Delta_{12}$ vanishes. The
anisotropy is then given by
\begin{eqnarray}\label{D26}
\Delta_{12}(G^{0},
G^{1};\omega)=\frac{-2c_{s}^{2}[m_{\sigma}^{2}(c_{s}^{2}-P')+R'_{\sigma}R''_{\sigma}w_{0}]+\omega^{2}
(2c_{s}^{2}G^{0}-2G^{1})(c_{s}^{2}-P')}
{2c_{s}^{2}[m_{\sigma}^{2}(c_{s}^{2}+P')-R'_{\sigma}R''_{\sigma}w_{0}]-\omega^{2}[2c_{s}^{4}G^{0}-6G^{1}P'+c_{s}^{2}(2G^{1}+2G^{0}P')]}.
\end{eqnarray}
Setting $G^{i}=1, i=0,1$, $P'=c_{s}^{2}$ and $B=0$ as well as
$R'_{\sigma}=R''_{\sigma}=0$, we get $\Delta_{12}=0$, as expected.
\subsubsection{Propagation in ${\mathbf{e}}_1-{\mathbf{e}}_3$
plane}\label{prop13}
\par\noindent
Setting $\theta\neq 0, \varphi=0$, the wave vector in the
${\mathbf{e}}_1-{\mathbf{e}}_3$ is given by
${\mathbf{k}}_{13}=\left(k_{13}\sin\theta,0,
k_{13}\cos\theta\right)$. Plugging now the approximation
$k_{13}=k_{0}+k'_{13}$, with $k_{0}=\frac{\omega}{c_{s}}$ in the
dispersion relation (\ref{D18}), and expanding the resulting
expression in the orders of $k'_{13}$, we arrive at
\begin{eqnarray}\label{D27}
k'_{13}=\frac{{\cal{N}}_{13}}{{\cal{D}}_{13}},
\end{eqnarray}
with the numerator
\begin{eqnarray}\label{D28}
{\cal{N}}_{13}&=&2c_{s}^{2}\omega
[m_{\sigma}^{2}(c_{s}^{2}-P')+R^{'}_{\sigma}R''_{\sigma}w_{0}]-\omega^{3}(2c_{s}^{2}G^{0}-G^{1}-G^{3})(c_{s}^{2}-P')\nonumber\\
&&-\omega^{3}(G^{1}-G^{3})(c_{s}^{2}-P')\cos(2\theta),
\end{eqnarray}
and the denominator
\begin{eqnarray}\label{D29}
{\cal{D}}_{13}&=&4c_{s}\omega^{2}(G^{1}+G^{3})P'+2c_{s}^{3}[2m_{\sigma}^{2}P'-2R^{'}_{\sigma}R''_{\sigma}w_{0}-\omega^{2}(G^{1}+G^{3}+2G^{0}P')]\nonumber\\
&&+2c_{s}\omega^{2}(G^{1}-G^{3})(c_{s}^{2}-2P')\cos(2\theta).
\end{eqnarray}
Setting $G^{i}=1, i=0,1,2$, $P'=c_{s}^{2}$ and $B=0$ as well as
$R'_{\sigma}=R''_{\sigma}=0$, we get $k'_{13}=0$, as expected. The
anisotropy $\Delta_{13}$ is then determined using the identity
$\omega\approx k_{13} v_{13}$ with $k_{13}=k_{0}+k_{13}$ and
$v_{13}=c_{s}\left(1+\Delta_{13}\right)$. Plugging $k'_{13}$ from
(\ref{D27})-(\ref{D29}) in $k_{13}$, the anisotropy reads
\begin{eqnarray}\label{D30}
\Delta_{13}(G^{0}, G^{1}, G^{3};
\theta,\omega)=\frac{\Delta_{n}}{\Delta_{d}},
\end{eqnarray}
with
\begin{eqnarray}\label{D31}
\Delta_{n}&=&-2c_{s}^{2}[m_{\sigma}^{2}(c_{s}^{2}-P')+R^{'}_{\sigma}R''_{\sigma}w_{0}]+\omega^{2}(2c_{s}^{2}G^{0}-G^{1}-G^{3})(c_{s}^{2}-P')
\nonumber\\
&&+\omega^{2}(G^{1}-G^{3})(c_{s}^{2}-P')\cos(2\theta),
\end{eqnarray}
and
\begin{eqnarray}\label{D32}
\Delta_{d}&=&2c_{s}^{2}[m_{\sigma}^{2}(c_{s}^{2}+P')-R^{'}_{\sigma}R''_{\sigma}w_{0}]-\omega^{2}[2c_{s}^{4}G^{0}-3(G^{1}+G^{3})P'+c_{s}^{2}(G^{1}+G^{3}+2G^{0}P')]\nonumber\\
&&+\omega^{2}(G^{1}-G^{3})(c_{s}^{2}-3P')\cos(2\theta).
\end{eqnarray}
Setting $G^{i}=1, i=0,1$, $P'=c_{s}^{2}$ and $B=0$ as well as
$R'_{\sigma}=R''_{\sigma}=0$, we get $\Delta_{13}=0$, as expected.
Here, in contrast to $\Delta_{12}$, the $\theta$-dependence remains
in the anisotropy $\Delta_{13}$. In the next section, we will
determine numerically the anisotropy functions $\Delta_{12}$ and
$\Delta_{13}$ for plane waves propagating in the transverse and
longitudinal directions with respect to the external magnetic
fields, respectively.
\section{Numerical results and discussion}\label{numerics}
\subsection{Energy density and pressure of a magnetized QGP near the chiral phase transition point}\label{num-energy}
\par
\begin{figure}[hbt]
\includegraphics[width=7.cm, height=5.cm]{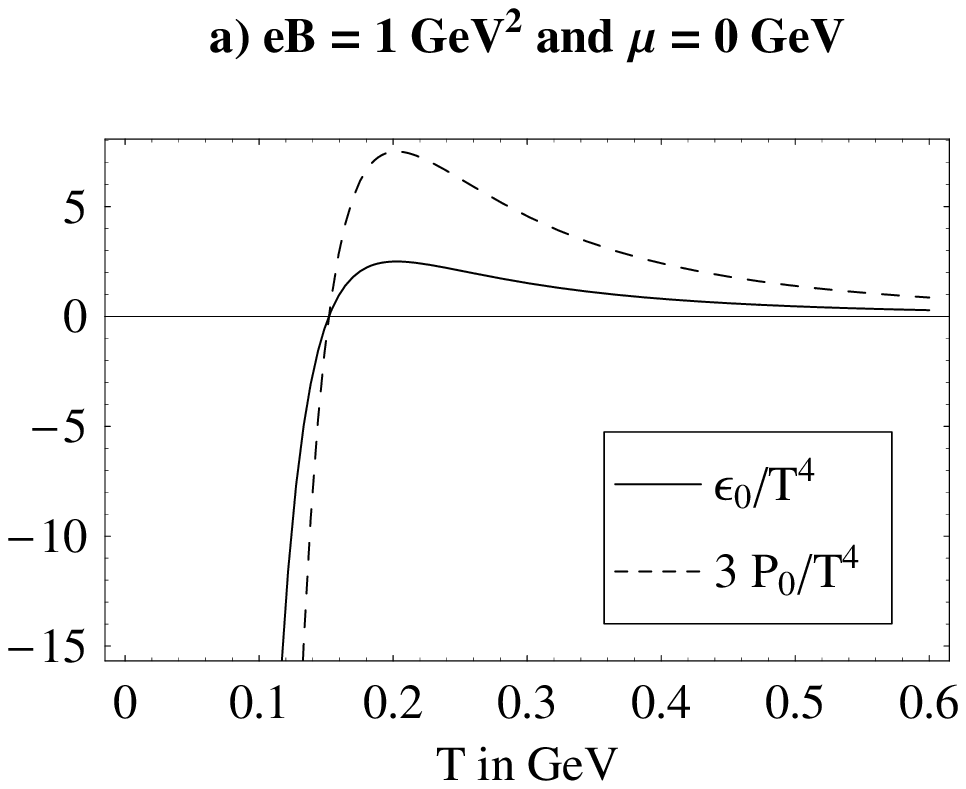}
\includegraphics[width=7.cm, height=5.cm]{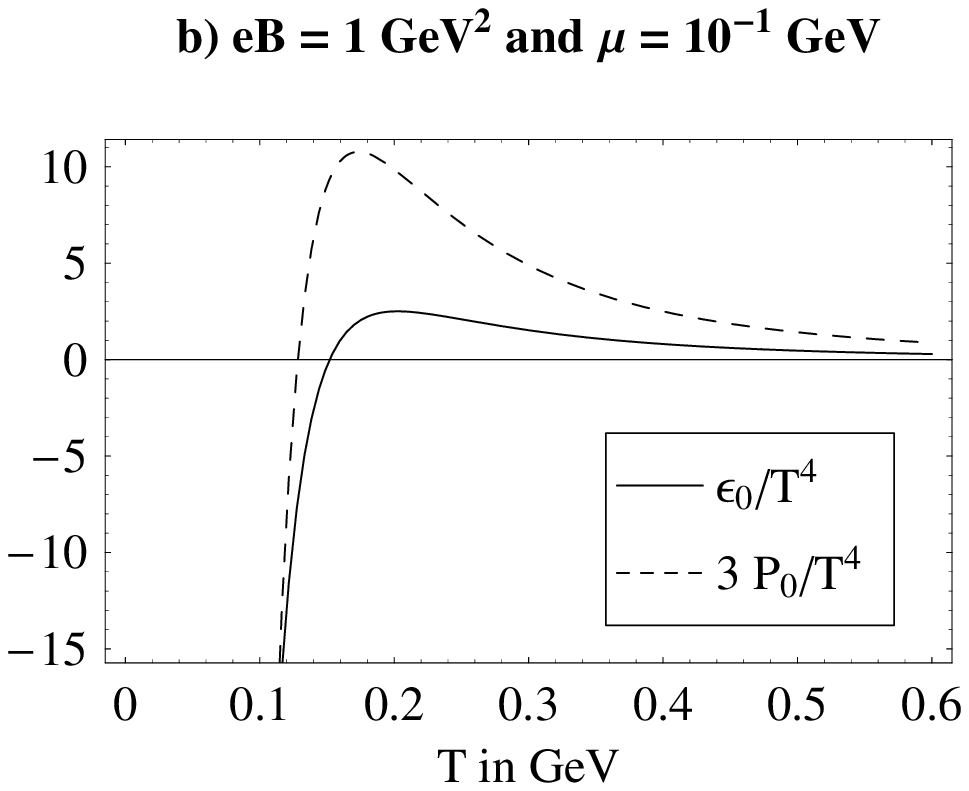}
\caption{The temperature dependence of the energy density
$\epsilon_{0}$ and pressure $P_{0}$ for $eB=1$ GeV$^2$ and $\mu=0$
GeV (a) as well as $\mu=10^{-1}$ GeV (b) at thermal equilibrium. At
$T_{c}\sim 100-120$ MeV, there is an increase of the energy density
as a consequence of a first order phase transition.
}\label{energydensity}
\end{figure}

\noindent
In this section, we will use the results from previous sections to
study the effect of a strong magnetic field on the sound velocity
$v_{s}$ of a plane wave propagating in an expanding
\textit{magnetized} QGP. The latter is modeled in this paper by the
NJL effective action in the presence of a constant magnetic field.
In previous section, using a linear approximation, we have defined
an anisotropy function $\Delta\equiv \frac{v_{s}}{c_{s}}-1$, where
$c_{s}=1/\sqrt{3}$ is the sound velocity in an ideal gas without
matter fields. In \cite{florkowski}, the temperature dependence of
$v_{s}$ in a hot QGP without magnetic field is studied. It is shown
that the sound velocity shows a minimum at the chiral critical
point. Whereas the results in \cite{florkowski} is found using
purely hydrodynamical arguments, we intend to present in this paper
a microscopic model which is merged carefully with hydrodynamics to
study the $(T,\mu)$ dependence of sound velocity in a magnetized
QGP. To this purpose, we have to know approximately at which
critical temperature, $T_{c}$, the expected chiral phase transition
may occur. The temperature dependence of the energy density of a
system can give us a useful hint in this direction. As it is known
from lattice QCD, we expect an increase in pressure $P$ and the
energy density $\epsilon$ at the chiral critical point as a
consequence of a first order phase transition. To determine $P$ in
our NJL model in a strong magnetic field, the relation (\ref{D10})
can be used between the thermodynamic pressure and the one-loop
effective potential $\Omega$ which is determined in Sect.
\ref{effpot}. The energy density $\epsilon$ can be determined using
the equation of state (\ref{D6}), i.e. $\epsilon=-P+n\mu+TS+BM$.
Here, the number and entropy densities $n$ and $s$, as well as the
magnetization density $M$ are given by thermodynamic relations
(\ref{D16}). To determine the thermodynamical quantities $\epsilon,
P, n$ and $s$ at thermal equilibrium, i.e. $\epsilon_{0}, P_{0},
n_{0}$ and $s_{0}$, we have to replace $\vec{\rho}=(\sigma,\pi)$
appearing in the thermodynamical potential (\ref{N23}) by the
dynamical mass $\sigma_{0}(eB,T,\mu)$ which is determined in
(\ref{N30}).\footnote{Let us remind that the configuration
$(\sigma_{0}\neq 0, \pi_{0}=0)$ builds the stationary point of the
effective potential or equivalently describes the thermodynamic
equilibrium.} As we have shown in Sect. \ref{dynamical}, the
dynamical mass can only be determined approximately. Using the same
approximations as described in the last paragraph of Sects.
\ref{effpot} and \ref{dynamical}, we have determined the energy
density $\epsilon_{0}/T^{4}$ and pressure $3P_{0}/T^{4}$ as a
function of $T$ for a fixed value of the magnetic field $eB=1$
GeV$^{2}$ and two different values of the chemical potential $\mu=0,
10^{-1}$ GeV  in Fig. \ref{energydensity}. As it is expected, the
energy density and pressure have a maximum at $T_{c}\sim 100-120$
MeV.
\par\noindent
We will next study the effect of the strong magnetic field fixed at
$eB=1$ GeV$^{2}$ on the sound velocity in the vicinity of the chiral
critical point $T_{c}\sim 100-120$ MeV. As we have seen in Sect.
\ref{prop12} and \ref{prop13}, the presence of an external magnetic
field leads to an anisotropy in the sound velocity, so that the
anisotropy function $\Delta$ for a plane wave propagating in the
transverse ($\Delta_{12}$) and longitudinal direction
($\Delta_{13}$) with respect to the external magnetic fields have
different dependence on the angles $(\theta,\varphi)$ of the
spherical coordinate systems. In what follows, we will study the
temperature dependence of $\Delta_{12}$ and $\Delta_{13}$
separately.
\subsection{Anisotropy function and sound velocity for a
propagation in ${\mathbf{e}}_1-{\mathbf{e}}_2$
plane}\label{num-sound-1}
\par
\begin{figure}[hbt]
\includegraphics[width=8.4cm, height=5.5cm]{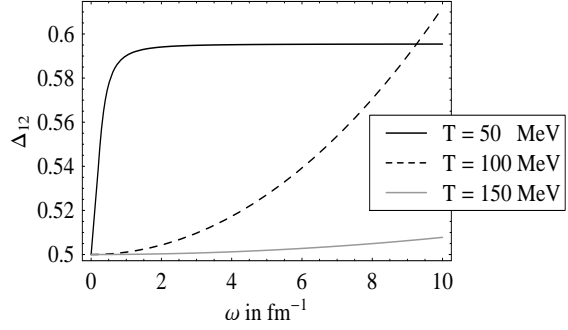}
\caption{The frequency dependence of $\Delta_{12}\equiv
\frac{v_{12}}{c_{s}}-1$ for three different temperature $T=50$ MeV
(below $T_{c}$), $T=100$ MeV ($\sim T_{c}$) and $T=150$ MeV (above
$T_{c}$) for $eB=1$ GeV$^{-2}$ and zero chemical
potential.}\label{omega}
\end{figure}

\noindent
As we have seen in Sect. \ref{prop12}, for a plane wave propagating
in ${\mathbf{e}}_1-{\mathbf{e}}_2$ plane, transverse to the magnetic
field ${\mathbf{B}}=B{\mathbf{e}}_{3}$, the anisotropy $\Delta_{12}$
from (\ref{D26}) does not depend on the angle $\varphi$ of the
spherical coordinate system. On the other hand, $\Delta_{12}$
depends on the frequency of the incident plane wave $\omega$. Fig.
\ref{omega} shows $\Delta_{12}$ as a function of $\omega$ for fixed
values of $eB=1$ GeV$^{2}$ and $\mu=0$ GeV and for $T=50$ MeV (below
$T_{c}$), $T=100$ MeV ($\sim T_{c}$) and $T=150$ MeV (above
$T_{c}$). The three curves show different slopes, but the maximum
change of $\Delta_{12}$ for the frequency $\omega\in [0,10]$
fm$^{-1}$ is small $\sim 0.2\%$.\footnote{In \cite{minami} for the
relevant frequency is taken in the interval $\omega\in [-0.1,+0.1]$
fm$^{-1}$.} In what follows, we will fix the frequency to be
$\omega=0.1$ fm$^{-1}$, where according to Fig. \ref{omega},
$\Delta_{12}$ remains approximately constant for $T\gtrsim T_{c}$.
\par
\begin{figure}[hbt]
\includegraphics[width=7cm, height=5.cm]{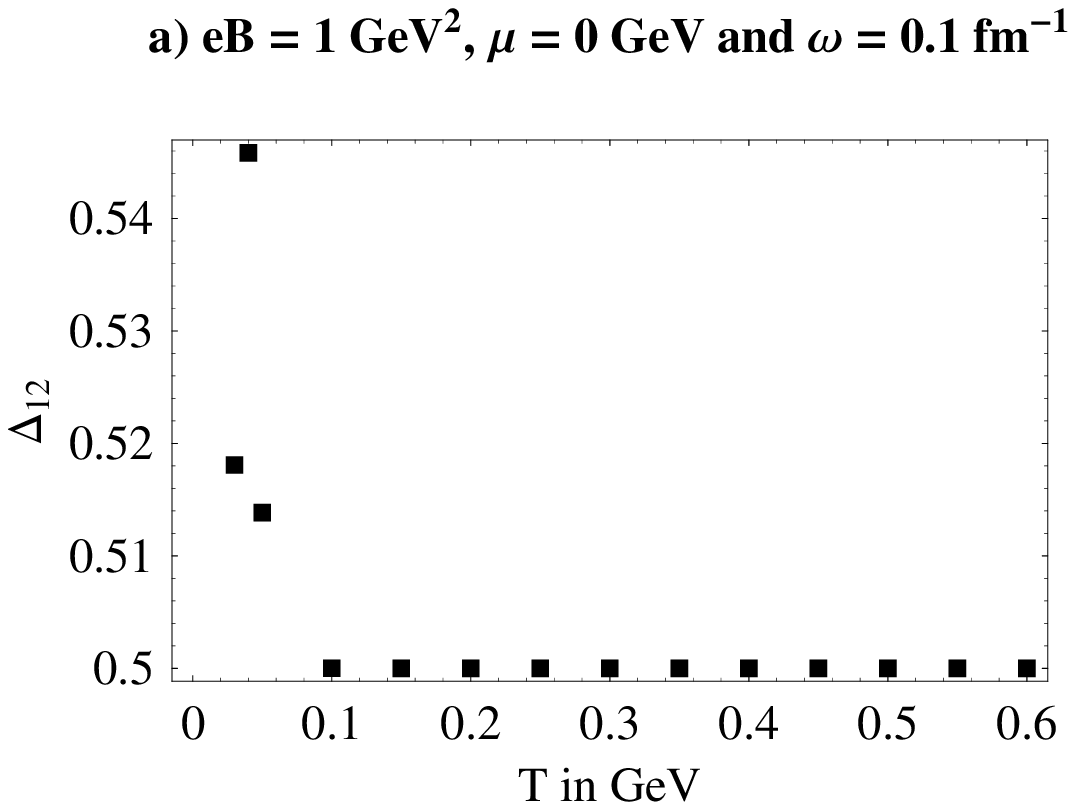}
\includegraphics[width=7cm, height=5.cm]{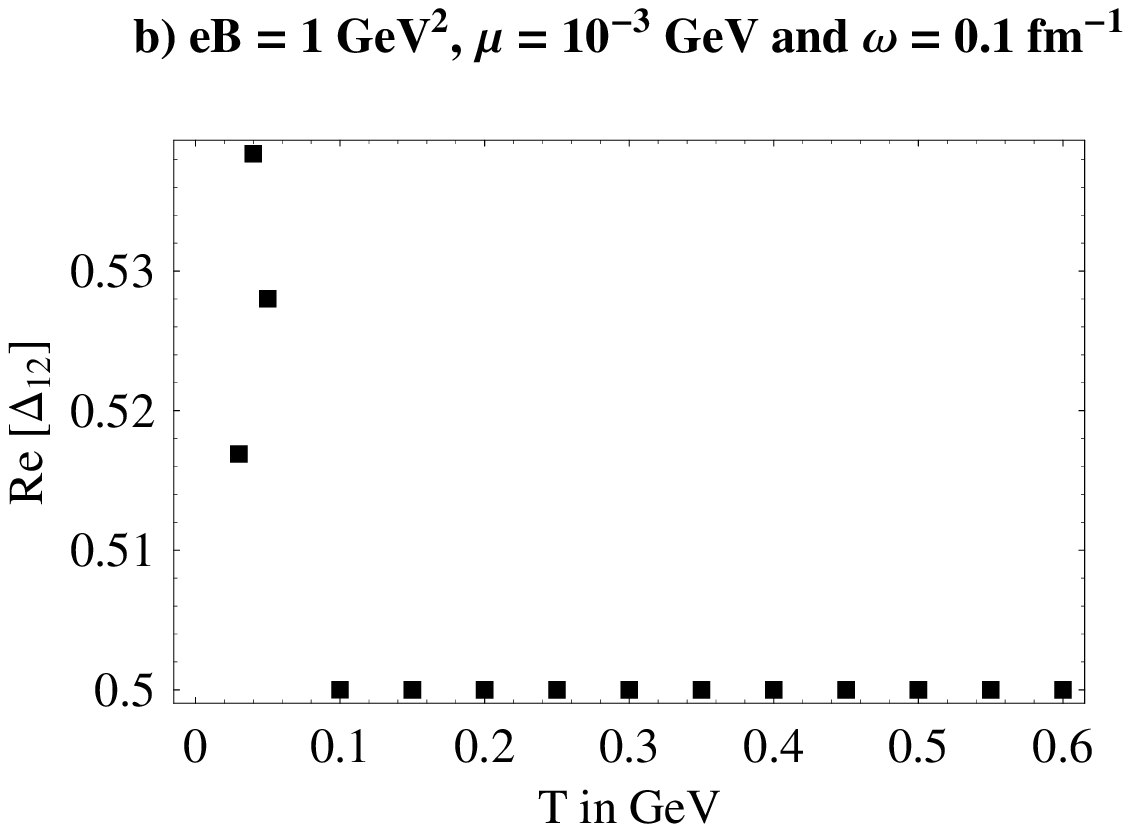}
\caption{The anisotropy $\Delta_{12}\equiv \frac{v_{12}}{c_{s}}-1$
as a function of temperature T for fixed value of $eB=1$ GeV$^{2}$,
$\omega=0.1$ fm$^{-1}$ as well as $\mu=0$ GeV (a) and $\mu=10^{-3}$
GeV (b). The sound velocity has a maximum at $T\sim 0.4 T_{c}-0.45
T_{c}$ and decreases at the critical temperature $T_{c}\sim 100-120$
MeV. After the phase transition at $T> 150$ MeV, it remains constant
$v_{12}\approx 1.5 c_{s}$.}\label{velocity}
\end{figure}

\noindent
In Table \ref{table2}, the anisotropy function $\Delta_{12}$ and the
velocity $v_{12}$ are determined for fixed $eB=1$ GeV$^{2}$ and
various  $T\in [30, 450]$ MeV and  $\mu=0, 10^{-3}$ GeV. Whereas for
$\mu=0$, the anisotropy $\Delta_{12}$ and the velocity $v_{12}$ are
real valued, they are imaginary for $\mu\neq 0$, e.g. $\mu=10^{-3}$
GeV. Qualitatively, the real part of $\Delta_{12}$ as well as
$v_{12}$ for $\mu=0$ and $\mu\neq 0$ has a maximum at $T\sim 40-45$
MeV$\sim 0.4-0.45 T_{c}$. In the transition regime $T_{c}\sim
100-150$ MeV, they arrive at their local minimum and remain constant
after phase transition $T> 150$ MeV (see Fig. \ref{velocity}). The
constant values of $v_{12}$ after the phase transition is $v_{12}=
1.5\ c_{s}\sim 0.866$. The temperature dependence of the sound
velocity is studied in \cite{florkowski} for $eB=0$ and $\mu=0$
using purely hydrodynamical methods: The authors use at high
temperature $T> 1.15 T_{c}$ the sound velocity function obtained
from lattice QCD \cite{lattice-sound}, whereas at low temperature
$T<0.15 T_{c}$ the result of the hadron gas model is used. At
moderate temperatures different interpolations between those two
results are employed. It is shown that the interpolating functions
have a local maximum at $T=0.4 T_{c}$ (corresponding to the maximal
value of the sound velocity in the hadron gas) and a local minimum
at $T=T_{c}$ (corresponding to the expected minimal value of the
sound velocity at the phase transition). Comparing to the results
from \cite{florkowski}, our results show that for a wave propagating
in the ${\mathbf{e}}_{1}-{\mathbf{e}}_{2}$ plane, transverse to the
direction of the magnetic field, the qualitative dependence of the
sound velocity $v_{12}$ on temperature does not change. As for the
imaginary part of $v_{12}$ for $\mu\neq 0$, as it can be seen in
Table \ref{table2}, they are several orders of magnitude smaller
than the real part of $v_{12}$. These kind of density fluctuations
are studied in \cite{minami}, and are interpreted as a possible
origin for the suppression of Mach cone at the chiral critical
point.
\begin{table}[h]
{{
\begin{tabular}{c||cc||ccc}\hline\hline
\multicolumn{3}{c||}{$eB=1$ GeV$^2$,$\ \mu=0$ GeV, $\
\omega=0.1$}&\multicolumn{3}{|c}{$eB=1$ GeV$^2$,$\ \mu=10^{-3}$ GeV,
$\ \omega=0.1$} \\
\hline
$T$ in MeV& $\Delta_{12}$&$v_{12}$&$\Delta_{12}$&&$v_{12}$\\
\hline
30&0.518087&0.876468&$0.516894+2.87\times 10^{-28}$i&$\qquad$&$0.875779-1.66\times 10^{-28}$i\\
40&0.54583&0.892486&$0.538411-6.81\times 10^{-31}$i&& $0.888202-3.93\times 10^{-31}$i\\
45&0.569558 &0.906184&$0.522862+1.17\times 10^{-32}$i&& $0.882020-3.93\times 10^{-31}$i \\
50&0.513846&0.874019&$0.528016-1.69\times 10^{-32}$i&& $0.882200-9.73\times 10^{-33}$i\\
100&0.500001&0.866032&$0.500005-3.31\times 10^{-27}$i&& $0.866029-1.91\times 10^{-27}$i\\
150&0.500001&0.866026&$0.5-2.67\times 10^{-28}$i&& $0.866026-1.54\times 10^{-28}$i\\
200&0.5&0.866025&$0.5-4.98\times 10^{-34}$i&&$0.866025-2.86\times 10^{-34}$i \\
250&0.5&0.866025&$0.5-2.46\times 10^{-33}$i&&$0.866025-1.42\times 10^{-33}$i \\
300&0.5&0.866025&$0.5-2.97\times 10^{-30}$i&&$0.866025-1.72\times 10^{-30}$i \\
350&0.5&0.866025&$0.5-8.57\times 10^{-31}$i&&$0.866025-4.95\times 10^{-31}$i \\
400&0.5&0.866025&$0.5+8.75\times 10^{-35}$i&&$0.866025+5.05\times 10^{-35}$i \\
450&0.5&0.866025&$0.5-1.52\times 10^{-34}$i&&$0.866025-8.75\times 10^{-32}$i \\
 \hline\hline
\end{tabular}
}} \caption{The anisotropy function $\Delta_{12}$ and the velocity
$v_{12}$ for $T\in [30, 450]$ MeV and fixed $eB=1$ GeV$^{2}$. For
nonzero baryon chemical potential $\Delta_{12}$ as well as $v_{12}$
are imaginary. At the transition temperature $\Delta_{12}$ as well
as $v_{12}$ reach their minimum value and remain constant after the
phase transition $T>T_{c}\approx 100-150$ MeV.}\label{table2}
\end{table}
\vspace{-1cm}
\subsection{Anisotropy function and sound velocity for a
propagation in ${\mathbf{e}}_1-{\mathbf{e}}_3$
plane}\label{num-sound-2}
\par
\begin{center}
\begin{figure}[hbt]
\includegraphics[width=5.5cm, height=5cm]{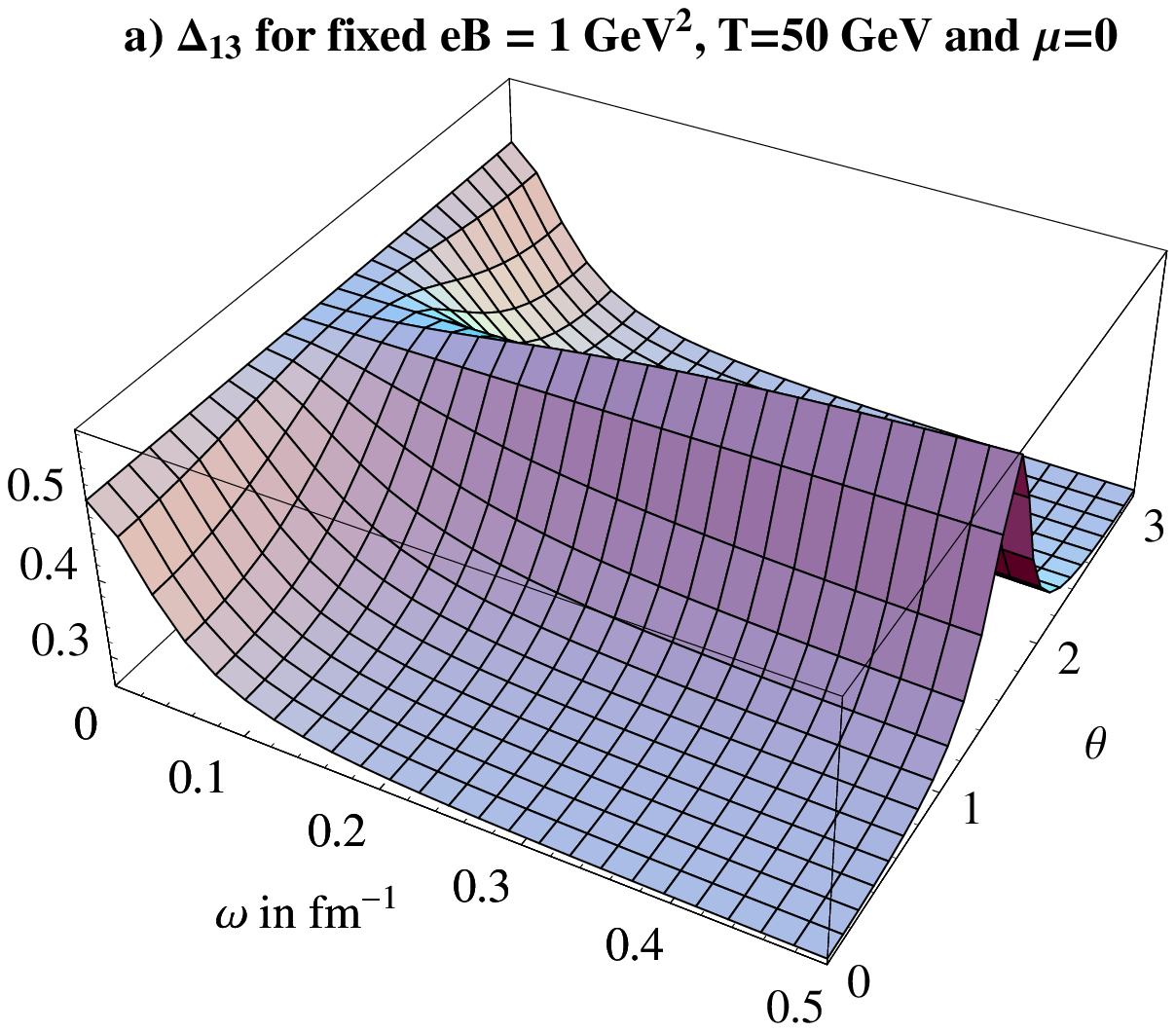}
\includegraphics[width=5.5cm, height=5cm]{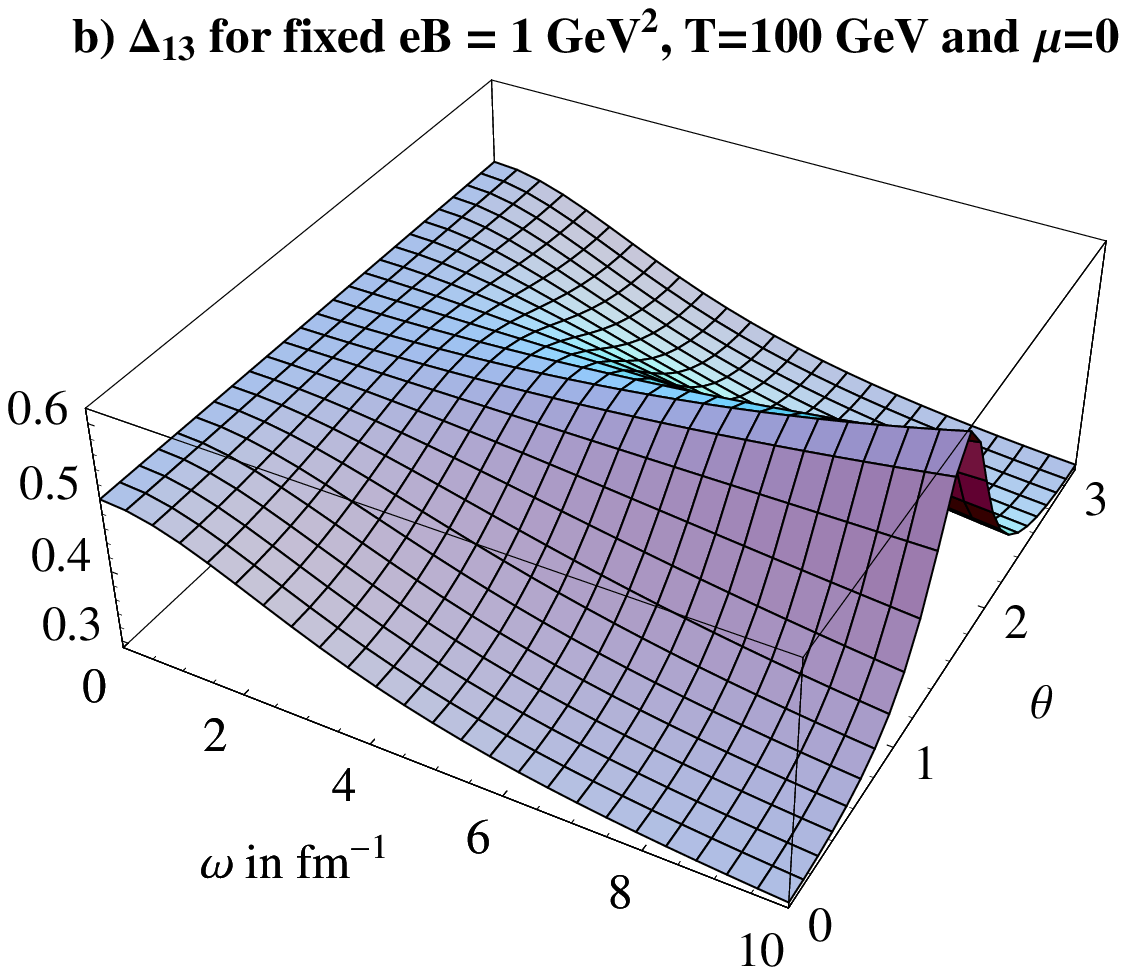}
\includegraphics[width=5.5cm, height=5cm]{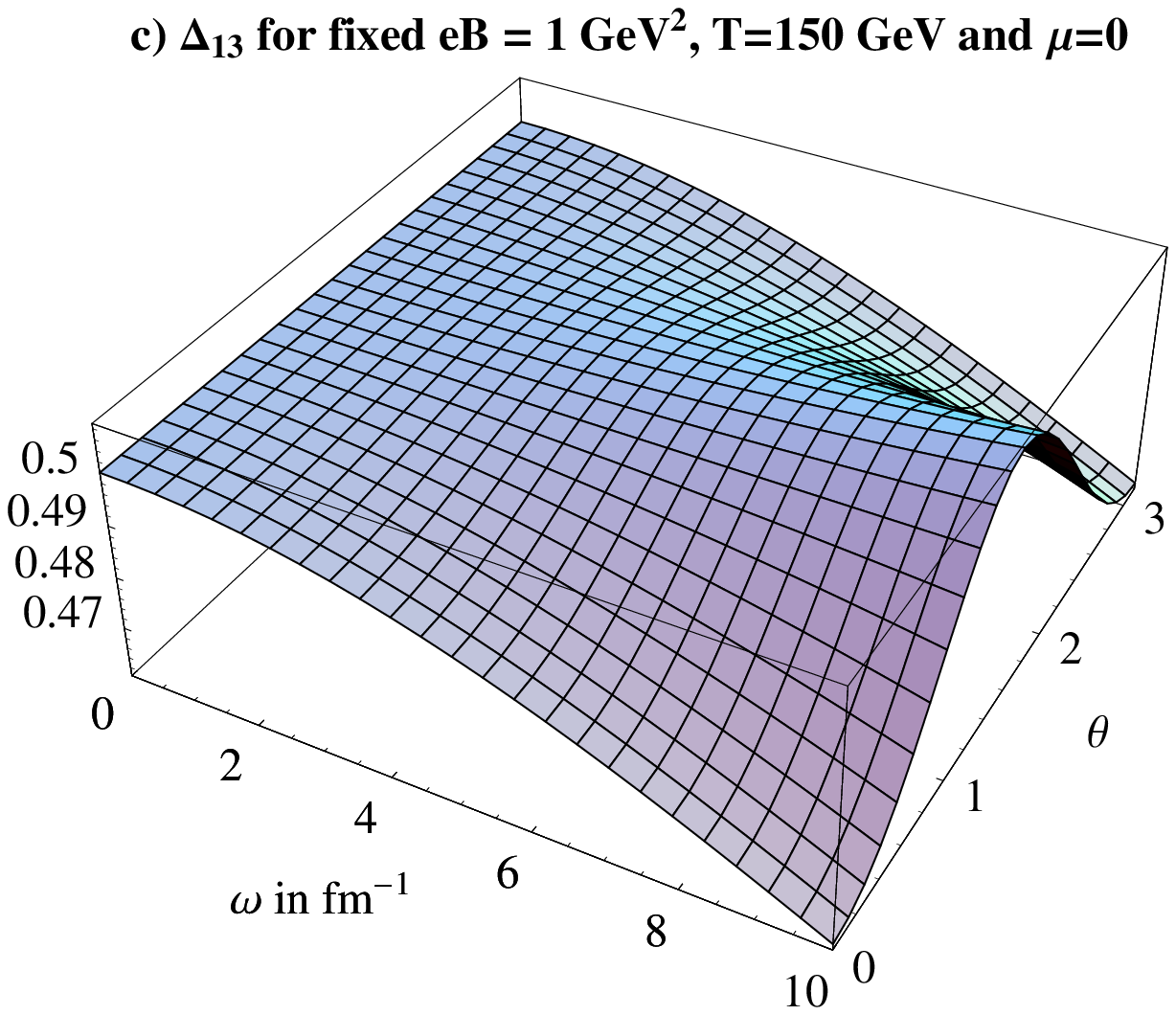}
\caption{Qualitative dependence of $\Delta_{13}$ on the frequency
$\omega$ and the angle $\theta$ for fixed values of $eB=1$ GeV$^2$,
$\mu=0$ as well as (a) $T=50$ MeV (below $T_{c}$), (b) $T=100$ MeV$
(\sim T_{c}$), and (c) $T=150$ MeV (above $T_{c}$). In contrast to
$\Delta_{12}$, $\Delta_{13}$ as well as the sound velocity for
$\theta\neq \frac{\pi}{2}$ decreases with $\omega$. For
$\theta=\frac{\pi}{2}$ $\Delta_{13}$ has the same behavior as
$\Delta_{12}$. }\label{omega-13}
\end{figure}
\end{center}

\noindent
For a wave propagating in the ${\mathbf{e}}_1-{\mathbf{e}}_3$ plane,
parallel to the external magnetic field, the anisotropy function
$\Delta_{13}$ from (\ref{D30}) as well as the velocity $v_{13}\equiv
c_{s}(1+\Delta_{13})$, depend on the angle $\theta$ of the spherical
coordinate system. It depends also on the frequency $\omega$, as for
$\Delta_{12}$. Fig. \ref{omega-13} shows the $(\omega,\theta)$
dependence of $\Delta_{13}$ for fixed values of $eB=1$ GeV$^2$,
$\mu=0$ as well as $T=50$ MeV (below $T_{c}$), $T=100$ MeV ($\sim
T_{c}$) and $T=150$ MeV (above $T_{c}$). Qualitatively, whereas
$\Delta_{12}$ increases with $\omega$ (see Fig. \ref{omega}),
$\Delta_{13}(\theta\neq \frac{\pi}{2})$ decreases with $\omega$ and
depends at the same time on the angle $\theta$ between the wave
vector ${\mathbf{k}}$ and the external magnetic field
${\mathbf{B}}$. Note that primarily for $\theta=\frac{\pi}{2}$, the
anisotropy is to be calculated from $\Delta_{12}$ (\ref{D26}). But,
it turns out that
$$\Delta_{12}(G^{0},G^{1};\omega)=\Delta_{13}(G^{0},G^{1},G^{3};\theta=\frac{\pi}{2},\omega),$$
where $\Delta_{12}$ is given in (\ref{D26}) and $\Delta_{13}$ in
(\ref{D30}). Thus, $\Delta_{13}\left(\theta=\frac{\pi}{2}\right)$
has the same $\omega$-dependence as $\Delta_{12}$.
\par
\begin{figure}[hbt]
\includegraphics[width=8.4cm, height=6.3cm]{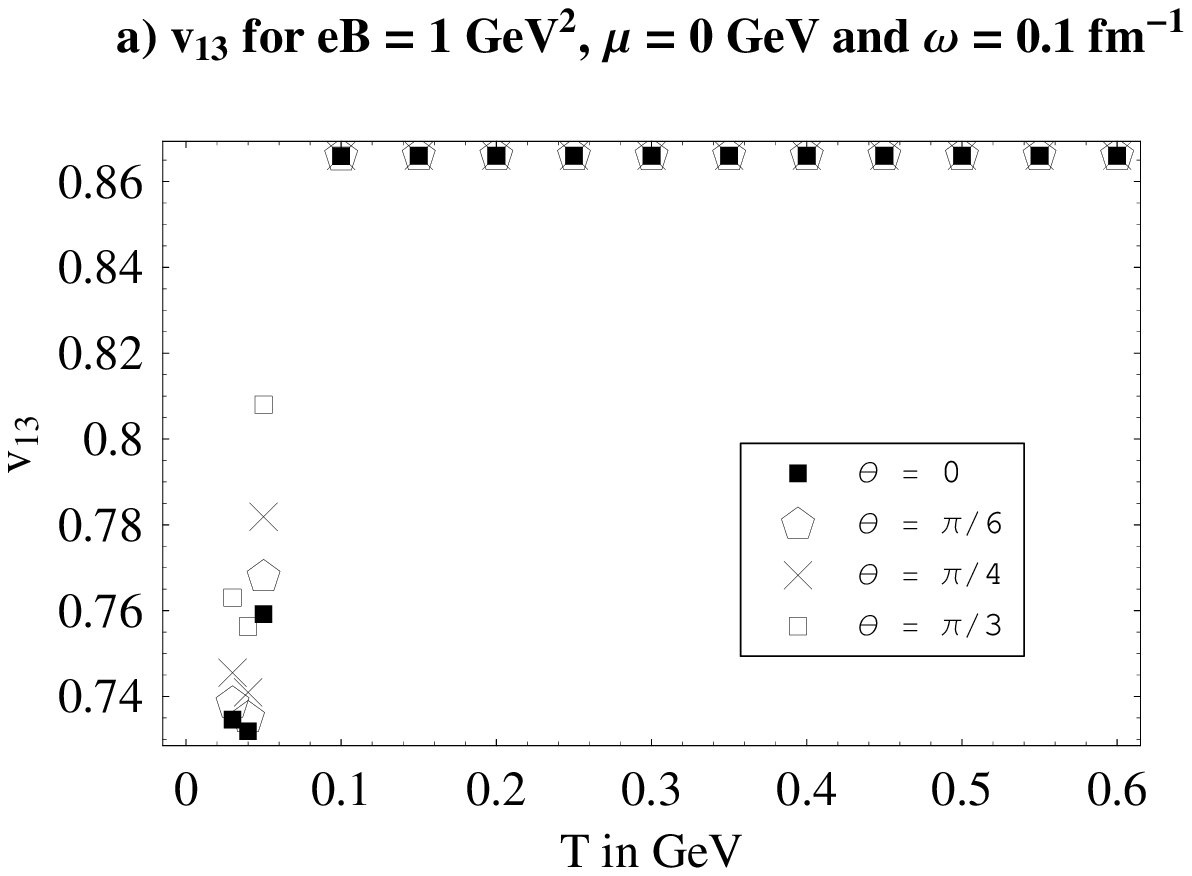}
\includegraphics[width=8.4cm, height=6.3cm]{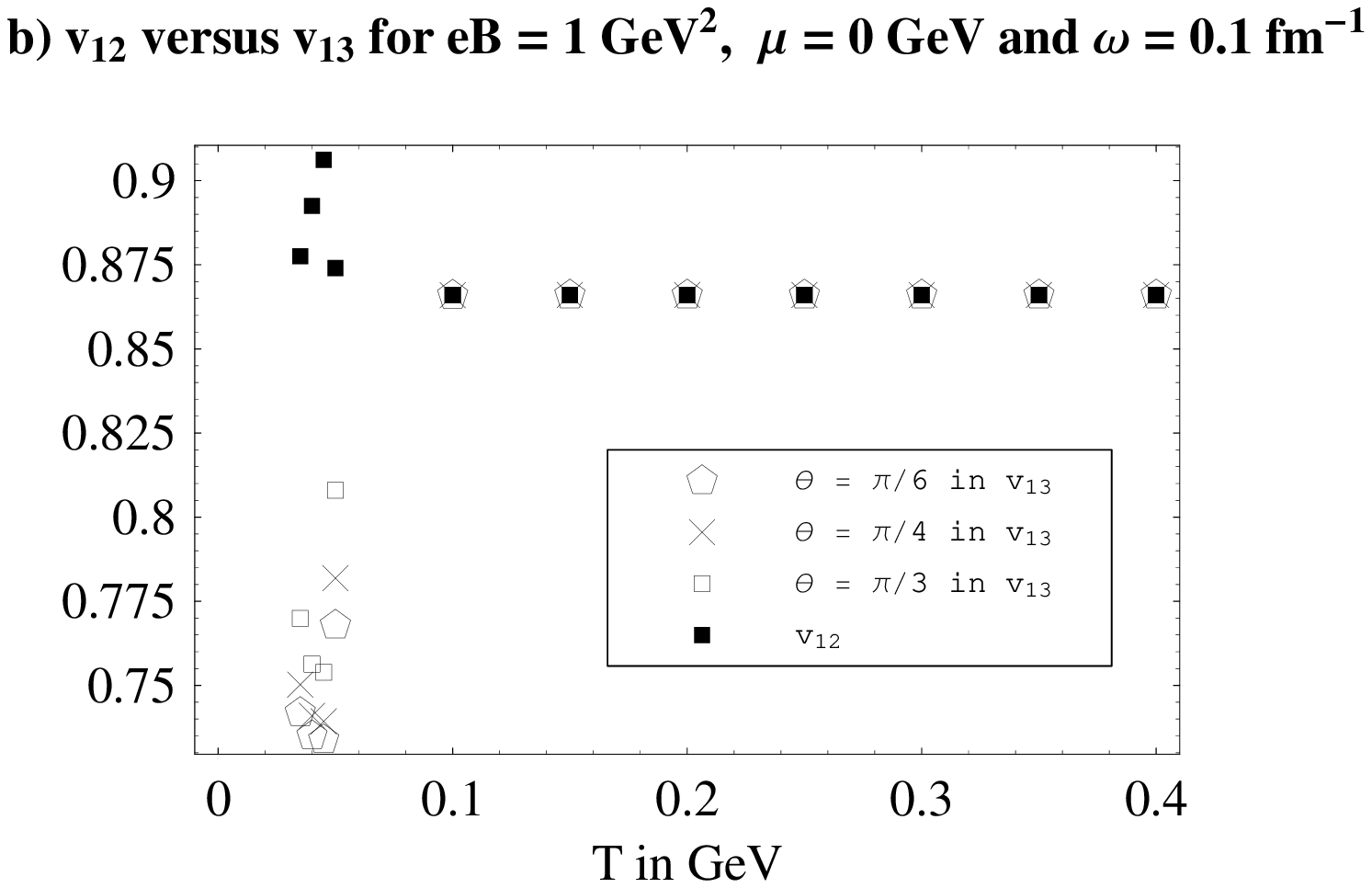}
\caption{a) Temperature dependence of the sound velocity $v_{13}$
for fixed values of
$\theta\in\{0,\frac{\pi}{6},\frac{\pi}{4},\frac{\pi}{6}\}$ and
$eB=1$ GeV$^{2}$ as well as $\mu=0$ GeV and $\omega=0.1$ fm$^{-1}$.
b) Temperature dependence of $v_{12}$ versus $v_{13}$ for
$\theta\in\{\frac{\pi}{6},\frac{\pi}{4},\frac{\pi}{6}\}$ and $eB=1$
GeV$^{2}$ as well as $\mu=0$ GeV and $\omega=0.1$ fm$^{-1}$. The
sound velocity $v_{12}$ ($v_{13}$) has a local maximum (minimum) at
$T<T_{c}$, reaches its local minimum (maximum) at $T\sim T_{c}$ and
remains constant for $T>T_{c}$. }\label{temperature-13}
\end{figure}

\noindent
In what follows, we will work with a fixed value of $\omega=0.1$
fm$^{-1}$, where for $T\gtrsim T_{c}$, $\Delta_{13}$ is
approximately constant (see Fig. \ref{omega-13}). In Fig.
\ref{temperature-13} a) the temperature dependence of the sound
velocity $v_{13}$ is presented for fixed values of
$\theta\in\{0,\frac{\pi}{6},\frac{\pi}{4},\frac{\pi}{6}\}$ and
$eB=1$ GeV$^{2}$ as well as $\mu=0$ GeV and $\omega=0.1$ fm$^{-1}$.
In contrast to $v_{12}$, the sound velocity $v_{13}$ increases with
temperature. It has therefore a local minimum at $T<T_{c}$, reaches
its local maximum at $T\sim T_{c}$ and remains constant for
$T>T_{c}$. Hence, the constant value $v_{13}=1.5\ c_{s}\approx
0.866$ of the sound velocity that arises in the above linear
approximation, is a lower bound for $v_{12}$ and an upper bound for
$v_{13}$ as the temperatures are higher than the critical
temperature $T\gtrsim T_{c}$. For $v_{13}$ this behavior is
independent on the angle $\theta$ [see Fig. \ref{temperature-13} b
for a comparison between $v_{12}$ and $v_{13}$ at various angles].
\begin{figure}[hbt]
\includegraphics[width=8.4cm, height=6.5cm]{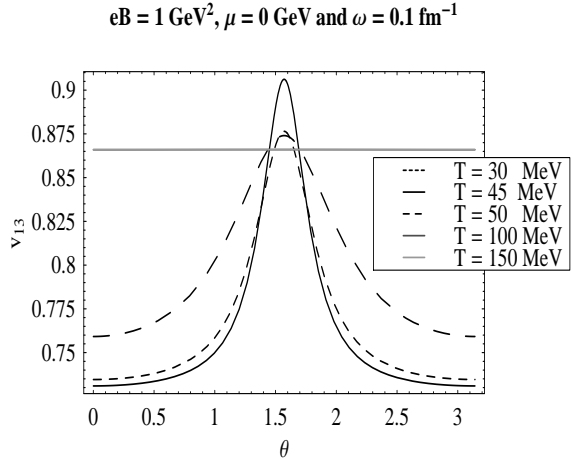}
\caption{The $\theta$-dependence of $v_{13}$ for  $T=30, 45,50$ MeV
($<T_{c}$) $T=100, 150$ ($\gtrsim T_{c}$) and for fixed values of
$eB=1$ GeV$^{2}$, $\mu=0$ GeV and $\omega=0.1$ fm$^{-1}$. Whereas
for $T<T_{c}$, $v_{13}$ is $\theta$ dependent and has its maximal
value at $\theta=\frac{1}{2}$, for $T\geq T_{c}$ it is almost
constant with $v_{13}=1.5\ c_{s}\approx 0.866$.}\label{velo-theta}
\end{figure}
\par\noindent
In Fig. \ref{velo-theta}, the $\theta$ dependence of $v_{13}$ is
plotted for $T=30, 45,50$ MeV ($<T_{c}$) $T=100, 150$ ($\gtrsim
T_{c}$) and for fixed values of $eB=1$ GeV$^{2}$, $\mu=0$ GeV and
$\omega=0.1$ fm$^{-1}$. As expected from Fig. \ref{temperature-13},
whereas for $T<T_{c}$, $v_{13}$ is $\theta$ dependent and has its
maximal value at $\theta=\frac{\pi}{2}$, for $T\geq T_{c}$ it is
almost constant with $v_{13}=0.5 c_{s}\approx 0.866$.
\section{Conclusion}\label{discussion}
\par\noindent
Strong magnetic fields play an important role in the physics of
non-central heavy ion collisions. They provide a possible signature
of the presence of CP-odd domains in the presumably formed QGP phase
\cite{warringa, kharzeev-chiral, kharzeev-summary}. Apart from other
mechanisms, relativistic shock waves, that are believed to be built
in the heavy ion collisions, can be made responsible for their
generation. In astrophysics, the strong magnetic fields generated in
collisionless relativistic shocks play an important role in the
fire-ball model for Gamma-ray Bursts \cite{wiersma}. Recent studies
on the effect of magnetic field in modifying the nature of the QCD
chiral phase transition in a linear $\sigma$-model indicate that for
high enough magnetic fields, comparable to the ones expected to be
created in the non-central heavy ion collisions at RHIC, the
original crossover is turned into a first order phase transition
\cite{fraga-magnetic, ayala09, campanelli}.
\par
In the present paper, we have studied the possible effects of strong
magnetic fields on the hydrodynamical signals of an expanding
perfect magnetized QGP coupled to chiral fields. In particular,
performing a first order stability analysis in a chiral
magnetohydrodynamic framework, the sound velocity of a propagating
plane wave in this medium is determined. To this purpose, we have
extended the variational method applied in \cite{fraga} on the
effective potential of a linear $\sigma$-model at finite $(T,\mu)$
and zero magnetic field ${\mathbf{B}}$ to the case of nonzero
magnetic field. As for the effective action, we have used the NJL
model of QCD at finite $(T,\mu)$, and in the presence of a strong
${\mathbf{B}}$ field. The magnetic field is assumed to be constant
and aligned in the third direction, i.e.
${\mathbf{B}}=B{\mathbf{e}}_{3}$. This model is known to exhibit a
dynamical chiral phase transition due to the phenomenon of magnetic
catalysis arising from the dynamically generated fermion mass
\cite{miransky-NPB-95}.
\par
In the first part of this paper, we have explicitly determined the
effective potential and effective kinetic term of the NJL model at
finite $(T,\mu)$ and in the presence of a strong $B$ field. In
particular, the effective kinetic term is determined in a derivative
expansion, where structure functions appear that depend on $(T,\mu)$
and the external magnetic field $B$. Using the effective Lagrangian
density, we have derived the dynamical mass generated in the regime
of LLL dominance. The dynamical mass, being the configuration that
minimizes the effective potential of the model, is used, in the
second part of the paper, as the equilibrium configuration once the
instabilities in the strongly magnetized QGP are set on. It depends
on $(T,\mu)$ and $B$ and vanishes once the QCD matter has passed the
chiral critical point (see Fig. 2 for its qualitative behavior for a
fixed value of the magnetic field).
\par
In the second part of the paper, a chiral magnetohydrodynamic
description of the strongly magnetized QGP is presented. We have, in
particular, compared the total energy-momentum $T^{\mu\nu}$ and the
polarization tensor $M^{\mu\nu}$, that arise field theoretically
from the Lagrangian density of the effective NJL model, with the
corresponding quantities from hydrodynamics. The polarization tensor
$M^{\mu\nu}$ appears in $T^{\mu\nu}$ and includes the magnetization
density $M$ of the medium. It is introduced in \cite{sachdev}, where
Nernst effect near the superfluid-insulator phase transition of
condensed matter physics is studied using an appropriate MHD
description. Note that in contrast to the work by Sachdev et al.
\cite{sachdev}, the magnetized fluid modeled in our paper is coupled
to chiral fields and involves the effects of a possible chiral phase
transition.
\par
The dispersion relation of the hot and dense medium in the presence
of a strong magnetic field is derived using a first order stability
analysis. As for the sound velocity of a plane wave propagating in
the magnetized QGP modeled in this paper, we have used a method
presented in \cite{parsons} and determined the anisotropy $\Delta$
in a linear approximation. Here, $\Delta\equiv
\frac{v_{s}}{c_{s}}-1$ with $v_{s}$ the sound velocity in the
medium, and $c_{s}=1/\sqrt{3}$ the sound velocity in an ideal gas.
In our model, $\Delta$ shows, as in the magnetic fluid considered in
\cite{parsons}, a non-trivial frequency dependence. Moreover, for a
plane wave propagating in the transverse
${\mathbf{e}}_{1}-{\mathbf{e}}_{2}$ plane with respect to the
external magnetic field ${\mathbf{B}}=B{\mathbf{e}}_{3}$,
$\Delta_{12}$ does not depend on the angle $\varphi$ between the
wave vector ${\mathbf{k}}$ and the external magnetic field
${\mathbf{B}}$. On the other hand, for a plane wave propagating
parallel to the magnetic field in the
${\mathbf{e}}_{1}-{\mathbf{e}}_{3}$ plane, $\Delta_{13}$ depends on
the angle $\theta$ between ${\mathbf{k}}$ and ${\mathbf{B}}$. Here,
$(\theta,\varphi)$ are the angles in the spherical coordinate
system. In Figs. \ref{temperature-13} b we have compared the sound
velocity $v_{12}$ and $v_{13}$ as a function of temperature for a
fixed values of $eB,\mu,\omega$ and various angles $\theta$. In
contrast to the sound velocity $v_{13}$ of a plane wave propagating
in the ${\mathbf{e}}_{1}-{\mathbf{e}}_{3}$ plane, the sound velocity
in the ${\mathbf{e}}_{1}-{\mathbf{e}}_{2}$ plane, $v_{12}$, has a
local maximum at $T<T_{c}$, reaches its local minimum at $T\sim
T_{c}$ and remains constant for $T>T_{c}$. The constant value
$v_{s}\sim 1.5\ c_{s}\approx 0.866$ seems therefore to be, in this
linear approximation, a \textit{lower bound} for $v_{12}$ and an
\textit{upper bound} for $v_{13}$ as the temperatures are higher
than the critical temperature $T>T_{c}$. For $v_{13}$ this behavior
is independent on the angle $\theta$.
\par
The universal properties of the equation of state $\epsilon(p)$, and
in particular the temperature dependence of the speed of sound at
$T\sim (2-3) T_{c}$, where the theory is close to being conformal,
is recently investigated in \cite{stephanov, cohen} using the
holographic principle in the framework of gauge-gravity duality of
string theory. It is shown, that for a general class of strongly
interacting theories at high temperatures the speed of sound
approaches the conformal value $c_{s}=1/\sqrt{3}$ from
\textit{below}. It would be interesting to extend the method used in
these papers to study the behavior of the sound velocity in a
strongly magnetized QGP.
\section{Acknowledgements}
\par\noindent
N.S. thanks F. Ardalan, Sh. Fayyazbakhsh, K. Sohrabi and Z. Davoudi
for useful discussions. This work is supported by the research
council of Sharif University of Technology and the Center of
Excellence in Physics (CEP).
\begin{appendix}
\section{High temperature expansion and Bessel function identities for finite $T$ and
$\mu$}\label{appA}
\setcounter{equation}{0}\par\noindent
As we have seen in Sect. \ref{njl}, the temperature dependent part
of the effective potential as well as the gap equation of the NJL
model of QCD consists of terms in the form
\begin{eqnarray}\label{AA1}
\sum\limits_{\ell=1}^{\infty}(-1)^{\ell}\frac{1}{\ell^{p}}K_{p}(\ell
z)\cosh(\mu\beta\ell).
\end{eqnarray}
Here, $z=m\beta$, with $m$ the dynamical mass,\footnote{In the case
of the NJL model, $m$ should be replaced by $\sigma_{0}$.} and $\mu$
the finite chemical potential. This expression can be written in the
generalized form
\begin{eqnarray}\label{AA2}
\sum\limits_{\ell=1}^{\infty}\frac{1}{\ell^{p}}K_{p}(\ell
z)\cos(\ell \phi),
\end{eqnarray}
with $\phi=\pi+i\mu\beta$. This identity is determined for $\mu=0$
in \cite{meisinger}. To derive the corresponding Bessel function
identities for summations of the form of (\ref{AA2}), the authors
starts in particular with the identity
\begin{eqnarray}\label{AA3}
\sum\limits_{\ell=1}^{\infty}K_{0}(\ell z)\cos(\ell
\phi)=\frac{1}{2}\bigg[\gamma+\ln\left(\frac{z}{4\pi}\right)\bigg]+\frac{\pi}{2}\sum_{\ell}^{\qquad
'}\bigg[\frac{1}{\sqrt{z^{2}+(\phi-2\pi\ell)^{2}}}-\frac{1}{2\pi|\ell|}\bigg],
\end{eqnarray}
where the notation $\sum_{\ell}^{\ '}$ is used to indicate that
singular terms in the summation over all values of
$\ell\in(-\infty,+\infty)$, here $\ell=0$ in $1/|\ell|$, are omitted
\cite{meisinger}. In this appendix, we will first evaluate the sum
over $\ell$ on the r.h.s. of (\ref{AA3}) and present it in terms of
a series expansion in $z=m\beta$. Then, using the result for
$\phi=\pi+i\mu\beta$, as it appears in the effective potential and
gap equation in Sect. \ref{njl}, we will determine (\ref{AA2}) for
$p=0$ and $p=1$. The method presented in this appendix can be used
to give and expansion of (\ref{AA2}) in terms of $z$ for all $p$.
This will be in particular useful if we are interested in a high
temperature expansion of the effective potential, as well as the gap
equation.
\subsection{High temperature expansion of (\ref{AA2}) for $p=0$}
\par\noindent
To start let us consider (\ref{AA3}) and denote the sum over
$\ell$ on the r.h.s. by $C(\phi,z,z')$, where $z=m\beta$ and
$z'=\mu\beta$
\begin{eqnarray}\label{AA4}
\sum\limits_{\ell=1}^{\infty}K_{0}(\ell z)\cos(\ell
\phi)=\frac{1}{2}\bigg[\gamma+\ln\left(\frac{z}{4\pi}\right)\bigg]+C(\phi,z,z'),
\end{eqnarray}
where in particular
\begin{eqnarray}\label{AA5}
C(\pi,z,z')=\frac{\pi}{2}\sum_{\ell}^{\qquad
'}\bigg[\frac{1}{\sqrt{z^{2}+(\pi+iz'-2\pi\ell)^{2}}}-\frac{1}{2\pi|\ell|}\bigg].
\end{eqnarray}
Expanding $C(\pi,z,z')$ in the orders of $z$, we arrive first at
\begin{eqnarray}\label{AA6}
C(\pi,z,z')&=&\sum\limits_{\ell=-\infty}^{\infty}{\cal{A}}_{\ell}(\pi,0,z')+
\sum_{\ell=-\infty}^{\infty}{\cal{B}}_{\ell}(\pi,z,z')+\sum\limits_{\ell=1}^{\infty}{\cal{C}}_{\ell},
\end{eqnarray}
where
\begin{eqnarray}\label{AA7}
{\cal{A}}_{\ell}(\pi,0,z')\equiv
\frac{1}{2|1-2\ell+\frac{iz'}{\pi}|},
\end{eqnarray}
is the zeroth order term in $z$, and
\begin{eqnarray}\label{AA8}
{\cal{B}}_{\ell}(\pi,z,z')=\bigg\{
-\frac{1}{4\pi^{2}}\frac{1}{|1-2\ell+\frac{iz'}{\pi}|^{3}}z^{2}+\frac{3}{16\pi^{4}}\frac{1}{|1-2\ell+\frac{iz'}{\pi}|^{5}}z^{4}
-\frac{5}{32\pi^{6}}\frac{1}{|1-2\ell+\frac{iz'}{\pi}|^{7}}z^{6} \pm
\cdots \bigg\},\nonumber\\
\end{eqnarray}
denotes the higher order terms in $z$. Further,
${\cal{C}}_{\ell}\equiv -\frac{1}{2\ell}$. Let us first consider the
sum over $\ell$ in ${\cal{A}}_{\ell}$
\begin{eqnarray}\label{AA9}
\sum\limits_{\ell=-\infty}^{\infty}{\cal{A}}_{\ell}(\pi,0,z')=\frac{1}{2}\sum\limits_{\ell=0}^{\infty}\left(\frac{1}{2\ell+1+\frac{iz'}{\pi}}+\frac{1}{2\ell+1-\frac{iz'}{\pi}}\right)=
\sum\limits_{\ell=0}^{\infty}\frac{(2\ell+1)}{(2\ell+1)^{2}+\frac{z'^{2}}{\pi^{2}}}.
\end{eqnarray}
Expanding ${\cal{A}}_{\ell}$ in the orders of
$\frac{z'}{\pi}=\frac{\mu\beta}{\pi}$, we arrive at
\begin{eqnarray}\label{AA10}
\sum\limits_{\ell=-\infty}^{\infty}{\cal{A}}_{\ell}(\pi,0,z')&=&\sum\limits_{k=0}^{\infty}\sum_{\ell=0}^{\infty}\frac{(-1)^{k}}{(2\ell+1)^{2k+1}}\left(\frac{z'}{\pi}\right)^{2k}\nonumber\\
&=&\sum\limits_{\ell=0}^{\infty}\frac{1}{(2\ell+1)}+
\sum\limits_{k=1}^{\infty}\frac{(-1)^{k}\left(2^{2k+1}-1\right)}{2^{2k+1}}\left(\frac{z'}{\pi}\right)^{2k}\zeta(2k+1),
\end{eqnarray}
where the definition of the Riemann $\zeta$-function
$$\zeta(z)\equiv \sum\limits_{\ell=1}^{\infty}\frac{1}{k^{z}},
\qquad z>1,$$ is used. Using now the identity
\begin{eqnarray}\label{AA11}
\sum\limits_{\ell=0}^{\infty}\frac{1}{2\ell+1}-\sum\limits_{\ell=1}^{\infty}\frac{1}{2\ell}=
\sum\limits_{\ell=1}^{\infty}\frac{(-1)^{\ell+1}}{\ell}=\ln
2,
\end{eqnarray}
we get
\begin{eqnarray}\label{AA12}
\sum\limits_{\ell=-\infty}^{\infty}{\cal{A}}_{\ell}(\pi,0,z')+\sum\limits_{\ell=1}^{\infty}{\cal{C}}_{\ell}=\ln
2+\sum\limits_{k=1}^{\infty}\frac{(-1)^{k}\left(2^{2k+1}-1\right)}{2^{2k+1}}\left(\frac{z'}{\pi}\right)^{2k}\zeta(2k+1).
\end{eqnarray}
To evaluate $\sum_{\ell}{\cal{B}}_{\ell}$ on the r.h.s. of
(\ref{AA6}) with ${\cal{B}}_{\ell}$ given in (\ref{AA8}), we split
the sum over $\ell\in (-\infty,+\infty)$ in a sum over $\ell\in
(-\infty,0]$ and $\ell\in[1,\infty)$. We arrive at
\begin{eqnarray}\label{AA13}
\sum\limits_{\ell=-\infty}^{0}{\cal{B}}_{\ell}(\pi,z,z')=\sum\limits_{n=1}^{\infty}\frac{(-1)^{n}}{4^{2n+1}
\left(n!\right)^{2}}\left(\frac{z}{\pi}\right)^{2n}\bigg|\psi^{(2n)}\left(\frac{1}{2}+\frac{iz'}{2\pi}\right)\bigg|,
\end{eqnarray}
as well as
\begin{eqnarray}\label{AA14}
\sum\limits_{\ell=1}^{\infty}{\cal{B}}_{\ell}(\pi,
z,z')=\sum\limits_{n=1}^{\infty}\frac{(-1)^{n}}{4^{2n+1}
\left(n!\right)^{2}}\left(\frac{z}{\pi}\right)^{2n}\bigg|\psi^{(2n)}\left(\frac{1}{2}-\frac{iz'}{2\pi}\right)\bigg|.
\end{eqnarray}
Here, $\phi^{(p)}(z)$ is $p$-th derivative of the polygamma-function
$\psi(z)\equiv \frac{d}{dz}\ln\Gamma(z)$ with respect to $z$.
Plugging (\ref{AA12})-(\ref{AA14}) in (\ref{AA6}), $C(\pi,z,z')$ is
given by
\begin{eqnarray}\label{AA15}
C(\pi,z,z')&=&\ln
2+\sum\limits_{k=1}^{\infty}\frac{(-1)^{k}\left(2^{2k+1}-1\right)}{2^{2k+1}}\left(\frac{z'}{\pi}\right)^{2k}\zeta(2k+1)\nonumber\\
&&+\sum\limits_{n=1}^{\infty}\frac{(-1)^{n}}{4^{2n+1}
\left(n!\right)^{2}}\left(\frac{z}{\pi}\right)^{2n}\left\{\bigg|\psi^{(2n)}\left(\frac{1}{2}+\frac{iz'}{2\pi}\right)\bigg|+
\bigg|\psi^{(2n)}\left(\frac{1}{2}-\frac{iz'}{2\pi}\right)\bigg|\right\}.
\end{eqnarray}
This leads to the first Bessel-function identity
\begin{eqnarray}\label{AA16}
\lefteqn{\hspace{-2cm}\sum\limits_{\ell=1}^{\infty}K_{0}(\ell
z)\cos(\ell
(\pi+iz'))=\frac{1}{2}\bigg[\gamma+\ln\left(\frac{z}{\pi}\right)\bigg]+\sum\limits_{k=1}^{\infty}\frac{(-1)^{k}\left(2^{2k+1}-1\right)}{2^{2k+1}}\left(\frac{z'}{\pi}\right)^{2k}\zeta(2k+1)
}
\nonumber\\
&&+\sum\limits_{n=1}^{\infty}\frac{(-1)^{n}}{4^{2n+1}
\left(n!\right)^{2}}\left(\frac{z}{\pi}\right)^{2n}\left\{\bigg|\psi^{(2n)}\left(\frac{1}{2}+\frac{iz'}{2\pi}\right)\bigg|+
\bigg|\psi^{(2n)}\left(\frac{1}{2}-\frac{iz'}{2\pi}\right)\bigg|\right\},
\end{eqnarray}
where (\ref{AA4}) is used.
\subsection{High temperature expansion of (\ref{AA2}) for $p=1$}
\par\noindent
In what follows, we will determine (\ref{AA2}) for $p=1$, and
$\phi=\pi+z'$ with $z'=\mu\beta$, by generalizing the method used in
\cite{meisinger}. Using the relation
\begin{eqnarray}\label{AA17}
\frac{d}{dz}\sum\limits_{\ell=1}^{\infty}\frac{z}{\ell}K_{1}(\ell
z)\cos(\ell\phi)=-z\sum\limits_{\ell=1}^{\infty}K_{0}(\ell
z)\cos(\ell\phi),
\end{eqnarray}
that follows from the recursion relation
\begin{eqnarray}\label{AA18}
\frac{d}{dz}K_{\nu}(z)=-K_{\nu-1}(z)-\frac{\nu}{z}K_{\nu}(z),
\end{eqnarray}
we have
\begin{eqnarray}\label{AA19}
\sum\limits_{\ell=1}^{\infty}\frac{1}{\ell}K_{1}(\ell
z)\cos(\ell\phi)=-\frac{1}{z}\int dz\left(
z\sum\limits_{\ell=1}^{\infty}K_{0}(\ell
z)\cos(\ell\phi)\right)+\frac{C(\phi)}{z}.
\end{eqnarray}
Here, $C(\phi)$ is an unknown function of $\phi$, that, in contrast
to the derivation presented in \cite{meisinger}, is a function of
$z'=\mu\beta$. In particular, we will set $\phi=\pi+iz'$ to evaluate
Bessel-function identities in the form (\ref{AA1}). To determine
$C(\phi)$, we multiply (\ref{AA19}) with $z$ and use the behavior
(\ref{AA21}) of the $K_{\nu}(z)$ in limit $z\to 0$,
\begin{eqnarray}\label{AA20}
K_{\nu}(z)\stackrel{z\to
0}{\longrightarrow}\frac{1}{z}\Gamma(\nu)\left(\frac{2}{z}\right)^{\nu},
\end{eqnarray}
to arrive at
\begin{eqnarray}\label{AA21}
C(\phi)=\sum\limits_{\ell=1}^{\infty}\frac{1}{\ell^{2}}\cos(\ell\phi)=\frac{1}{2}[\mbox{Li}_{2}(e^{+i\phi})+\mbox{Li}_{2}(e^{-i\phi})],
\end{eqnarray}
where the polylogarithm-function $\mbox{Li}_{p}(z)$ is defined, in
general, as
\begin{eqnarray}\label{AA22}
\mbox{Li}_{p}(z)=\sum\limits_{k=1}^{\infty}\frac{z^{k}}{k^{p}}.
\end{eqnarray}
The second Bessel-function identity can therefore be derived from
(\ref{AA19}) in combination with (\ref{AA16}) and (\ref{AA21}) as
\begin{eqnarray}\label{AA23}
\lefteqn{\sum\limits_{\ell=1}^{\infty}\frac{1}{\ell}K_{1}(\ell
z)\cos(\ell\phi)=\frac{1}{8}z\bigg[1-2\gamma_{E}-2\ln\left(\frac{z}{\pi}\right)}\nonumber\\
&&-\sum\limits_{n=1}^{\infty}
\frac{(-1)^{n}(1+n)}{(\Gamma(2+n))^{2}}
\left(\frac{z}{4\pi}\right)^{2n}\left\{
\bigg|\psi^{(2n)}\left(\frac{1}{2}-\frac{iz'}{2\pi}\right)\bigg|+
\bigg|\psi^{(2n)}\left(\frac{1}{2}+\frac{iz'}{2\pi}\right)\bigg|
\right\}\bigg]\nonumber\\
&&-\frac{z^{2}}{3}\sum\limits_{k=1}^{\infty}\frac{(-1)^{k}(2^{2k+1}-1)}{2^{2k+1}}\left(\frac{z'}{\pi}\right)^{2k}\zeta(2k+1)
+\frac{1}{2z}\left\{\mbox{Li}_{2}\left(e^{i(\pi+iz')}\right)+\mbox{Li}_{2}\left(e^{-i(\pi+iz')}\right)\right\}.
\end{eqnarray}
Same method can be used to determine (\ref{AA2}) for arbitrary
$p>1$.
\section{Mellin transformation and the summation over Matsubara frequencies at finite $T$ and
$\mu$}\label{mellin}
\setcounter{equation}{0}\par\noindent
As we have shown in Sect. \ref{kinetic}, the integrals arising in
the expressions for the structure functions $G^{\mu},\mu=0,\cdots,
4$ from (\ref{N47}), have the general form
\begin{eqnarray}\label{AB1}
I=\frac{1}{\beta}\sum\limits_{\ell=-\infty}^{+\infty}f(\omega_{\ell}-i\mu)=
\frac{1}{\beta}\sum\limits_{\ell=-\infty}^{+\infty}\int\frac{d^{d}k}{(2\pi)^{d}}\
\frac{({\mathbf{k}}^{2})^{a}\tilde{k}_{0}^{2t}}{({\mathbf{k}}^{2}+\tilde{k}_{0}^{2}+m^{2})^{\alpha}},
\end{eqnarray}
where, $\tilde{k}_{0}\equiv (\omega_{\ell}-i\mu)$ with
$\omega_{\ell}\equiv \frac{2\pi}{\beta}(2\ell+1)$ the Matsubara
frequencies, and $\mu$ the chemical potential. In \cite{mellin}, a
similar integral as in (\ref{AB1}), with $a=0, t=0$ and $\mu=0$
potential is evaluated using the Mellin transformation
\begin{eqnarray}\label{AB2}
f(x)&=&\frac{1}{2\pi i}\int_{c-i\infty}^{c+i\infty}
x^{-s}{\cal{M}}[f;s]ds,
\end{eqnarray}
and its inverse transformation
\begin{eqnarray}\label{AB3}
{\cal{M}}[f;s]&=&\int_{0}^{\infty}x^{s-1}f(x) dx.
\end{eqnarray}
As it is denoted in \cite{mellin}, the above transformation normally
exists only in a strip $\alpha<\mathfrak{Re}[s]<\beta$, and the
inversion contour must lie in this strip $\alpha<c<\beta$. In this
appendix, we will present a generalization of the results in
\cite{mellin} for arbitrary $a,t$ and nonzero chemical potential
$\mu\neq 0$. To this purpose, let us start with
$\sum_{\ell}f(\omega_{\ell}-i\mu)$ on the l.h.s. of (\ref{AB1}).
Using (\ref{AB2}), we arrive first at
\begin{eqnarray}\label{AB4}
I&=&\frac{1}{\beta}\sum\limits_{\ell=-\infty}^{+\infty}f(\omega_{\ell}-i\mu)=\frac{1}{2\pi
i\beta}\sum\limits_{\ell=-\infty}^{+\infty}\int_{c-i\infty}^{c+i\infty}ds\left(\omega_{\ell}-i\mu\right)^{-s}
{\cal{M}}[f;s]\nonumber\\
&=&\frac{1}{2\pi
i\beta}\int_{c-i\infty}^{c+i\infty}ds\left(\frac{2\pi}{\beta}\right)^{-s}\sum
\limits_{\ell=-\infty}^{+\infty}\left(\ell+\frac{1}{2}-\frac{i\mu\beta}{2\pi}\right)^{-s}
{\cal{M}}[f;s],
\end{eqnarray}
where the definition of $\omega_{\ell}$ is used. Performing now the
sum over $\ell$, as
\begin{eqnarray}\label{AB5}
\sum\limits_{\ell=-\infty}^{+\infty}\left(\ell+\frac{1}{2}-\frac{i\mu\beta}{2\pi}\right)^{-s}=
\zeta\left(s;\frac{1}{2}-\frac{i\mu\beta}{2\pi}\right)+\zeta\left(s;\frac{1}{2}+\frac{i\mu\beta}{2\pi}\right),
\end{eqnarray}
and plugging this expression on the r.h.s. of (\ref{AB4}), the
integral $I$ is given by $I=I_{-\mu}+I_{+\mu}$, with
\begin{eqnarray}\label{AB6}
I_{\pm\mu}=\frac{1}{2\pi i\beta}\int_{c-i\infty}^{c+i\infty}ds
\left(\frac{2\pi}{\beta}\right)^{-s}\zeta\left(s;\frac{1}{2}\pm\frac{i\mu\beta}{2\pi}\right)
{\cal{M}}[f;s].
\end{eqnarray}
On the other hand, comparing (\ref{AB3}) with (\ref{AB1}),
${\cal{M}}[f;s]$ can be determined. It is given by
\begin{eqnarray}\label{AB7}
{\cal{M}}[f;s]=\int y^{s-1}
dy\int\frac{d^{d}k}{(2\pi)^{d}}\frac{k^{2a}y^{2t}}{\left({\mathbf{k}}^{2}+{\mathbf{y}}^{2}+
m^{2}\right)^\alpha}.
\end{eqnarray}
To evaluate this integral, we combine two vectors ${\mathbf{k}}$ and
${\mathbf{y}}$ to a $D\equiv s+d+2(t+a)$-dimensional Euclidean
vector. Evaluating now the $D$-dimensional integral using standard
identities from dimensional regularization \cite{peskin},
${\cal{M}}[f;s]$ is given by
\begin{eqnarray}\label{AB8}
{\cal{M}}[f;s]&=&\frac{(2\pi)^{s+2t}}{d\Omega_{s+2t}}\frac{d\Omega_{d}(2\pi)^{2a}}{d\Omega_{d+2a}}\int\frac{d^{d+2(a+t)+s}k}{(2\pi)^{d+2(a+t)+s}}
\frac{1}{({\mathbf{k}}^{2}+m^{2})^{\alpha}}\nonumber\\
&=&\frac{1}{2(4\pi)^{d/2}}\frac{\Gamma\left(\frac{d}{2}+a\right)\Gamma\left(\frac{s}{2}+t\right)
\Gamma\left(\alpha-\frac{s}{2}-\frac{d}{2}-t-a\right)}{\Gamma\left(\frac{d}{2}\right)\Gamma(\alpha)\
m^{2(\alpha-t-a)-s-d}},
\end{eqnarray}
where $d\Omega_{d}=\frac{2\pi^{d/2}}{\Gamma(d/2)}$ is used. Plugging
now this expression in $I_{\pm}$ from (\ref{AB6}) and evaluating the
contour integration over $s$ using the residue theorem, we arrive at
\begin{eqnarray}\label{AB9}
\lefteqn{I=\frac{1}{2
(4\pi)^{d/2}\Gamma(\alpha)\beta}\frac{\Gamma\left(\frac{d}{2}+a\right)}{\Gamma\left(\frac{d}{2}\right)}
\left(\frac{2\pi}{\beta}\right)^{-2\alpha+d+2(t+a)} }\nonumber\\
&&\times\sum\limits_{k=0}^{\infty}\frac{(-1)^{k}}{k!}\Gamma\left(\alpha-a+k-\frac{d}{2}\right)\bigg[
\zeta\left(2\left(\alpha+k-t-a\right)-d;\frac{1}{2}-
\frac{i\mu\beta}{2\pi}\right)+(\mu\to
-\mu)\bigg]\left(\frac{m\beta}{2\pi}\right)^{2k}.\nonumber\\
\end{eqnarray}
Here, $k=0,1,2,\cdots,\infty$ labels the pole of the
$\Gamma$-function at $s=2(\alpha-t-a+k)-d$. Being a power series in
$(m\beta)$, the result from (\ref{AB9}) can be easily used whenever
a high temperature expansion of Feynman integrals at finite
temperature and density is possible. Note that a similar result is
also presented in \cite{ayala}, where the integral in (\ref{AB1}) is
evaluate for arbitrary $a, t$ and zero chemical potential $\mu=0$.
\section{The derivation of the equations in (\ref{D14})}\label{derivation}
\subsection{Derivation of the second equation in (\ref{D14})}
\setcounter{equation}{0}\par\noindent
The first equation in (\ref{D14}) is already derived in (\ref{D6}).
To derive the second equation in (\ref{D14}), let us consider
(\ref{D5}), that is derived from (\ref{D2}) by setting $k=\sigma$
and using the definition (\ref{D3})
\begin{eqnarray}\label{AC1}
[G^{0}\partial_{0}^{2}-G^{i}\partial_{i}^{2}+m_{\sigma}^{2}]\sigma_{1}=-n_{1}\left(\frac{\partial
R_{\sigma}}{\partial
n}\right)_{s_{0},M,\sigma_{0}}+s_{1}\left(\frac{\partial
R_{\sigma}}{\partial s}\right)_{n_{0},M,\sigma_{0}}.
\end{eqnarray}
Plugging $\xi_{1}=\tilde{\xi}_{1}e^{-ikx}$ with
$\xi_{1}=\{n_{1},s_{1},\sigma_{1}, {\mathbf{v}}_{1}\}$ and
$k^{\mu}=(\omega,{\mathbf{k}})$, we get first
\begin{eqnarray}\label{AC2}
\left(G^{0}\omega^{2}-G^{i}k_{i}^{2}-m_{\sigma}^{2}\right)\tilde{\sigma}_{1}&=&\frac{{\mathbf{k}}\cdot
\tilde{\mathbf{v}}_{1}}{\omega}\bigg[n_{0}\left(\frac{\partial
R_{\sigma}}{\partial
n}\right)_{s_{0},M,\sigma_{0}}+s_{0}\left(\frac{\partial
R_{\sigma}}{\partial s}\right)_{n_{0},M,\sigma_{0}}\bigg],
\end{eqnarray}
where (\ref{D16}) is used. Defining now
\begin{eqnarray}\label{AC3}
R'_{\sigma}\equiv \frac{1}{W_{0}}\bigg[n_{0}\left(\frac{\partial
R_{\sigma}}{\partial
n}\right)_{s_{0},M,\sigma_{0}}+s_{0}\left(\frac{\partial
R_{\sigma}}{\partial s}\right)_{n_{0},M,\sigma_{0}}\bigg],
\end{eqnarray}
we arrive at the second equation in (\ref{D14})
\begin{eqnarray}\label{AC4}
\left(G^{0}\omega^{2}-G^{i}k_{i}^{2}-m_{\sigma}^{2}\right)\tilde{\sigma}_{1}&=&\frac{W_{0}}{\omega}R'_{\sigma}
{\mathbf{k}}\cdot \tilde{\mathbf{v}}_{1}.
\end{eqnarray}
Let us now consider (\ref{AC3}) and write it as
\begin{eqnarray}\label{AC5}
R'_{\sigma}\equiv
\frac{1}{W_{0}}\left\{\frac{\partial(T,\mu,B,\sigma)}{\partial
(n,s,M,\sigma)}\bigg[n\frac{\partial
\left(R_{\sigma},s,M,\sigma\right)}{\partial
\left(T,\mu,B,\sigma\right)}-s\frac{\partial
\left(R_{\sigma},n,M,\sigma\right)}{\partial
\left(T,\mu,B,\sigma\right)}\bigg]\right\}_{0},
\end{eqnarray}
where the notation $\left(\frac{\partial A}{\partial
B}\right)_{C}\equiv \frac{\partial (A,C)}{\partial (B,C)}$ is used.
Generalizing now the identity $\frac{\partial (A,C)}{\partial
(M,N)}=\mbox{det}\left(
\begin{array}{cc}
\frac{\partial A}{\partial M}&\frac{\partial A}{\partial N}\\
\frac{\partial B}{\partial M}&\frac{\partial B}{\partial N}
\end{array}
\right)$ from \cite{schwabl} for a case with four variables, and
replacing the Jacobian $\frac{\partial(T,\mu,B,\sigma)}{\partial
(n,s,M,\sigma)}$ on the r.h.s. of (\ref{AC5}) by
$\left(\frac{\partial(n,s,M,\sigma)}{\partial
(T,\mu,B,\sigma)}\right)^{-1}$, we get first
\begin{eqnarray}\label{AC6}
R'_{\sigma}=\frac{1}{W_{0}}\left\{\left({\left|
\begin{array}{cccc}
\frac{\partial n}{\partial T}&\frac{\partial n}{\partial
\mu}&\frac{\partial n}{\partial B}&\frac{\partial n}{\partial
\sigma}\\
\frac{\partial s}{\partial T}&\frac{\partial s}{\partial
\mu}&\frac{\partial s}{\partial B}&\frac{\partial s}{\partial
\sigma}\\
\frac{\partial M}{\partial T}&\frac{\partial M}{\partial
\mu}&\frac{\partial M}{\partial B}&\frac{\partial M}{\partial
\sigma}\\
\frac{\partial \sigma}{\partial T}&\frac{\partial \sigma}{\partial
\mu}&\frac{\partial \sigma}{\partial B}&\frac{\partial
\sigma}{\partial
\sigma}\\
\end{array}
\right|}\right)^{-1}\left(n \left|
\begin{array}{cccc}
\frac{\partial R_{\sigma}}{\partial T}&\frac{\partial
R_{\sigma}}{\partial \mu}&\frac{\partial R_{\sigma}}{\partial
B}&\frac{\partial R_{\sigma}}{\partial
\sigma}\\
\frac{\partial s}{\partial T}&\frac{\partial s}{\partial
\mu}&\frac{\partial s}{\partial B}&\frac{\partial s}{\partial
\sigma}\\
\frac{\partial M}{\partial T}&\frac{\partial M}{\partial
\mu}&\frac{\partial M}{\partial B}&\frac{\partial M}{\partial
\sigma}\\
\frac{\partial \sigma}{\partial T}&\frac{\partial \sigma}{\partial
\mu}&\frac{\partial \sigma}{\partial B}&\frac{\partial
\sigma}{\partial
\sigma}\\
\end{array}
\right| - s \left|
\begin{array}{cccc}
\frac{\partial R_{\sigma}}{\partial T}&\frac{\partial
R_{\sigma}}{\partial \mu}&\frac{\partial R_{\sigma}}{\partial
B}&\frac{\partial R_{\sigma}}{\partial
\sigma}\\
\frac{\partial n}{\partial T}&\frac{\partial n}{\partial
\mu}&\frac{\partial n}{\partial B}&\frac{\partial n}{\partial
\sigma}\\
\frac{\partial M}{\partial T}&\frac{\partial M}{\partial
\mu}&\frac{\partial M}{\partial B}&\frac{\partial M}{\partial
\sigma}\\
\frac{\partial \sigma}{\partial T}&\frac{\partial \sigma}{\partial
\mu}&\frac{\partial \sigma}{\partial B}&\frac{\partial
\sigma}{\partial
\sigma}\\
\end{array}
\right| \right)\right\}_{0},
\end{eqnarray}
where the subscript $0$ means that the $(n,s,\sigma)$ in the final
result of the determinants are to be replaced by
$(n_{0},s_{0},\sigma_{0})$ from the thermal equilibrium. At this
stage the thermodynamic relations from (\ref{E16}) can be used to
get the final expression for $R'_{\sigma}$
\begin{eqnarray}\label{AC7}
R'_{\sigma}&=&\frac{{\cal{J}}}{W_{0}} \left\{\frac{\partial
P_{0}}{\partial \mu} \left|
\begin{array}{cccc}
\frac{\partial^2 P_{0}}{\partial T\partial
\sigma_{0}}&\frac{\partial^2 P_{0}}{\partial \mu\partial
\sigma_{0}}&\frac{\partial^{2}P_{0}}{\partial
B\partial\sigma_{0}}&\frac{\partial^2 P_{0}}{\partial\sigma_{0}^{2}}\\
\frac{\partial^2 P_{0}}{\partial^2
T}&\frac{\partial^{2}P_{0}}{\partial \mu\partial T}&\frac{\partial^2
P_{0}}{\partial B\partial T}&\frac{\partial^{2}P_{0}}{\partial
\sigma_{0}\partial T}\\
\frac{\partial^{2} P_{0}}{\partial T\partial
B}&\frac{\partial^{2}P_{0}}{\partial \mu\partial B}&\frac{\partial
^{2}P_{0}}{\partial^2 B}&\frac{\partial^{2}P_{0}}{\partial
\sigma_{0}\partial B}\\
\frac{\partial \sigma_{0}}{\partial T}&\frac{\partial
\sigma_{0}}{\partial \mu}&\frac{\partial \sigma_{0}}{\partial
B}&1\\
\end{array}
\right| - \frac{\partial P_{0}}{\partial T}  \left|
\begin{array}{cccc}
\frac{\partial^2 P_{0}}{\partial T\partial
\sigma_{0}}&\frac{\partial^2 P_{0}}{\partial \mu\partial
\sigma_{0}}&\frac{\partial^{2}P_{0}}{\partial
B\partial\sigma_{0}}&\frac{\partial^2 P_{0}}{\partial\sigma_{0}^{2}}\\
\frac{\partial^2 P_{0}}{\partial
T\partial\mu}&\frac{\partial^{2}P_{0}}{\partial^{2}\mu}&\frac{\partial^2
P_{0}}{\partial B\partial \mu}&\frac{\partial^{2}P_{0}}{\partial
\sigma_{0}\partial \mu}\\
\frac{\partial^{2} P_{0}}{\partial T\partial
B}&\frac{\partial^{2}P_{0}}{\partial \mu\partial B}&\frac{\partial
^{2}P_{0}}{\partial^2 B}&\frac{\partial^{2}P_{0}}{\partial
\sigma_{0}\partial B}\\
\frac{\partial \sigma_{0}}{\partial T}&\frac{\partial
\sigma_{0}}{\partial \mu}&\frac{\partial \sigma_{0}}{\partial
B}&1\\
\end{array}
\right|\right\},
\end{eqnarray}
where $P_{0}\equiv P_{0}(T,\mu,B,\sigma_{0})$, and the Jacobian
${\cal{J}}$ is defined by
\begin{eqnarray}\label{AC8}
{\cal{J}}\equiv \left|
\begin{array}{cccc}
\frac{\partial^2 P_{0}}{\partial T\partial \mu}&\frac{\partial^2
P_{0} }{\partial^2 \mu}&\frac{\partial^2 P_{0}}{\partial
B\partial\mu}&\frac{\partial^2 P_{0}}{\partial
\sigma_{0}\partial\mu}\\
\frac{\partial^2 P_{0}}{\partial^2 T}&\frac{\partial^2
P_{0}}{\partial \mu\partial T}&\frac{\partial^2 P_{0}}{\partial
B\partial T}&\frac{\partial^2 P_{0}}{\partial
\sigma_{0}\partial T}\\
\frac{\partial^2 P}{\partial T\partial B}&\frac{\partial^2
P_{0}}{\partial \mu\partial B}&\frac{\partial^2 P_{0}}{\partial^2
B}&\frac{\partial^2 P_{0}}{\partial
\sigma_{0}\partial B}\\
\frac{\partial \sigma_{0}}{\partial T}&\frac{\partial
\sigma_{0}}{\partial \mu}&\frac{\partial \sigma_{0}}{\partial
B}&1\\
\end{array}
\right|^{-1}.
\end{eqnarray}
\subsection{Derivation of the third equation in (\ref{D14})}
\par\noindent
To derive the third equation in (\ref{D14}), let us start with
(\ref{D9}), where $P_{1}$ is defined in (\ref{D11}). Plugging, as
above, $\xi_{1}=\tilde{\xi}_{1}e^{-ikx}$ with
$\xi_{1}=\{n_{1},s_{1},\sigma_{1}, {\mathbf{v}}_{1}\}$, we get first
\begin{eqnarray}\label{AC9}
W_{0}\omega\tilde{\mathbf{v}}_{1}=\bigg[\tilde{n}_{1}\left(\frac{\partial
P}{\partial
n}\right)_{s_{0},M,\sigma_{0}}+\tilde{s}_{1}\left(\frac{\partial
P}{\partial
s}\right)_{n_{0},M,\sigma_{0}}+\tilde{\sigma}_{1}\left(\frac{\partial
P}{\partial \sigma}\right)_{n_{0},s_{0},M}\bigg].
\end{eqnarray}
Using now (\ref{D16}) and defining $P'$ as in (\ref{D15})
\begin{eqnarray}\label{AC10}
P'\equiv \frac{1}{W_{0}}\bigg[n_{0}\left(\frac{\partial P}{\partial
n}\right)_{s_{0},M,\sigma_{0}}+s_{0}\left(\frac{\partial P}{\partial
s}\right)_{n_{0},M,\sigma_{0}}\bigg],
\end{eqnarray}
we get
\begin{eqnarray}\label{AC11}
\left(\omega^{2}-{\mathbf{k}}^{2}P'\right)\left({\mathbf{k}}\cdot
\tilde{\mathbf{v}}_{1}\right)=\frac{{\mathbf{k}}^{2}\omega}{W_{0}}\left(\frac{\partial
P}{\partial \sigma}\right)_{n_{0},s_{0},M}\tilde{\sigma}_{1}.
\end{eqnarray}
Defining $\left(\frac{\partial P}{\partial
\sigma}\right)_{n_{0},s_{0},M}=R''_{\sigma}\left(W_{0}+BM\right)$
with $R''_{\sigma}$ given in (\ref{D13}), we arrive at
\begin{eqnarray}\label{AC12}
\left(\omega^{2}-{\mathbf{k}}^{2}P'\right)\left({\mathbf{k}}\cdot
\tilde{\mathbf{v}}_{1}\right)=\frac{{\mathbf{k}}^{2}\omega}{W_{0}}R''_{\sigma}
\left(W_{0}+BM\right)\tilde{\sigma}_{1},
\end{eqnarray}
as expected. In what follows, we present the final results for $P'$
and $R''_{\sigma}$ from (\ref{AC10}) and (\ref{D13}), respectively.
Using the same method leading from (\ref{AC5}) to (\ref{AC7}), $P'$
from (\ref{AC10}) can be first given as
\begin{eqnarray}\label{AC13}
P'=\frac{1}{W_{0}}\left\{\left({\left|
\begin{array}{cccc}
\frac{\partial n}{\partial T}&\frac{\partial n}{\partial
\mu}&\frac{\partial n}{\partial B}&\frac{\partial n}{\partial
\sigma}\\
\frac{\partial s}{\partial T}&\frac{\partial s}{\partial
\mu}&\frac{\partial s}{\partial B}&\frac{\partial s}{\partial
\sigma}\\
\frac{\partial M}{\partial T}&\frac{\partial M}{\partial
\mu}&\frac{\partial M}{\partial B}&\frac{\partial M}{\partial
\sigma}\\
\frac{\partial \sigma}{\partial T}&\frac{\partial \sigma}{\partial
\mu}&\frac{\partial \sigma}{\partial B}&\frac{\partial
\sigma}{\partial
\sigma}\\
\end{array}
\right|}\right)^{-1}\left(n \left|
\begin{array}{cccc}
\frac{\partial P}{\partial T}&\frac{\partial P}{\partial
\mu}&\frac{\partial P}{\partial B}&\frac{\partial P}{\partial
\sigma}\\
\frac{\partial s}{\partial T}&\frac{\partial s}{\partial
\mu}&\frac{\partial s}{\partial B}&\frac{\partial s}{\partial
\sigma}\\
\frac{\partial M}{\partial T}&\frac{\partial M}{\partial
\mu}&\frac{\partial M}{\partial B}&\frac{\partial M}{\partial
\sigma}\\
\frac{\partial \sigma}{\partial T}&\frac{\partial \sigma}{\partial
\mu}&\frac{\partial \sigma}{\partial B}&\frac{\partial
\sigma}{\partial
\sigma}\\
\end{array}
\right| - s \left|
\begin{array}{cccc}
\frac{\partial P}{\partial T}&\frac{\partial P}{\partial
\mu}&\frac{\partial P}{\partial B}&\frac{\partial P}{\partial
\sigma}\\
\frac{\partial n}{\partial T}&\frac{\partial n}{\partial
\mu}&\frac{\partial n}{\partial B}&\frac{\partial n}{\partial
\sigma}\\
\frac{\partial M}{\partial T}&\frac{\partial M}{\partial
\mu}&\frac{\partial M}{\partial B}&\frac{\partial M}{\partial
\sigma}\\
\frac{\partial \sigma}{\partial T}&\frac{\partial \sigma}{\partial
\mu}&\frac{\partial \sigma}{\partial B}&\frac{\partial
\sigma}{\partial
\sigma}\\
\end{array}
\right| \right)\right\}_{0}.
\end{eqnarray}
Using now the thermodynamic relations from (\ref{E16}) to replace
$n_{0},s_{0},M,\sigma_{0}$ by the derivative of $P_{0}$ with respect
to $\mu,T,B$ respectively, we get
\begin{eqnarray}\label{AC14}
P'&=&\frac{\cal{J}}{W_{0}}\left\{\frac{\partial P_{0}}{\partial \mu}
\left|
\begin{array}{cccc}
\frac{\partial P_{0}}{\partial T}&\frac{\partial P_{0}}{\partial
\mu}&\frac{\partial P_{0}}{\partial
B}&\frac{\partial P_{0}}{\partial\sigma_{0}}\\
\frac{\partial^2 P_{0}}{\partial^2
T}&\frac{\partial^{2}P_{0}}{\partial \mu\partial T}&\frac{\partial^2
P_{0}}{\partial B\partial T}&\frac{\partial^{2}P_{0}}{\partial
\sigma_{0}\partial T}\\
\frac{\partial^{2} P_{0}}{\partial T\partial
B}&\frac{\partial^{2}P_{0}}{\partial \mu\partial B}&\frac{\partial
^{2}P_{0}}{\partial^2 B}&\frac{\partial^{2}P_{0}}{\partial
\sigma_{0}\partial B}\\
\frac{\partial \sigma_{0}}{\partial T}&\frac{\partial
\sigma_{0}}{\partial \mu}&\frac{\partial \sigma_{0}}{\partial
B}&1\\
\end{array}
\right| - \frac{\partial P_{0}}{\partial T}  \left|
\begin{array}{cccc}
\frac{\partial P_{0}}{\partial T}&\frac{\partial P_{0}}{\partial
\mu}&\frac{\partial P_{0}}{\partial
B}&\frac{\partial P_{0}}{\partial\sigma_{0}}\\
\frac{\partial^2 P_{0}}{\partial
T\partial\mu}&\frac{\partial^{2}P_{0}}{\partial^{2}\mu}&\frac{\partial^2
P_{0}}{\partial B\partial \mu}&\frac{\partial^{2}P_{0}}{\partial
\sigma_{0}\partial \mu}\\
\frac{\partial^{2} P_{0}}{\partial T\partial
B}&\frac{\partial^{2}P_{0}}{\partial \mu\partial B}&\frac{\partial
^{2}P_{0}}{\partial^2 B}&\frac{\partial^{2}P_{0}}{\partial
\sigma_{0}\partial B}\\
\frac{\partial \sigma_{0}}{\partial T}&\frac{\partial
\sigma_{0}}{\partial \mu}&\frac{\partial \sigma_{0}}{\partial
B}&1\\
\end{array}
\right| \right\},
\end{eqnarray}
with the Jacobian ${\cal{J}}$ defined in (\ref{AC8}). As next, using
the definition of $R''_{\sigma}$ from (\ref{D13}),
\begin{eqnarray}\label{AC15}
R''_{\sigma}\equiv
\frac{1}{(W_{0}+BM)}\bigg[n_{0}\left(\frac{\partial
R_{\sigma}}{\partial
n}\right)_{s_{0},M,\sigma_{0}}+s_{0}\left(\frac{\partial
R_{\sigma}}{\partial
s}\right)_{n_{0},M,\sigma_{0}}+M\left(\frac{\partial
R_{\sigma}}{\partial M}\right)_{n_{0},s_{0},\sigma_{0}}\bigg],
\end{eqnarray}
and following the same method as described above, we arrive at
\begin{eqnarray}\label{AC16}
R''_{\sigma}&=&\frac{{\cal{J}}}{(W_{0}+BM)}\left\{\frac{\partial
P_{0}}{\partial \mu} \left|
\begin{array}{cccc}
\frac{\partial^2 P_{0}}{\partial T\partial
\sigma_{0}}&\frac{\partial^2 P_{0}}{\partial \mu\partial
\sigma_{0}}&\frac{\partial^{2}P_{0}}{\partial
B\partial\sigma_{0}}&\frac{\partial^2 P_{0}}{\partial\sigma_{0}^{2}}\\
\frac{\partial^2 P_{0}}{\partial^2
T}&\frac{\partial^{2}P_{0}}{\partial \mu\partial T}&\frac{\partial^2
P_{0}}{\partial B\partial T}&\frac{\partial^{2}P_{0}}{\partial
\sigma_{0}\partial T}\\
\frac{\partial^{2} P_{0}}{\partial T\partial
B}&\frac{\partial^{2}P_{0}}{\partial \mu\partial B}&\frac{\partial
^{2}P_{0}}{\partial^2 B}&\frac{\partial^{2}P_{0}}{\partial
\sigma_{0}\partial B}\\
\frac{\partial \sigma_{0}}{\partial T}&\frac{\partial
\sigma_{0}}{\partial \mu}&\frac{\partial \sigma_{0}}{\partial
B}&1\\
\end{array}
\right| - \frac{\partial P_{0}}{\partial T}  \left|
\begin{array}{cccc}
\frac{\partial^2 P_{0}}{\partial T\partial
\sigma_{0}}&\frac{\partial^2 P_{0}}{\partial \mu\partial
\sigma_{0}}&\frac{\partial^{2}P_{0}}{\partial
B\partial\sigma_{0}}&\frac{\partial^2 P_{0}}{\partial\sigma_{0}^{2}}\\
\frac{\partial^2 P_{0}}{\partial
T\partial\mu}&\frac{\partial^{2}P_{0}}{\partial^{2}\mu}&\frac{\partial^2
P_{0}}{\partial B\partial \mu}&\frac{\partial^{2}P_{0}}{\partial
\sigma_{0}\partial \mu}\\
\frac{\partial^{2} P_{0}}{\partial T\partial
B}&\frac{\partial^{2}P_{0}}{\partial \mu\partial B}&\frac{\partial
^{2}P_{0}}{\partial^2 B}&\frac{\partial^{2}P_{0}}{\partial
\sigma_{0}\partial B}\\
\frac{\partial \sigma_{0}}{\partial T}&\frac{\partial
\sigma_{0}}{\partial \mu}&\frac{\partial \sigma_{0}}{\partial
B}&1\\
\end{array}
\right| \right.\nonumber\\
&&+\left.\frac{\partial P_{0}}{\partial B} \left|
\begin{array}{cccc}
\frac{\partial^{2}P_{0}}{\partial
T\partial\sigma_{0}}&\frac{\partial^{2}P_{0}}{\partial
\mu\partial\sigma_{0}}&\frac{\partial^{2}P_{0}}{\partial
B\partial\sigma_{0}}&\frac{\partial^{2}P_{0}}{\partial
\sigma_{0}^{2}}\\
\frac{\partial^{2}P_{0}}{\partial
T\partial\mu}&\frac{\partial^{2}P_{0}}{\partial^{2}\mu}&\frac{\partial^{2}P_{0}}{\partial
B\partial\mu}&\frac{\partial^{2}P_{0}}{\partial
\sigma_{0}\partial\mu}\\
\frac{\partial^{2}P_{0}}{\partial^{2}T}&\frac{\partial^{2}P_{0}}{\partial
\mu\partial T}&\frac{\partial^{2}P_{0}}{\partial B\partial
T}&\frac{\partial^{2}P_{0}}{\partial
\sigma\partial T}\\
\frac{\partial \sigma_{0}}{\partial T}&\frac{\partial
\sigma_{0}}{\partial \mu}&\frac{\partial \sigma_{0}}{\partial B}& 1\\
\end{array}
\right| \right\},
\end{eqnarray}
where the Jacobian ${\cal{J}}$ defined in (\ref{AC8}).
\end{appendix}


\begin{thebibliography}{99}
\bibitem{ollitrault}
  J.~Y.~Ollitrault, \textit{Relativistic hydrodynamics},
  Eur.\ J.\ Phys.\  {\bf 29}, 275 (2008), arXiv: 0708.2433 [nucl-th].
\par
  P.~Romatschke,
  \textit{New Developments in Relativistic Viscous Hydrodynamics}, arXiv: 0902.3663 [hep-ph].
\bibitem{vanderkolk}
  T.~Hirano, N.~van der Kolk and A.~Bilandzic,
  \textit{Hydrodynamics and flow}, arXiv: 0808.2684 [nucl-th].
\bibitem{schaefer}
  T.~Schaefer and D.~Teaney,
  \textit{Nearly Perfect Fluidity: From cold atomic gases to hot quark gluon plasmas}, arXiv: 0904.3107 [hep-ph].
\bibitem{shuryak80}
  E.~V.~Shuryak,
  \textit{Quantum Chromodynamics and the theory of superdense matter}, Phys.\ Rept.\  {\bf 61} (1980) 71.
\bibitem{hydro}
e.g.  C.~M.~Hung and E.~V.~Shuryak,
  \textit{Hydrodynamics near The QCD phase transition: Looking or the longest lived fireball},
  Phys.\ Rev.\ Lett.\  {\bf 75}, 4003 (1995), arXiv: hep-ph/9412360.
\par
  D.~Teaney, J.~Lauret and E.~V.~Shuryak,
  \textit{Hydro+cascade, flow, the equation of state, predictions and data},
  Nucl.\ Phys.\  A {\bf 698}, 479 (2002), arXiv: nucl-th/0104041.
\par
  D.~Teaney, J.~Lauret and E.~V.~Shuryak,
  \textit{A hydrodynamic description of heavy ion collisions at the SPS and RHIC}, arXiv: nucl-th/0110037.
\par
  J.~Casalderrey-Solana, E.~V.~Shuryak and D.~Teaney,
  \textit{Hydrodynamic flow from fast particles}, arXiv: hep-ph/0602183.
\bibitem{kodama}
e.g.
  F.~Karsch, D.~Kharzeev and K.~Tuchin,
  \textit{Universal properties of bulk viscosity near the QCD phase transition},
  Phys.\ Lett.\  B {\bf 663}, 217 (2008), arXiv: 0711.0914 [hep-ph].
\par
  G.~S.~Denicol, T.~Kodama, T.~Koide and Ph.~Mota,
  \textit{Effect of bulk viscosity on elliptic flow near QCD phase transition}, arXiv: 0903.3595 [hep-ph].
\par
B.~C.~Li and M.~Huang,
  \textit{Thermodynamic properties and bulk viscosity near phase transition in the
  Z(2) and O(4) models}, arXiv: 0903.3650 [hep-ph].
\bibitem{iwazaki09}
  H.~Fujii, K.~Itakura and A.~Iwazaki,
  \textit{Instabilities in non-expanding glasma}, arXiv: 0903.2930 [hep-ph].
\bibitem{pufu3-4}
  H.~Nastase,
  \textit{On high energy scattering inside gravitational backgrounds}, arXiv: hep-th/0410124.
\par
  H.~Nastase,
  \textit{The RHIC fireball as a dual black hole}, arXiv: hep-th/0501068.
\par
  E.~Shuryak, S.~J.~Sin and I.~Zahed,
  \textit{A Gravity Dual of RHIC Collisions},
  J.\ Korean Phys.\ Soc.\  {\bf 50}, 384 (2007), arXiv: hep-th/0511199.
\par
  A.~J.~Amsel, D.~Marolf and A.~Virmani,
  \textit{Collisions with Black Holes and Deconfined Plasmas},
  JHEP {\bf 0804}, 025 (2008), arXiv: 0712.2221 [hep-th].
\par
  D.~Grumiller and P.~Romatschke,
  \textit{On the collision of two shock waves in AdS5},
  JHEP {\bf 0808}, 027 (2008), arXiv: 0803.3226 [hep-th].
\bibitem{gravity-shock}
  S.~S.~Gubser, S.~S.~Pufu and A.~Yarom,
  \textit{Entropy production in collisions of gravitational shock waves and of heavy
  ions},
  Phys.\ Rev.\  D {\bf 78}, 066014 (2008), arXiv: 0805.1551 [hep-th].
\par
  S.~Lin and E.~Shuryak,
  \textit{Grazing collisions of gravitational shock waves and entropy production in
  heavy ion collision}, arXiv: 0902.1508 [hep-th].
\par
  J.~L.~Albacete, Y.~V.~Kovchegov and A.~Taliotis,
  \textit{Asymmetric collision of two shock waves in AdS$_5$},
  arXiv: 0902.3046 [hep-th].
\par
  S.~S.~Gubser, S.~S.~Pufu and A.~Yarom,
  \textit{Off-center collisions in AdS$_5$ with applications to multiplicity estimates in heavy-ion collisions},
  arXiv: 0902.4062 [hep-th].
\bibitem{karsch}
  A.~Bazavov {\it et al.},
  \textit{Equation of state and QCD transition at finite temperature},
  arXiv: 0903.4379 [hep-lat].
\bibitem{kolb}
  E.~W.~.~Kolb and M.~S.~.~Turner,
  \textit{The Early Universe}, Redwood City, USA, Addison-Wesley (1988).
\bibitem{fraga}
  C.~E.~Aguiar, E.~S.~Fraga and T.~Kodama,
  \textit{Hydrodynamical instabilities beyond the chiral critical point},
  J.\ Phys.\ G {\bf 32}, 179 (2006), arXiv: nucl-th/0306041.
\bibitem{sudden-hadron}
J.~Rafelski and J.~Letessier,
  \textit{Sudden hadronization in relativistic nuclear collisions},
  Phys.\ Rev.\ Lett.\  {\bf 85}, 4695 (2000), arXiv: hep-ph/0006200.
\par
  O.~Scavenius, A.~Dumitru and A.~D.~Jackson,
  \textit{Explosive decomposition in ultrarelativistic heavy ion collision},
  Phys.\ Rev.\ Lett.\  {\bf 87}, 182302 (2001), arXiv: hep-ph/0103219.
\par
  A.~Dumitru and R.~D.~Pisarski,
  \textit{Explosive collisions at RHIC?},
  Nucl.\ Phys.\  A {\bf 698}, 444 (2002), arXiv: hep-ph/0102020.
\bibitem{paech}
  K.~Paech, H.~Stoecker and A.~Dumitru,
  \textit{Hydrodynamics near a chiral critical point},
  Phys.\ Rev.\  C {\bf 68}, 044907 (2003), arXiv: nucl-th/0302013.
\bibitem{koide}
  H.~T.~Elze, T.~Kodama, Y.~Hama, M.~Makler and J.~Rafelski,
  \textit{Variational approach to hydrodynamics: From QGP to general relativity},
  arXiv:hep-ph/9809570.
\par
  H.~T.~Elze, Y.~Hama, T.~Kodama, M.~Makler and J.~Rafelski,
  \textit{Variational Principle for Relativistic Fluid Dynamics},
  J.\ Phys.\ G {\bf 25}, 1935 (1999), arXiv: hep-ph/9910208.
\bibitem{parsons}
  J.~D.~Parsons,
  \textit{Sound velocity in a magnetic field},
  J.\ Phys.\ D {\bf 8}, 1219, (1975).
\bibitem{njl}
  Y.~Nambu and G.~Jona-Lasinio,
  \textit{Dynamical model of elementary particles based on an analogy with
  superconductivity. I},
  Phys.\ Rev.\  {\bf 122}, 345 (1961).
\par
  Y.~Nambu and G.~Jona-Lasinio,
  \textit{Dynamical model of elementary particles based on an analogy with
  superconductivity. II},
  Phys.\ Rev.\  {\bf 124}, 246 (1961).
\par
  U.~Vogl and W.~Weise,
  \textit{The Nambu and Jona-Lasinio model: Its implications for hadrons and
  nuclei},
  Prog.\ Part.\ Nucl.\ Phys.\  {\bf 27}, 195 (1991).
\par
  S.~P.~Klevansky,
  \textit{The Nambu-Jona-Lasinio model of quantum chromodynamics},
  Rev.\ Mod.\ Phys.\  {\bf 64}, 649 (1992).
\bibitem{klevansky}
  T.~M.~Schwarz, S.~P.~Klevansky and G.~Papp,
  \textit{The phase diagram and bulk thermodynamical quantities in the NJL model  at
  finite temperature and density},
  Phys.\ Rev.\  C {\bf 60}, 055205 (1999), arXiv: nucl-th/9903048.
\bibitem{miransky-NPB-95}
  V.~P.~Gusynin, V.~A.~Miransky and I.~A.~Shovkovy,
  \textit{Dimensional reduction and catalysis of dynamical symmetry breaking by a
  magnetic field},
  Nucl.\ Phys.\  B {\bf 462}, 249 (1996), arXiv: hep-ph/9509320.
\bibitem{cosmology}
  E.~Elizalde, E.~J.~Ferrer and V.~de la Incera,
  \textit{Neutrino propagation in a strongly magnetized medium},
  Phys.\ Rev.\  D {\bf 70}, 043012 (2004).
\par
  E.~J.~Ferrer and V.~de la Incera,
  \textit{Neutrino propagation and oscillations in a strong magnetic field},
  Int.\ J.\ Mod.\ Phys.\  A {\bf 19}, 5385 (2004).
\bibitem{condensed}
  K.~Farakos, G.~Koutsoumbas and N.~E.~Mavromatos,
  \textit{Dynamical flavour symmetry breaking by a magnetic field in lattice
  QED(3)},
  Phys.\ Lett.\  B {\bf 431}, 147 (1998).
\par
  K.~Farakos and N.~E.~Mavromatos,
  \textit{Hidden non-Abelian gauge symmetries in doped planar antiferromagnets},
  Phys.\ Rev.\  B {\bf 57}, 3017 (1998).
\par
  G.~W.~Semenoff, I.~A.~Shovkovy and L.~C.~R.~Wijewardhana,
  \textit{Phase transition induced by a magnetic field},
  Mod.\ Phys.\ Lett.\  A {\bf 13}, 1143 (1998).
\par
  E.~J.~Ferrer, V.~P.~Gusynin and V.~de la Incera,
  \textit{Magnetic field induced gap and kink behavior of thermal conductivity in
  cuprates},
  Mod.\ Phys.\ Lett.\  B {\bf 16}, 107 (2002).
\par
  E.~J.~Ferrer, V.~P.~Gusynin and V.~de la Incera,
  \textit{Thermal conductivity in 3D NJL model under external magnetic field},
  Eur.\ Phys.\ J.\  B {\bf 33}, 397 (2003).
\par
  E.~V.~Gorbar, V.~P.~Gusynin, V.~A.~Miransky and I.~A.~Shovkovy,
  \textit{Dynamics in the quantum Hall effect and the phase diagram of graphene},
  Phys.\ Rev.\  B {\bf 78}, 085437 (2008), arXiv: 0806.0846 [cond-mat.mes-hall].
\bibitem{warringa}
  D.~E.~Kharzeev, L.~D.~McLerran and H.~J.~Warringa,
  \textit{The effects of topological charge change in heavy ion collisions: 'Event by
  event P and CP violation'},
  Nucl.\ Phys.\  A {\bf 803}, 227 (2008), arXiv: 0711.0950 [hep-ph].
\par
  H.~J.~Warringa,
  \textit{Implications of CP-violating transitions in hot quark matter on heavy ion
  collisions},
  J.\ Phys.\ G {\bf 35}, 104012 (2008), arXiv: 0805.1384 [hep-ph].
\par
  A.~J.~Mizher and E.~S.~Fraga,
  \textit{CP Violation in the Linear Sigma Model},
  Nucl.\ Phys.\  A {\bf 820}, 247 (2009), arXiv: 0810.4115 [hep-ph].
\par
  A.~J.~Mizher and E.~S.~Fraga,
  \textit{CP violation and chiral symmetry restoration in the hot linear sigma model
  in a strong magnetic background}, arXiv: 0810.5162 [hep-ph].
\bibitem{kharzeev-chiral}
  K.~Fukushima, D.~E.~Kharzeev and H.~J.~Warringa,
  \textit{The Chiral Magnetic Effect},
  Phys.\ Rev.\  D {\bf 78}, 074033 (2008), arXiv: 0808.3382 [hep-ph].
\bibitem{kharzeev-summary}
  D.~E.~Kharzeev,
  \textit{Hot and dense matter: from RHIC to LHC: Theoretical overview},
  arXiv: 0902.2749 [hep-ph].
\bibitem{kharzeev50-voloshin}
  S.~A.~Voloshin,
  \textit{Parity violation in hot QCD: How to detect it},
  Phys.\ Rev.\  C {\bf 70}, 057901 (2004), arXiv: hep-ph/0406311.
\bibitem{kharzeev51-STAR}
  I.~V.~Selyuzhenkov  [STAR Collaboration],
  \textit{Global polarization and parity violation study in Au + Au collisions},
  Rom.\ Rep.\ Phys.\  {\bf 58}, 049 (2006), arXiv: nucl-ex/0510069.
\bibitem{sakharov}
  A.~D.~Sakharov,
  \textit{Violation of CP invariance, C asymmetry, and baryon asymmetry of the
  universe},
  Pisma Zh.\ Eksp.\ Teor.\ Fiz.\  {\bf 5}, 32 (1967)
  [JETP Lett.\  {\bf 5}, 24 (1967\ SOPUA,34,392-393.1991\ UFNAA,161,61-64.1991)].
\bibitem{shaposhnikov}
  M.~Giovannini and M.~E.~Shaposhnikov,
  \textit{Primordial magnetic fields, anomalous isocurvature fluctuations and big
  bang nucleosynthesis},
  Phys.\ Rev.\ Lett.\  {\bf 80}, 22 (1998).
\bibitem{wiersma}
  J.~Wiersma and A.~Achterberg,
  \textit{Magnetic field generation in relativistic shocks - An early end of the
  exponential Weibel instability in electron-proton plasmas},''
  Astron.\ Astrophys.\  {\bf 428}, 365 (2004), arXiv: astro-ph/0408550.
\bibitem{weibel}
  E.~S.~Weibel,
  \textit{Spontaneously growing transverse waves in a plasma due to ananisotropic
  velocity distribution},
  Phys.\ Rev.\ Lett.\  {\bf 2}, 83 (1959).
\par
  A.~Rebhan,
  \textit{Hard loop effective theory of the (anisotropic) quark gluon plasma}, arXiv: 0811.0457 [hep-ph].
\bibitem{iwazaki2}
  H.~Fujii, K.~Itakura and A.~Iwazaki,
  \textit{Instabilities in non-expanding glasma},
  arXiv:0903.2930 [hep-ph].
\bibitem{nielsen-olesen}
  N.~K.~Nielsen and P.~Olesen,
  \textit{An Unstable Yang-Mills Field Mode},
  Nucl.\ Phys.\  B {\bf 144}, 376 (1978).
\par
  N.~K.~Nielsen and P.~Olesen,
  \textit{Electric Vortex Lines From The Yang-Mills Theory},
  Phys.\ Lett.\  B {\bf 79}, 304 (1978).
\bibitem{fraga-magnetic}
  N.~O.~Agasian and S.~M.~Fedorov,
  \textit{Quark-hadron phase transition in a magnetic field},
  Phys.\ Lett.\  B {\bf 663}, 445 (2008), arXiv: 0803.3156 [hep-ph].
\par
  E.~S.~Fraga and A.~J.~Mizher,
  \textit{Chiral transition in a strong magnetic background},
  Phys.\ Rev.\  D {\bf 78}, 025016 (2008), arXiv: 0804.1452 [hep-ph].
\par
  E.~S.~Fraga and A.~J.~Mizher,
  \textit{Can a strong magnetic background modify the nature of the chiral transition
  in QCD?},
  Nucl.\ Phys.\  A {\bf 820}, 103C (2009), arXiv: 0810.3693 [hep-ph].
\bibitem{ayala09}
  A.~Ayala, A.~Bashir, A.~Raya and A.~Sanchez,
  \textit{Chiral phase transition in relativistic heavy-ion collisions with weak
  magnetic fields: ring diagrams in the linear sigma model}, arXiv: 0904.4533 [hep-ph].
\bibitem{campanelli}
  L.~Campanelli and M.~Ruggieri,
  \textit{Probing the QCD vacuum with an abelian chromomagnetic field: A study within
  an effective model}, arXiv: 0905.0853 [hep-ph].
\bibitem{sadooghi-sohrabi}
  N.~Sadooghi and K.~Sohrabi~Anaraki,
  \textit{Improved ring potential of QED at finite temperature and in the presence of
  weak and strong magnetic fields},
  Phys.\ Rev.\  D {\bf 78}, 125019 (2008), arXiv: 0805.0078 [hep-ph].
\bibitem{florkowski}
  M.~Chojnacki and W.~Florkowski,
  \textit{Temperature dependence of sound velocity and hydrodynamics of
  ultra-relativistic heavy-ion collisions},
  Acta Phys.\ Polon.\  B {\bf 38}, 3249 (2007), arXiv: nucl-th/0702030.
\bibitem{minami}
  C.~Sasaki, B.~Friman and K.~Redlich,
  \textit{Density fluctuations as signature of a non-equilibrium first order phase
  transition},
  J.\ Phys.\ G {\bf 35}, 104095 (2008), arXiv: 0804.3990 [hep-ph].
\par
  Y.~Minami and T.~Kunihiro,
  \textit{Dynamical Density Fluctuations around QCD Critical Point Based on
  Dissipative Relativistic Fluid Dynamics},
  arXiv:0904.2270 [hep-th].
\bibitem{ebert}
  D.~Ebert and V.~C.~Zhukovsky,
  \textit{Chiral phase transitions in strong chromomagnetic fields at finite
  temperature and dimensional reduction},
  Mod.\ Phys.\ Lett.\  A {\bf 12}, 2567 (1997).
\bibitem{meisinger}
  P.~N.~Meisinger and M.~C.~Ogilvie,
  \textit{Complete high temperature expansions for one-loop finite temperature
  effects},
  Phys.\ Rev.\  D {\bf 65}, 056013 (2002).
\bibitem{mellin}
  D.~J.~Bedingham,
  \textit{Dimensional regularization and Mellin summation in high-temperature
  calculations}, arXiv: hep-ph/0011012.
\bibitem{miransky-old}
  V.~A.~Miransky,
  \textit{On the generating functional for proper vertices of local composite
  operators in theories with dynamical symmetry breaking},
  Int.\ J.\ Mod.\ Phys.\  A {\bf 8}, 135 (1993).
\bibitem{de-groot}
  S.~R.~ de Groot, \textit{The Maxwell equations: Non-relativistic
  and relativistic derivations from electron theory}, North-Holland
  Pub. Co., Amsterdam (1969).
\bibitem{sachdev}
  S.~A.~Hartnoll, P.~K.~Kovtun, M.~Muller and S.~Sachdev,
  \textit{Theory of the Nernst effect near quantum phase transitions in condensed
  matter, and in dyonic black holes},
  Phys.\ Rev.\  B {\bf 76}, 144502 (2007), arXiv:0706.3215 [cond-mat.str-el].
\bibitem{buchel}
  E.~I.~Buchbinder, A.~Buchel and S.~E.~Vazquez,
  \textit{Sound Waves in (2+1) Dimensional Holographic Magnetic Fluids},
  JHEP {\bf 0812}, 090 (2008), arXiv: 0810.4094 [hep-th].
\par
  E.~I.~Buchbinder and A.~Buchel,
  \textit{The Fate of the Sound and Diffusion in Holographic Magnetic Field},
  arXiv: 0811.4325 [hep-th].
\par
  E.~I.~Buchbinder and A.~Buchel,
  \textit{Relativistic Conformal Magneto-Hydrodynamics from Holography},
  arXiv:0902.3170 [hep-th].
\bibitem{schwinger}
  J.~S.~Schwinger, \textit{On gauge invariance and vacuum
  polarization}, Phys.\ Rev.\  {\bf 82}, 664 (1951).
\bibitem{miransky-1}
  V.~P.~Gusynin, V.~A.~Miransky and I.~A.~Shovkovy,
  \textit{Dynamical flavor symmetry breaking by a magnetic field in
  (2+1)-dimensions},
  Phys.\ Rev.\  D {\bf 52}, 4718 (1995).
\bibitem{sato}
  S.~Kanemura, H.~T.~Sato and H.~Tochimura,
  \textit{Thermodynamic Gross-Neveu model under constant electromagnetic field},
  Nucl.\ Phys.\  B {\bf 517}, 567 (1998).
\bibitem{costa}
  P.~Costa, M.~C.~Ruivo and C.~A.~de Sousa,
  \textit{Thermodynamics and critical behavior in the Nambu-Jona-Lasinio model of
  QCD},
  Phys.\ Rev.\  D {\bf 77}, 096001 (2008), arXiv:0801.3417 [hep-ph].
\bibitem{shovkovy}
  J.~L.~Noronha and I.~A.~Shovkovy,
  \textit{Color-flavor locked superconductor in a magnetic field},
  Phys.\ Rev.\  D {\bf 76}, 105030 (2007), arXiv:0708.0307 [hep-ph].
\bibitem{son-superfluid}
  C.~P.~Herzog, P.~K.~Kovtun and D.~T.~Son,
  \textit{Holographic model of superfluidity},
  arXiv:0809.4870 [hep-th].
\bibitem{lattice-sound}
  Y.~Aoki, Z.~Fodor, S.~D.~Katz and K.~K.~Szabo,
  \textit{The equation of state in lattice QCD: With physical quark masses towards
  the continuum limit},
  JHEP {\bf 0601}, 089 (2006), arXiv:hep-lat/0510084.
\bibitem{ayala}
  A.~Sanchez, A.~Ayala and G.~Piccinelli,
  \textit{Effective potential at finite temperature in a constant hypermagnetic
  field: Ring diagrams in the standard model},
  Phys.\ Rev.\  D {\bf 75}, 043004 (2007), arXiv: hep-th/0611337.
\par
  A.~Ayala, G.~Piccinelli, A.~Sanchez and M.~E.~Tejeda-Yeomans,
  \textit{Feynman parametrization and Mellin summation at finite temperature},
  Phys.\ Rev.\  D {\bf 78}, 096001 (2008), arXiv: 0804.3414 [hep-ph].
\bibitem{stephanov}
  P.~M.~Hohler and M.~A.~Stephanov,
  \textit{Holography and the speed of sound at high temperatures},
  arXiv: 0905.0900 [hep-th].
\bibitem{cohen}
  A.~Cherman, T.~D.~Cohen and A.~Nellore,
  \textit{A bound on the speed of sound from holography},
  arXiv: 0905.0903 [hep-th].
\bibitem{peskin}
  M.~E.~Peskin and D.~V.~Schroeder,
  \textit{An Introduction To Quantum Field Theory}, Reading, USA: Addison-Wesley
  (1995).
\bibitem{schwabl}
  F.~Schwabl, \textit{Statistical Mechanics}, Second Edition,
  Springer Verlag, Berlin, Heidelberg, New York (2006).
\end{thebibliography}
\end{document}